Mestrado em Engenharia Informática
Dissertação
Relatório Final

# Computational Model Of Axon Guidance

Rui André Ponte Costa
racosta@student.dei.uc.pt

Orientadores:
Dr. Luís Macedo, Dr. João Malva e Dr. Ernesto Costa
Data: 10 de Julho de 2009

FCTUC DEPARTAMENTO
DE ENGENHARIA INFORMÁTICA
FACULDADE DE CIÊNCIAS E TECNOLOGIA
UNIVERSIDADE DE COIMBRA

# Abstract

Axon guidance (AG) towards their target during embryogenesis or after injury is an important issue in the development of neuronal networks. During their growth, axons often face complex decisions that are difficult to understand when observing just a small part of the problem. In this work we propose a computational model of AG based on activity-independent mechanisms that takes into account the most important aspects of AG. This model may lead to a better understanding of the AG problem in several systems (e.g. midline, optic pathway, olfactory system) as well as the general mechanisms involved.

The computational model that we propose is strongly based on the experimental evidences available from Neuroscience studies, and has a three-dimensional representation. The model includes the main elements (neurons, with soma, axon and growth cone; glial cells acting as guideposts) and mechanisms (attraction/repulsion guidance cues, growth cone adaptation, tissue-gradient intersections, axonal transport, changes in the growth cone complexity and a range of responses for each receptor).

The growth cone guidance is defined as a function that maps the receptor activation by ligands into a repulsive or attractive force. This force is then converted into a turning angle using spherical coordinates. A regulatory network between the receptors and the intracellular proteins is considered, leading to more complex and realistic behaviors. The ligand diffusion through the extracellular environment is modeled with linear or exponential functions.

Based on this model we developed yArbor, the first AG simulator, which can be used in the future by neuroscientists interested in a better comprehension of this phenomenon.

Concerning experimentation, it was developed the first computational model and a new theoretical model of the midline crossing of Drosophila axons that focus all the decision points. The computational model created allows describing to a great extent the behaviors that have been reported in the literature, for three different pathfinding scenarios: (i) normal, (ii) comm mutant and (iii) robo mutant. Moreover, this model suggests new hypotheses, being the most relevant the existence of an inhibitory link between the DCC receptor and the Comm protein that is Netrin-mediated or mediated by a third unkown signal.

The following parameters are considered as evaluation measures: (i) decisions on critical points, (ii) the concentrations and activities of receptors and proteins, (iii) the total axon length and number of simulations, (iv) the final topographic map and finally (v) the visual results in three dimensions. These metrics are then compared with experimental evidences.

In conclusion, in our approach, AG is an emergent behavior of the system as a whole, with realistic rules and elements that together could lead to the behaviors observed in Neurobiology experimental studies.

# Contents

# Chapter 1

# Introduction

## 1.1 Motivation

There are over a trillion cells (neurons) in human brain that make connections with, on average, over a thousand target neurons [87]. How these connections (made by axons) are generated is one of the greatest questions in neuroscience that are not yet fully understood. Apparently the problem is so simple as to know how axons travel from one source neuron to its target neuron (axon guidance problem, Figure 1.1). This phenomenon was discovered more than a hundred years ago by the famous Spanish neuroanatomist, Ramón y Cajal [98]. Despite the apparent simplicity of this question, it is a difficult problem. It is known that there are thousands of actors (e.g. Axons, Neurons, Glia cells, Guideposts, tissues, etc. [74]) and that these actors interact with each other using signaling molecules (i.e. guidance cues [87, 12]), constituting a very complex system in which interesting behaviors can be found (e.g. cooperation (fasciculation) [40], adaptation [74] and branching [62]). The study of axon guidance (AG, also called axon pathfinding) is made difficulty by the time required for *in vivo*/*vitro* experiments, ethic aspects and by problems arising from the interpretation of non-controlled side effects [62].

The precise patterns of connections found in the nervous system are formed using a set of cues, some of which are molecular (activity-independent) while others depend directly on the neural activity (activity-dependent) [34]. It has been shown that the activity-independent mechanisms are more relevant for axon guidance.

The answer to the AG problem is of high importance because it could lead to new insights in neuronal regeneration, which is relevant to the improvement of the life quality of millions of people over the world. As referred in [62], it is estimated that in United States about 200 000 people live with disabling spinal cord injury, with approximately 30 added to this number every day. This is just an example, considering that there are much more diseases related with this issue, such as Parkinson, Alzheimer and Epilepsy [42]. As suggested by [62, 67, 17] combined computational-experimental collaborations can potentially make inroads that neither could achieve alone. In the words of Robert Millikan [65], "Science walks forward on two feet, namely theory and experiment. Sometimes it is one foot which is put forward first, sometimes the other, but continuous progress is only made by the use of both." The comprehension of this problem brings clear bene-fits for neuroscience, but it can also inspire the creation of new algorithms for routing [15], autonomous wiring, machine learning or even a better comprehension of physics phenomena where attractive and repulsive forces are present (e.g. magnetism).

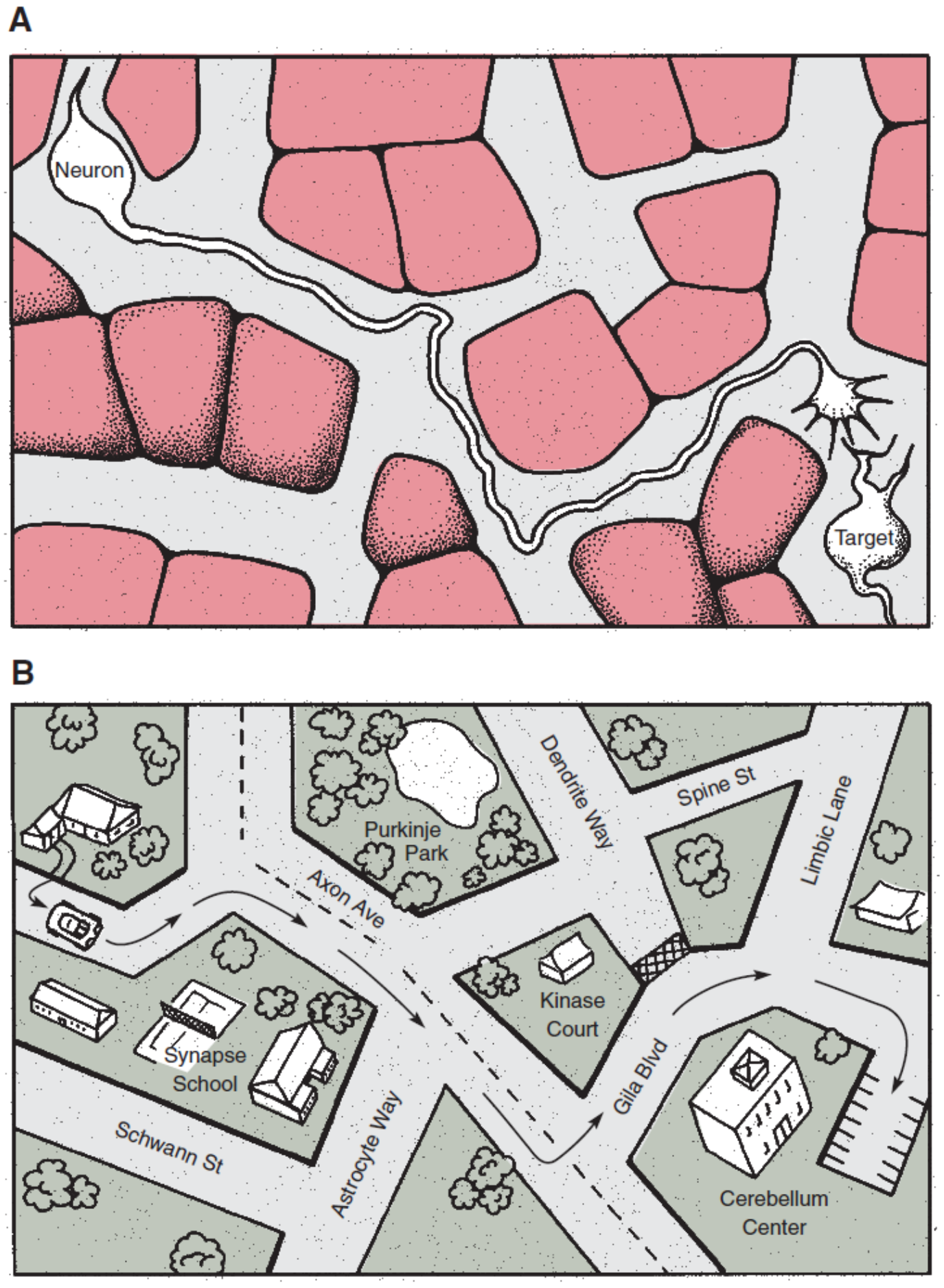

Figure 1.1: Metaphor between axon guidance (A) and a driver navigating in the streets (B) [74]. Both have a distant target, must navigate through a complex environment and perform difficult decisions. Most humans make mistakes during this journey, but axons do not under normal physiological conditions. How can axons accomplish such complex tasks so accuratly?

The detection and response to a chemical gradient is ubiquitous in biology (e.g. reproductive cells, cell migration, immune response, worms movement and predator-prey relationships in some species). Therefore this problem is of high importance in several other systems and a better comprehension in one can lead to new insights in the others. One example of this ubiquity is given by a study that established an analogy between the chemotaxis in AG and the chemotaxis in neutrophils, a type of white blood cell [67]. In this case the whole cell moves in response to a molecular gradient, rather than just the growth cone, but their behaviors are in many ways similar. Both seem to have three main components: directional sensing, motility and polarisation. This analogy provides some clues towards a better comprehension of this phenomenon. Furthermore in [86] a comparison between the guidance mechanisms and molecules used during neuron migrarion and AG is drawn. Neuron migration is the process by which the neurons migrate from their birthplace to their correct positions. There are in AG and neuron migrarion four basic guidance principles (contact attraction, chemoattraction, contact repulsion and chemorepulsion), similar guidance cues and the same diverse guidance responses based on these guidance cues.

From a more macroscopic level in nature we can find similar phenomena in other fields, such as in plant development. The sea rocket is a plant that can distinguish between plants that are related to it and those that are not , giving a good or bad treatment to their family or enimies (see a movie and details in [1]). This behavior shows that plants act in a similar way when compared with AG because they are attracted and repelled. In our daily life we can find several similar examples, such as cell phones connection to antennas based on the location (functioning the antennas as attractive gradients) or the typical problems that we face every day when trying to reach somewhere (target) using guideposts (intermediate targets), see Figure 1.1. For all these different areas the understanding of this complex journey performed by the axon could be inspiring.

## 1.2 Contributions

In this thesis we introduce a computational model of AG that is the basis for yArbor, the first AG simulator. The computational model is based on the main elements and on activity-independent mechanisms found in experimental studies. This model lead us to develop a three-dimensional simulator with a user-friendly and flexible graphical interface. In order to evaluate and validate our model two sets of experiments are presented: the first concerns simple studies and the second the AG in the *Drosophila* midline. Based on the second experimentation we introduce the first computational mode of AG in *Drosophila* midline that focus all the decision points, which led us to propose a new theoretical explanation by introducing new hypotheses.

## 1.3 Organization

The organization of this thesis is as follows. The next chapter gives an overview of the current neurobiological knowledge in AG. Chapter 3 analyses the state of the art concerning AG modeling. In chapter 4 the computational model is presented followed by a description of the simulator in chapter 5. Finally, in chapter 6 we address and discuss the experimentation and conclude the thesis in chapter 7.

[1]http://www.nytimes.com/2008/06/10/science/10plant.html

## Chapter 2

# Neurobiological knowledge

During embryogenesis or after injury axons need to be guided from the soma of the source neuron (represented by the blue ball in Figure 2.1) to the dendrites of target neurons (represented by the green ball in Figure 2.1). In this phase a particular and dynamic structure exists in the growth part of the axon, known as growth cone (GC, see Figure2.2). This very peculiar structure contains several picks (filopodia) linked by lamellipodia; both are actin-rich structures and their dynamics are driven by F-actin polymerization and disassembly (Figure 2.2). Microtubules grow from the axon shaft into the GC body to anchor stable filopodia and extend the axon (Figure 2.2). Extracellular guidance cues bias growth GC dynamics to stabilize more filopodia on one side, steering the axon towards local gradients [20, 75].

In axon guidance there are several substrates to which GCs are exposed: the cell membranes of other neurons, glia cells, undifferentiated neuroepithelial cells and a variety of axons traveling in different directions, extracellular matrix proteins, glycoproteins, cell adhesion molecules and various diffusible growth-promoting or growth-inhibiting factors [74]. However, in this work only the elements thought to be more relevant are contemplated. During this chapter all these elements are introduced and briefly explained.

## 2.1 Neuron

The neurons are the most relevant cells in axon guidance (Figure 2.3). Neuronal axons grow by sensing the extracellular environment; however there are evidences showing an influence of neurons in the environment in which they navigate towards their targets, i.e. neurons seem to have a pro-active role [14]. Neurons are highly polarized cells, containing dentrites, a cell body (soma) where the nucleus is found and axon(s). These cellular compartments are filled with intracellular structures, including organelles and soluble proteins. The axons of some neurons are wrapped by specialized glial cells, forming the myelin sheets. In axon guidance the most important elements are mainly the axons and the dentrites (the final target of axons).

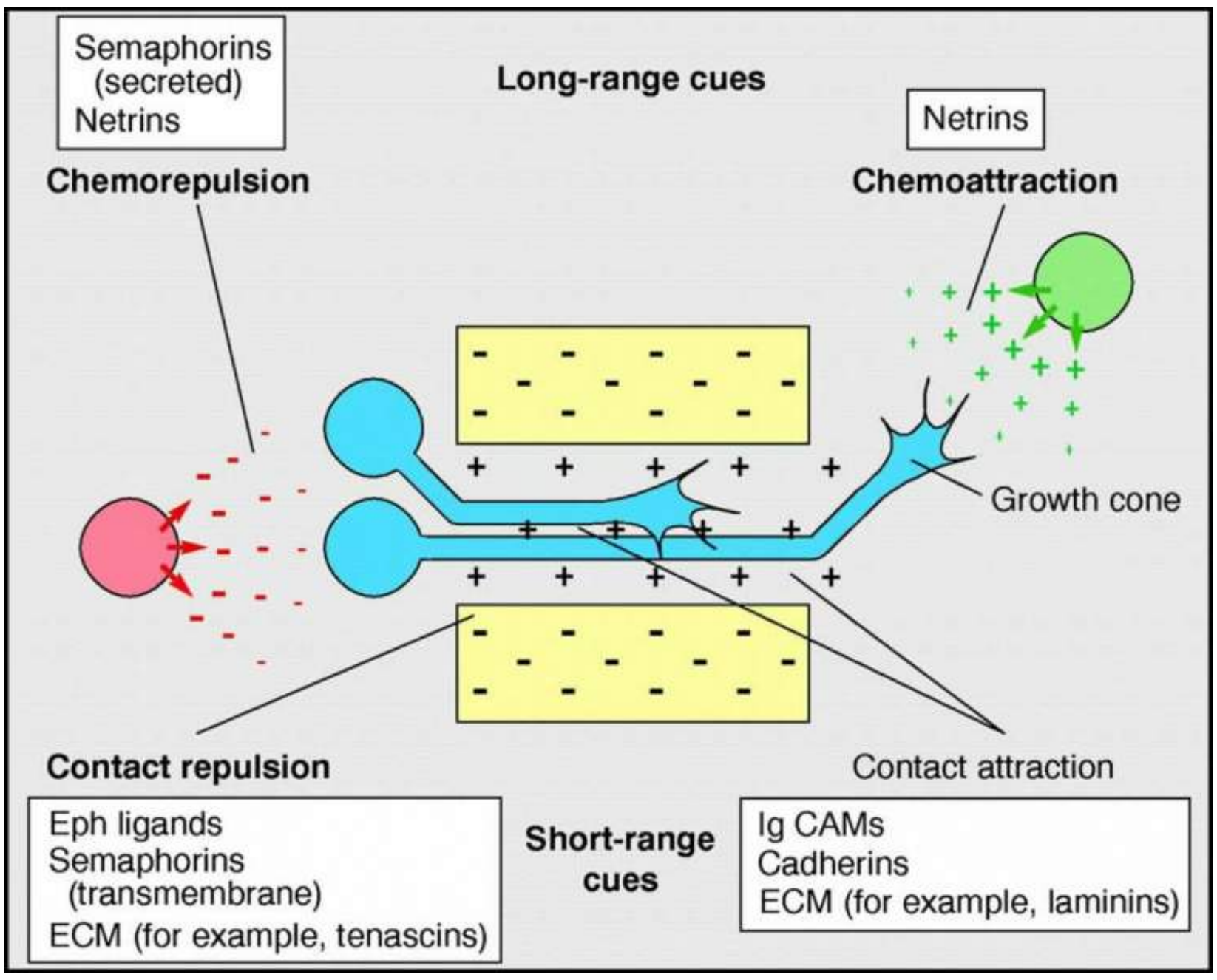

Figure 2.1: Guidance forces [87].

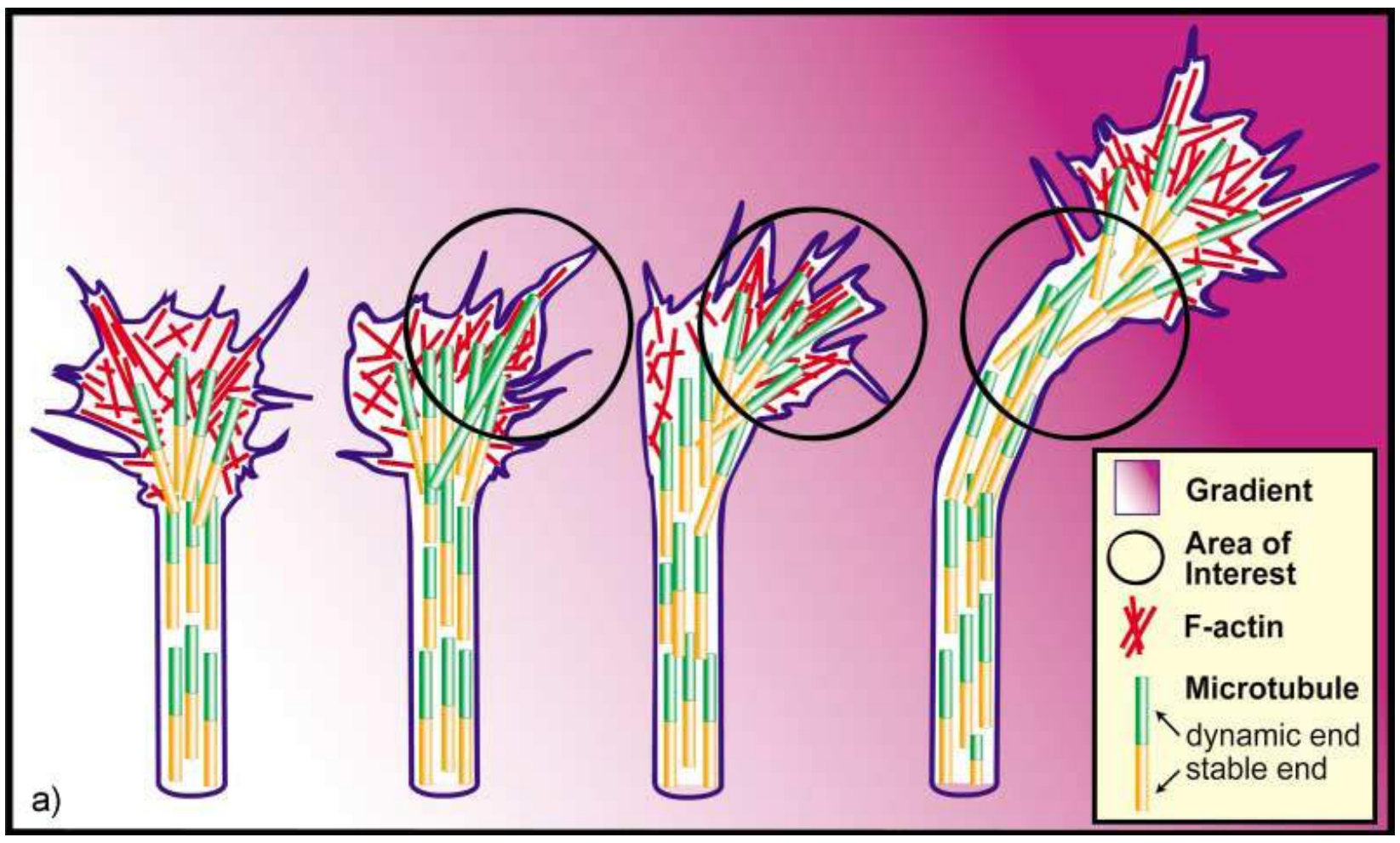

Figure 2.2: Hypothetical model for the cytoskeletal reorganization underlying growth cone turning [20].

### 2.1.1 Dendrites

Dendrites are branches that project from the neuron cell body; these structures conduct to the soma the electrochemical signal generated upon stimulation by neighbor neurons. The stimulation by upstream neurons is received by dendrites via synapses that are located throughout the dendritic arbor [47]. At the end of the axon guidance process the axons must connect directly to the dendrites (axodendritic synapses), to the soma (axosomatic synapses) or to other axons (axo-axonic synapses).

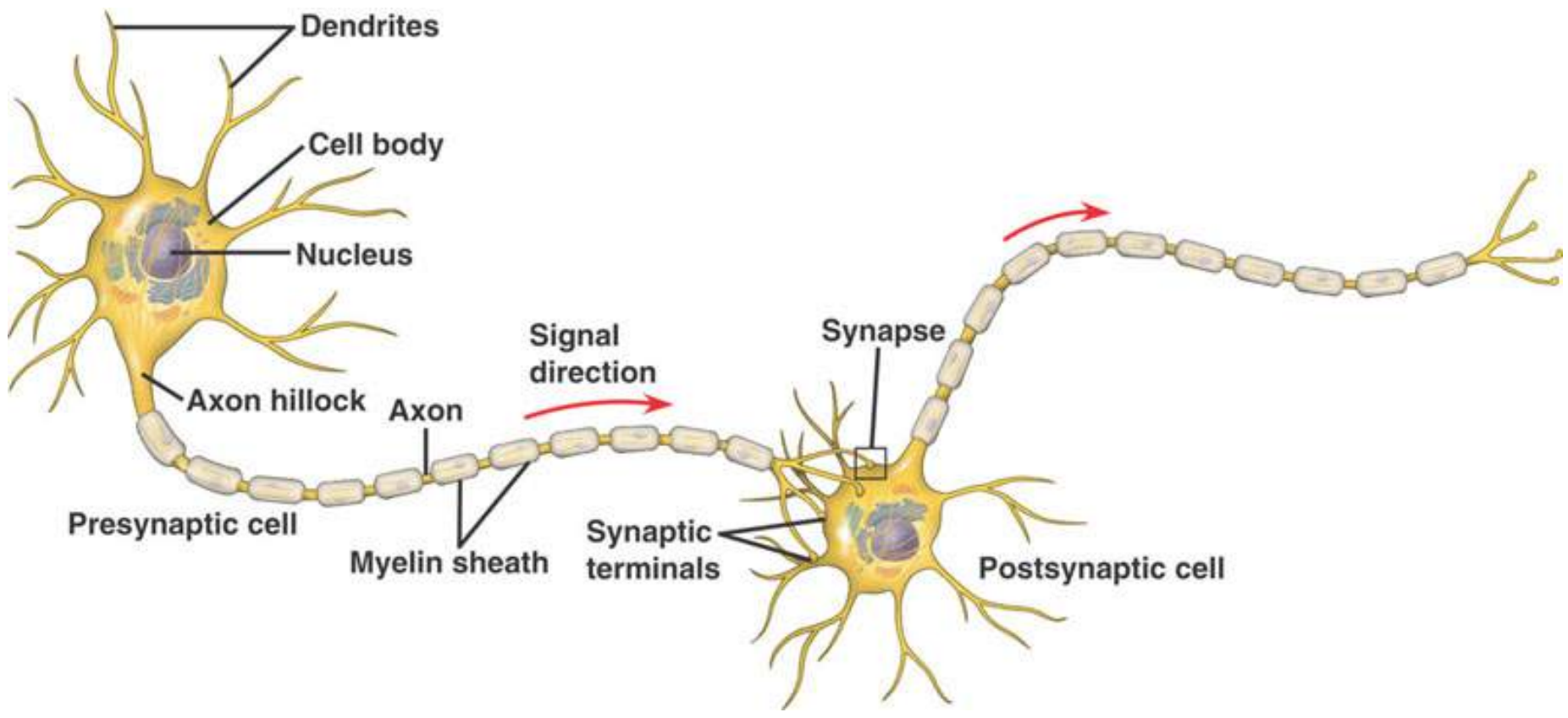

Figure 2.3: Typical neuron structure.

### 2.1.2 Cell body (Soma)

The cell body (also known as soma) is the central structure of every neuron because it contains all the main organelles, including the nucleus (where the genetic code is located), endoplasmic reticulum, the golgi complex (where the proteins are processed) and mitochondria [47]. The size of the soma can range from about 3 micrometers to over 1 millimeter. The soma is one of the most relevant structures in axon guidance because it contains all the information (genetic code) and machinery that define how the axon and GC should interpret the guidance cues. It is the cell body that determines most of the intracellular aspects during axon guidance. This is done for instance by producing new receptors and, other proteins that are needed at the GC.

### 2.1.3 Axon

The axon is responsible for transporting the electrochemical signal from the soma towards the nerve ending, where the release of neurotransmitters propagates the information to the neighbor cell or to the GC during the axonal development. Sometimes the axons are one meter long therefore the transportation process must be clever and effective [47].

Axons contain microtubule-associated motor proteins that transport protein vesicles containing proteins between the soma and the synapses at the axon terminals, and in the opposite direction. Such transport of molecules towards and away from the soma maintains critical cell functions, such as receptor endocytosis (Section 2.4.3) and protein recycling.

#### Growth Cone

The GC is the main element in axon guidance. During neuronal development it has an hand-like structure that constantly perceives the external environment. After reaching nearby its target the GC connects the axon to the soma or the dendrites of the target neuron.

Growth cones are specialized sensory-motor structures at the extending axonal tip [35]. They are highly motile, constantly changing their morphology by ”extending or retracting finger-like filopodia” in order to perceive their surrounding environment. Growth cone motility is driven by the coordinated regulation of cytoskeletal dynamics, specifically actin and tubulin, and substrate adhesion (see Figure 2.2).

An important structure of the GC is its membrane where the receptors are integrated and bind extracellular ligands. The distribution, the type and concentration of these receptors are major players in determining the GC behavior.

Most of the intracellular activity of the GC remains unclear. However in the recent years the number of findings has been growing. $Ca^{2+}$ (Calcium) plays an important role in the GC guidance [28]. Elevation of $Ca^{2+}$ has been associated with the slow down, stop or retraction of GCs, while its reduction typically promotes outgrowth of the axon (negative regulation). Studies support the idea that the GC is attracted when affected by modest elevations of $Ca^{2+}$, i.e. the GC just has the ability to respond in a specific range of concentration. Other intracellular regulatory components are the cyclic nucleotides [74, 4], more specifically the cAMP that have been related with the axon choices.

### 2.1.4 Source

In this work we define the source neuron as the neuron where the axon guidance journey begins. It is from its soma that the axon starts this amazing growing process. The genetic composition of this cell defines how the axon will respond to the cues perceived through its journey.

### 2.1.5 Target

The target neurons are the final target of the axons. Each target neuron contains several dendrites with which axons form synapses (the communication interface between neurons) [74]. Each branched axon may also form several synapses. Axons pass by several intermediate targets (e.g. glial cells and guideposts) but target neurons are their main objective. Therefore, these targets are the most important diffusion gradients and consequently guidance cues.

How the axons known the path that they must follow in order to establish a connection with their target neuron is not well understood. However, two main components seem to be relevant to this process: (i) environmental (spatial molecular guidance) and (ii) genetic (the ligands that the target neuron express). When the axon is reaching its target it should connect to it, but how this happens is another issue that remains to be elucidated (target innervation problem) [87]. For instance, BDNF is one of the factors that seems to be important for a correct target innervation in Xenopus (i.e. genus of highly aquatic frog) embryos [43]. Experiments focusing on multiple targets can be found in [9]. From a simple perspective the axon connects to its target when it passes nearby and the idea that each target-axon pair have unique labels is refuted by the amount of different labels that would be needed to accomplish this theory. Furthermore, this process of positional information might be encoded in the form of gradients of signaling molecules.

## 2.2 Glial Cells

Glial cells are the most abundant brain cells; There are about 10 glia cells for each neuron in human brain. They are non-neuronal cells that provide support and nutrition, maintain homeostasis (stable and constant condition) and some of them wrap axons forming the myelin sheets. Axon myelination allows faster signal transmission.

Over the last 30 years several studies have revealed important roles of glial cells in axon guidance [4, 69], and showed a cooperation between axons and glia [14]. Guidance molecules such as the chemoattractant Netrin and the chemorepellent Slits are secreted by glia cells [4]. They function as intermediate targets that help axons in fasciculation, targeting and turning. Interestingly, the survival and migration of glial cells are sometimes regulated by the neurons. For instance, neurons use the signalling molecule Neuregulins to regulate Oligodendrocytes (a class of glial cells that form myelin in the central nervous system) [25]. Glia cells are generated in excess, and only the Oligodendrocytes that provide correct enwrapment to axons survive.

In addition to the effects of the interaction between neurons and glia in axon guidance glial cells are also important in other topics, such as aging and disease [59]. In the following subsections the most relevant types of glial cells are introduced.

### 2.2.1 Microglia

Microglia cells They resemble macrophages and perform phagocytosis (cell digestion) in order to protect neurons of the Central Nervous System (CNS). These cells correspond approximately to 15% of the total cells of the CNS and can be found in all areas of the brain and spinal cord. They are smaller than macroglia, have their own movement and are activated and proliferate when the brain is damaged [53].

### 2.2.2 Macroglia

#### Astrocytes

Astrocytes are the most frequent macroglial cells. They link the neurons to blood vessels by multiple projections; therefore astrocytes provide neurons with nutrients (e.g. glucose). Furthermore, they regulate the external chemical environment of neurons by removing excess ions (mainly potassium) and recycling neurotransmitters among other functions [72]. Abnormal accumulation of extracellular potassium can result in epileptic neuronal activity [73].

#### Oligodentrocytes/Schwann cells

This type of cells coat axons with their cell membrane providing a specialized membrane called myelin. This myelin sheath provides electrical insulation to the axons allowing them to propagate electrical signals more efficiently (faster action potential) [6].

## 2.3 Guidance cues

AG is controlled by extracellular receptors and their ligands, which together form the guidance complexes. Several families of guidance complexes have been identified (see the following subsections), and they may have bifunctional effects. For example, Netrins and their receptors DCC/UNC5 can either act to attract or repel axons, depending on the cellular context. The differential effects of guidance complexes may be due to regulation between receptors [37] or by internal factors [12]. In the Figure 2.4, the most important guidance cues and their guidance effects are shown in a simplified diagram.

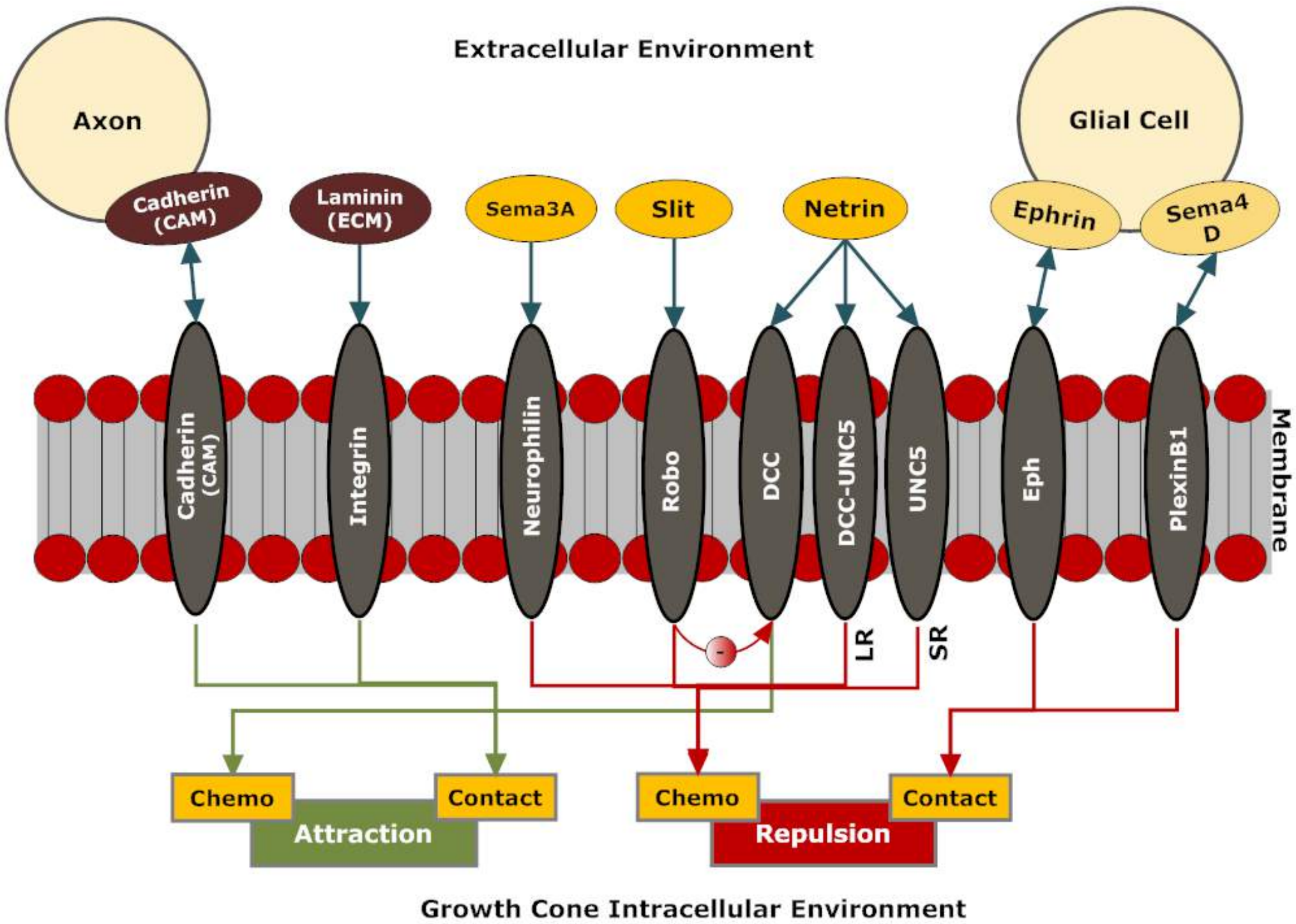

Figure 2.4: Diagram with a simplified representation of the most relevant guidance cues. LR means Long-range and SR Short-Range.

Figure 2.1 illustrates the four forces applied in the axon guidance: contact attraction, chemoattraction, contact repulsion and chemorepulsion. Individual GCs might be pushed from behind by a chemorepellent (red), pulled from afar by a chemoattractant (green), and hemmed in by attractive (gray) and repulsive (yellow) contact cues [87].

### 2.3.1 Traditional

In this section the most studied guidance complexes are introduced.

#### Complex Netrins/DCC-UNC5

Netrins are a conserved family of secreted proteins that guide migrating cells and axonal GCs [21, 12, 4]. There is one type of netrin in worms, two in flies, and at least two in chicks, mice and men. It was initially found in the midline crossing but nowadays it is known that this complex is revelant in many other systems. Their receptors are DCC that leads to attraction, DCC/UNC5 that leads to long range repulsion and UNC5 that leads to short range repulsion (see Figure 2.4). These different responses depend on the quantity of receptors expressed in the GC membrane and on the affinity of the receptors to the ligand.

#### Complex Ephrins/Eph

Both Ephrins and their receptors, Eph, are divided into two families, A and B [12, 4, 68, 93]. This complex is essential to a correct retinotectal topographic map formation (see Section 2.6.2 for more details). There are 14 Eph receptors in vertebrates (7 for each family). Ephrins bind preferentially to their correspondent receptors, i.e. Ephrin-A bind to Eph-A and Ephrin-B bind to Eph-B. This complex functions on a contact basis, because both ligands and receptors are membrane attached. It can produce contactattraction, contactrepulsion and inhibition depending on the context [45]. The Ephrins are frequently expressed in glia cells surface.

#### Complex Slits/Robo

Several studies indicate a conserved role for Slit proteins as repellent guidance cues expressed by glia cells for commissural axons (axons that generally cross the midline) [48, 22]. This ligand is diffusible through the extracellular environment.

Robo (short for roundabout), the receptor for Slit ligands, when removed from embryos, lead to an increase in the number of axons that cross the midline, including both ipsilateral axons that aberrantly cross and commissural axons that recross. These studies demonstrated the importance of this complex in midline crossing as a repellent guidance cue. Robo can bind to the netrin-1 receptor DCC leading to the inhibition of netrin-1 attractive activity (intracellular regulation) [4]. The regulation of Robo expression by Comm protein in *Drosophila* midline crossing is a good example of the existence of internal regulation of receptors by proteins.

#### Complex Semaphorins/Plexins-Neuropilins

Semaphorins represent a large family of more than 20 members divided into eight classes: classes I and II are found in invertebrates, classes III, IV and VII in vertebrates and class V are present in both [4]. Some semaphorins are diffusible (Classes II and III), acting has chemo cues (long range), and others are integrated in the membrane (Classes I, IV, V, VI and VII), acting as contact cues (short range). Most of the semaphorins identified so far mediate axon repulsion, but some of them are bifunctional, attractive or inhibitory. Their main receptors belong to Neurophilin and Plexin families, but several others have been identified.

### 2.3.2 Others

Many of the guidance events observed during development of nervous system do not appear to be governed by the traditional guidance cues. At the same time the low number of molecules involved in these events were few when looking into the immense complexity of the nervous system. Therefore, the development of neuronal networks certainly requires additional guidance cues and receptors that remain to be discovered. Some recent discoveries of new guidance cues are presented in this section.

### Cell adhesion molecules (CAMs)

CAMs play a central role in mediating contact-dependent regulation of axon behavior (contact-attraction) [4]. They can be found on the surface of axons and have been implicated axon fasciculation.

Three main classes of cell adhesion molecules have been identified: integrins, cadherins, and the immunoglobulin superfamily. Many CAMs share the ability to bind homophilically. Homophilic binding means that proteins of the same type bind to each other [74]. Furthermore, it was shown that CAMs can interact with axon guidance receptors.

### Extracellular Matrix (ECM)

The Extracellular Matrix is the extracellular part of tissue that provides structural support to cells and other important functions, including axonal growth [74]. Thus, axonal growth is promoted by some types of ECM, such as laminin (essential in the centrifugal navigation of axons within retina) and fibronectin.

### Morphogens

Morphogens are substances that govern tissue development and, more specifically, the positions of the cell types within a tissue [4]. The three major families are Hedgehog, TGF-$\beta$/BMP and Wnt. Recent studies have demonstrated that morphogens also have a role in axon guidance. During the midline crossing BMP appears to repell commissural axons that are later attracted by Sonic hedgehog (shh) and Netrin-1. The same Shh is present in the optic chiasm where it repels the commissural axons leading to a correct crossing (see Section 2.6.2).

### Neurotransmitters

Neurotransmitters are used for communication between neurons, but recent studies have also demonstrated their function during nervous system development, and more specifically in axon guidance [4]. The neurotransmitters Ach, GABA and glutamate can produce an attractive turning response by the GC [4]. This role of neurotransmitters was reinforced by simulation studies [40].

Some studies reveal that is neurotransmitters influence GCs and which have been implicated as chemoattractants. Moreover simulations made by Hentschel [40] confirm that this is possible.

For instance, some neurotransmitters (ACh, GABA and glutamate) can produce an attractive turning response in the GC. However more studies are needed in order to explain this guidance process.

### Neurotrophins

Neurotrophins play several roles during development, including effects on axonal growth and orientation [4]. They are important for cell survival (trophic effects) and axonal growth, but their role in axon guidance needs further investigation. The two most studied ligands are Nerve Growth Factor (NGF) and Brain-derived neurotrophic factor (BDNF). Their role in axon guidance may arise from the fact that they activate signaling similar to those used by traditional guidance cues. Neurotrophins also play a role in axon branching.

## 2.4 Mechanisms

In this complex system there are many phenomena happening simultaneously. The principal phenomenon is the guidance mechanism, whereby GCs appear to be guided by two opposite mechanisms, repulsion and attraction [87, 12]. This simple behavior in a complex environment allows the axon to reach its target.

### 2.4.1 Axonal

Throughout this section we describe four axonal mechanisms directly related with AG.

#### Turning/Branching/Pruning

There are four axonal steering mechanisms [62] (see Figure 2.5):

1. *veil extension* in which the lamellipodial veil near an attractive cue is extended to support further growth in that direction;
2. *filopodial dilation*, depicting the growth of filopodia in the direction of an attractive cue;
3. the combination of *veil extension* and *filopodial dilation* results in the GC migration in the direction of an attractive cue (or away from a repulsive one);
4. *branching and pruning* happens when for example near a boundary the axon branches itself and after that the inappropriate or nonfunctional branches are pruned away;
5. branching can occur after a neurite has passed a target and this is called *back-branching*.

#### Fasciculation

Growing axons often form fascicles, a process called fasciculation [40]. Contact attraction mediated be molecules (e.g. CAMs) on the surface of the axons has been implicated in fasciculation although is not enough to keep axons together. When the axons reach near the target region they do the inverse behavior, i.e. defasciculate, allowing them to innervate their targets. Various mechanisms have been suggested for these processes:

1. axons come together as a result of their random movements;
2. contact repulsive signals (e.g. semaphorins [87]) from surrounding cells push axons together;
3. axons are attracted/repelled by diffusible molecules that they themselves secrete (e.g. neurotransmitters [44]).

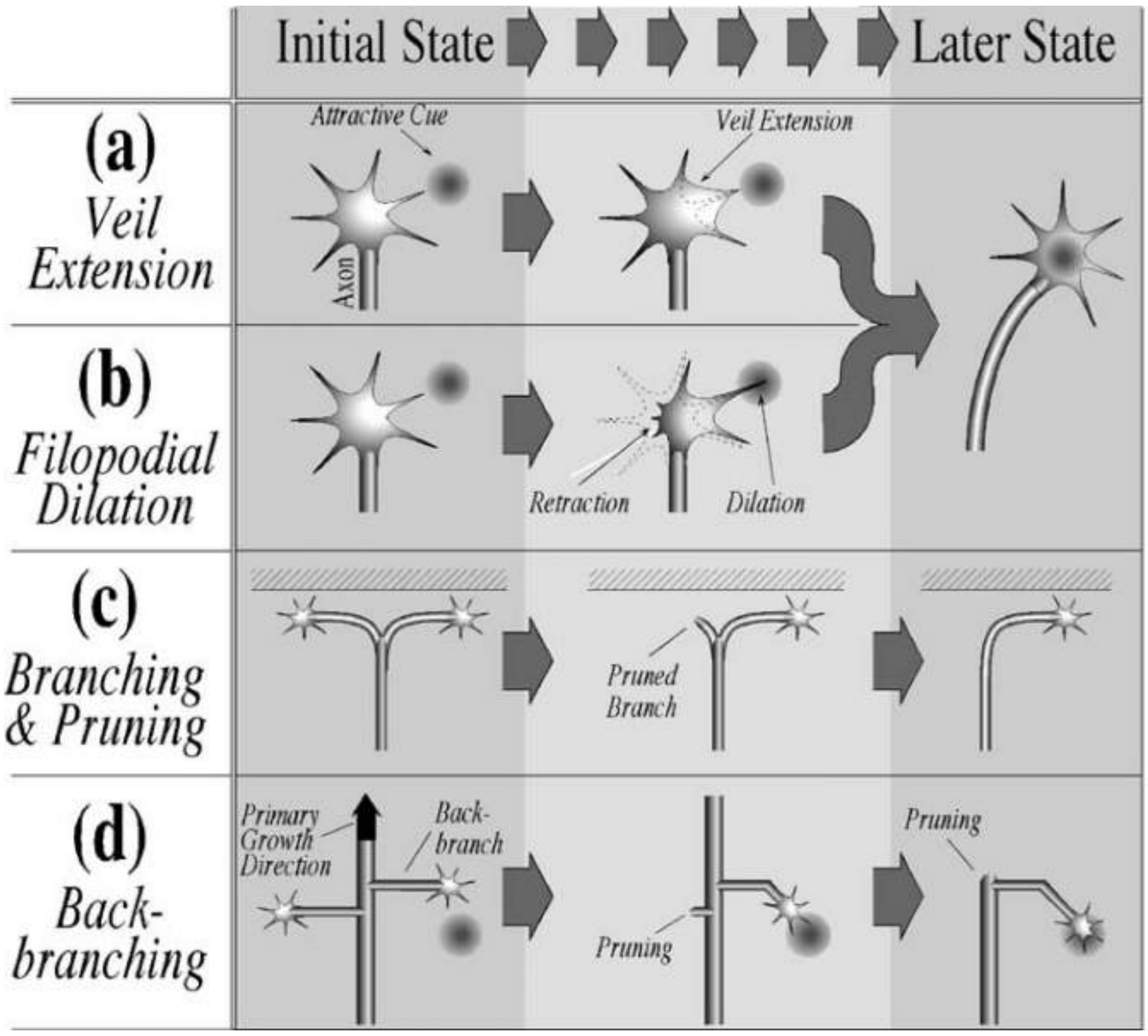

Figure 2.5: Caricatures of GC and axonal steering mechanisms [62].

4. axons are attracted/repelled by contact molecules (e.g. CAMs) or the extracellular-matrix.

This process becomes even more interesting because the axons switch from one fascicle to another at specific choice points. This selective fasciculation simplifies the assembly of large nervous systems (e.g. human) [87].

**Transport**

Axons can not synthesize proteins and are too long for diffusion of products synthesized in soma [94]. Therefore, specialized transport systems exist that move materials along axon. There are two types of transport, fast and slow. The fast transport contains anterograde movement (towards axons terminal) at a velocity of 400 mm/day and retrograde movement (towards soma), 30%-60% slower than anterograde. The slow transport is represented by a anterograde movement with the speed of 0.5-5mm/day. The anterograde transport direct newly sintetized elements (e.g. receptors) to the end tip of the axon, while the retrograde transport moves vesicular structures towards the soma.

Axonal transport is very important for axon elongation, but less important for axon guidance [60]. For axon guidance purposes it is relevant the axonal transport of recep-tors produced in the soma which are later incorporated in the membrane of the GC.

#### Growth cone complexification

GCs become more complex (with more receptors) and slow at points where they must interpret complex guidance information (i.e. choice points with several ligands) [63]. When travelling along a well-defined path they are fast and become (bullet shaped) and are less sensitive to guidance cues [67].

### 2.4.2 Adaptation

Similarly to what happens in the immune system, during axon guidance the GC also adapts to guidance cues that it senses during its journey. The adaptation process is usually associated with a fast desensitization (e.g. 20s) response and a slower resensitization response (e.g. 5m) [70, 55]. This phenomemon of GC desensitization was identified when chick retinal GCs failed to respond to a repulsive cue after being repeatedly presented to it. This was also demonstracted for attractive cues, with axons that moved on from these cues, such as Netrin-1 in the midline. Most of the evidences available show that the GC adapts to background levels of specific cues and after being resensitizated its response to guidance cues becomes different, probably due to a lower number of receptors (due to receptor endocytosis). Has also been demonstrated that the desensitization is independent of the receptor type. The resensitization may be obtained when the number of receptors is enough to unleash a response due to protein synthesis [70].

### 2.4.3 Receptors

The receptors function as the GC sensors; therefore their role is essential for a correct pathfinding. In this section we describe two mechanisms that influence the GC response mediated by the receptors.

#### Endocytosis

This is the process by which the GC receptors are removed from the membrane. Receptor endocytosis might function as a form of global inhibition, whereas local excitation and amplification might be achieved by a switch mechanism in which downstream signalling occurs only when the local density of activated receptors exceeds a specific threshold [67]. This process is essential for GC adaptation and it is usually followed by a slower production of the removed receptors.

#### Threshold/Saturation

There are evidences [67] for receptor activation based on a threshold. Therefore, the GC displays a response only when a certain percentage of its receptors are activated. However, with endocytosis and receptor recycling the average density of surface receptors in the GC tends to be maintained slightly below the threshold, and therefore the GC can be influenced by weak gradients. This range of response is increased when adaptation occurs [70].

It should be noted that if the chemoattractant concentration is either too large or too small, a "real" GC will not be able to sense a concentration gradient [29]. If the chemoattractant concentartion is too large, the receptors become saturated; if it is too small, noise effects become dominant.

### 2.4.4 Attraction/Repulsion

These two behaviors are the most relevant for axon guidance. The four guidance forces (chemoattraction, chemorepulsion, contact attraction and contact repulsion) are the main cues that help the GC to find its target [64]. Some cues can perform both attraction and repulsion depending on the context. The new direction of the GC depends on the receptors activated and the type of responses that these receptors induces. The receptors expressed in a specific GC depend on its initial location and on its guidance history. Besides these two behaviors sometimes guidance cues can lead to GC inhibition and ultimately it can collapse [18].

Polarization is the ability of a cell to arrange key signaling components into persistent and distinct front and rear regions [67]. This mechanism acts as a memory that allows the cell to avoid continuos changes in its behavior.

## 2.5 Hypotheses

In this section we present two of the most important hypotheses regarding AG.

### 2.5.1 Stereotropism

This is one of the most simple hypothesis that helps to explaining pathfinding. Tunnels are formed using the spaces between neuroepithelial cells (subtype of stem cells) or glial cells which act as a street that guide the GC [35, p. 87]. This mechanism lead Singer [80] to propose the Blueprint hypothesis according to which neuroepithelial cells form intercellular spaces or tunnels that help the GCs in their orientation.

### 2.5.2 Imprint-matching

Experimental findings have suggested that retinal axons might grow into the tectum until they have reached a ligand concentration matching that of their site of origin (see Section 2.6.2). This is the imprint-matching concept of retinotectal guidance [87, 58]. However this hypothesis can be applied in several topographic maps in CNS.

## 2.6 Systems

In this section we present the two most studied guidance systems, (i) midline crossing and guidance in the (ii) optic pathway. However, there are many other interesting systems (e.g. olfactory system [13]).

### 2.6.1 The Midline Crossing

The evolution of a bilateral nervous system allowed the development of species with more complex behaviors and, ultimately the evolution of humans [49]. This division is known as midline and the understanding of how and why the axons (not) cross it is of great importance for a deeper comprehension of evolution and development of the animal kingdom.

The guidance of commissural axons in *Drosophila* and vertebrates is one of the best-understood models of axon pathfinding, and the studies using this model have provided a great contribution to the understanding of AG [22]. This is an interesting model because the comissural axons usually only cross the midline once, what is at the first glance strange because the midline seems to only display an attractive behavior. In this system the most studied animals are the *Drosophila* (Figure 2.6 d,e) and *Vertebrates* species (Figure 2.6 a,b,c). This system has commissural (white in Figure 2.6) and ipsilateral (black in Figure 2.6) axons. The first tend to cross the midline while the second remain in the same side. The three main questions that investigators have been addressing are [22]:

1. Why do commissural axons, but not other axons, cross the midline? There are clear indications that ipsilateral axons are repelled by Slit, whereas commissural axons, initially, are not.

2. Why, if the midline is so attractive, do these axons then leave it again?

3. Once they emerge on the opposite side of the midline, why do commissural axons turn longitudinally into specific lateral pathways rather than being guided back into the midline by the same cues that got them there in the first place?

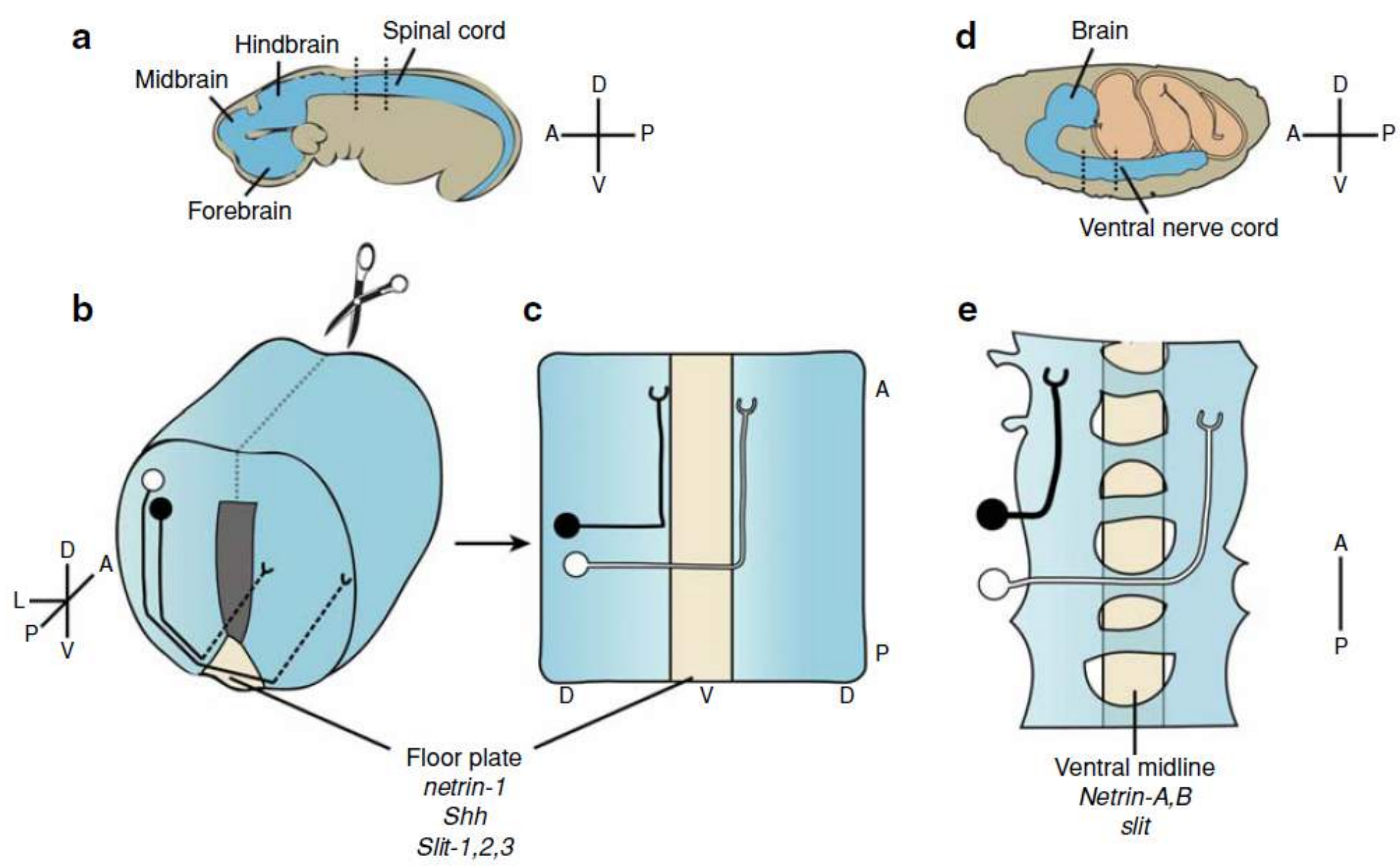

Figure 2.6: Commissural axon pathfinding in the mouse spinal cord (a,b,c) and *Drosophila* ventral nerve cord (d,e) [22].

**Drosophila**

One of the most used animal models to study this topic is the midline crossing by the *Drosophila* axons.

In the *Drosophila* the midline is represented by the ventral nerve cord and crossing of this region by commissural axons requires the lack of surface expression of Robo. (Figure 2.7). The protein Comm blocks the action of the receptors Robo by binding to them (i.e. Comm is an endosomal sorting receptor for Robo) and once the commissural axons crossed the midline, Comm protein levels are reduced and Robo returns to the surface preventing the axons from crossing [22]. The importance of Robo receptors (Figure 2.8 C) and Comm proteins (Figure 2.8 B) has been tested in experiments using mutants lacking these proteins. This system is discussed in more detail in section 6.2.6 considering our simulation results.

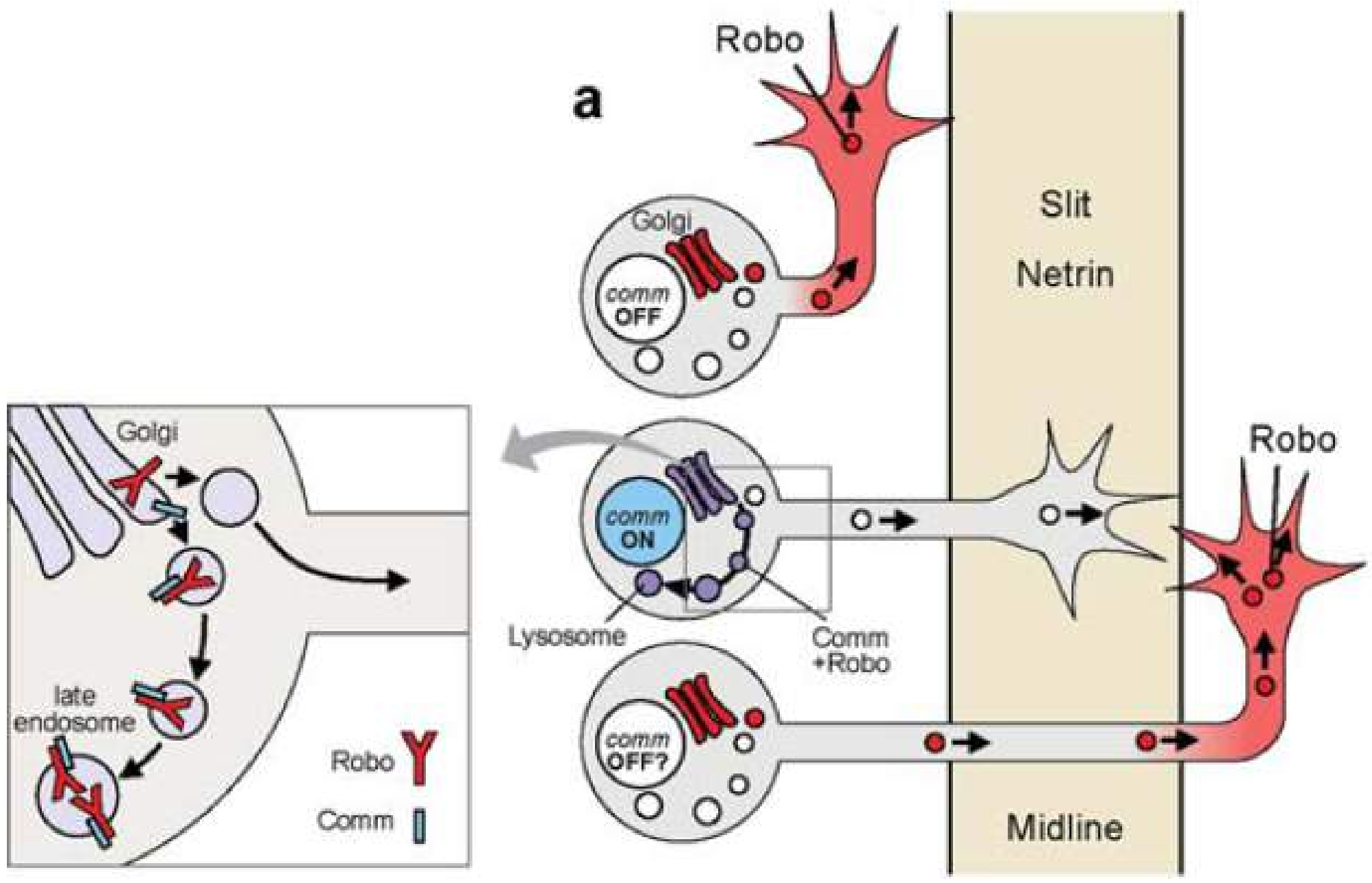

Figure 2.7: Drosophila midline crossing [22].

Two models have been proposed for the regulation made by Comm protein: the "sorting" model, that argues that the neuron by itself blocks the Robo action, using intracellular Comm and the "clearance" model that consider the midline as a source of Comm protein. The sorting model is currently the most accepted [48].

**Vertebrate**

In the Vertebrates (e.g. in mice) a secondary Robo receptor, Robo3, antagonizes Robo1 to allow the midline crossing. When Robo1 is inhibit the commissural axons are attracted by Netrin-1 (Figure 2.9) and sonic hedgehog [22]. Before midline crossing, Robo3 levels are high and Robo1 levels are low. After crossing, Robo3 levels are low and Robo1 levels high; therefore the commissural axons are now repelled by Slit through Robo1. Moreover, attraction to netrin-1 may be downregulated, due to a regulatory link between Robo1 and the netrin receptor, the DCC [82].

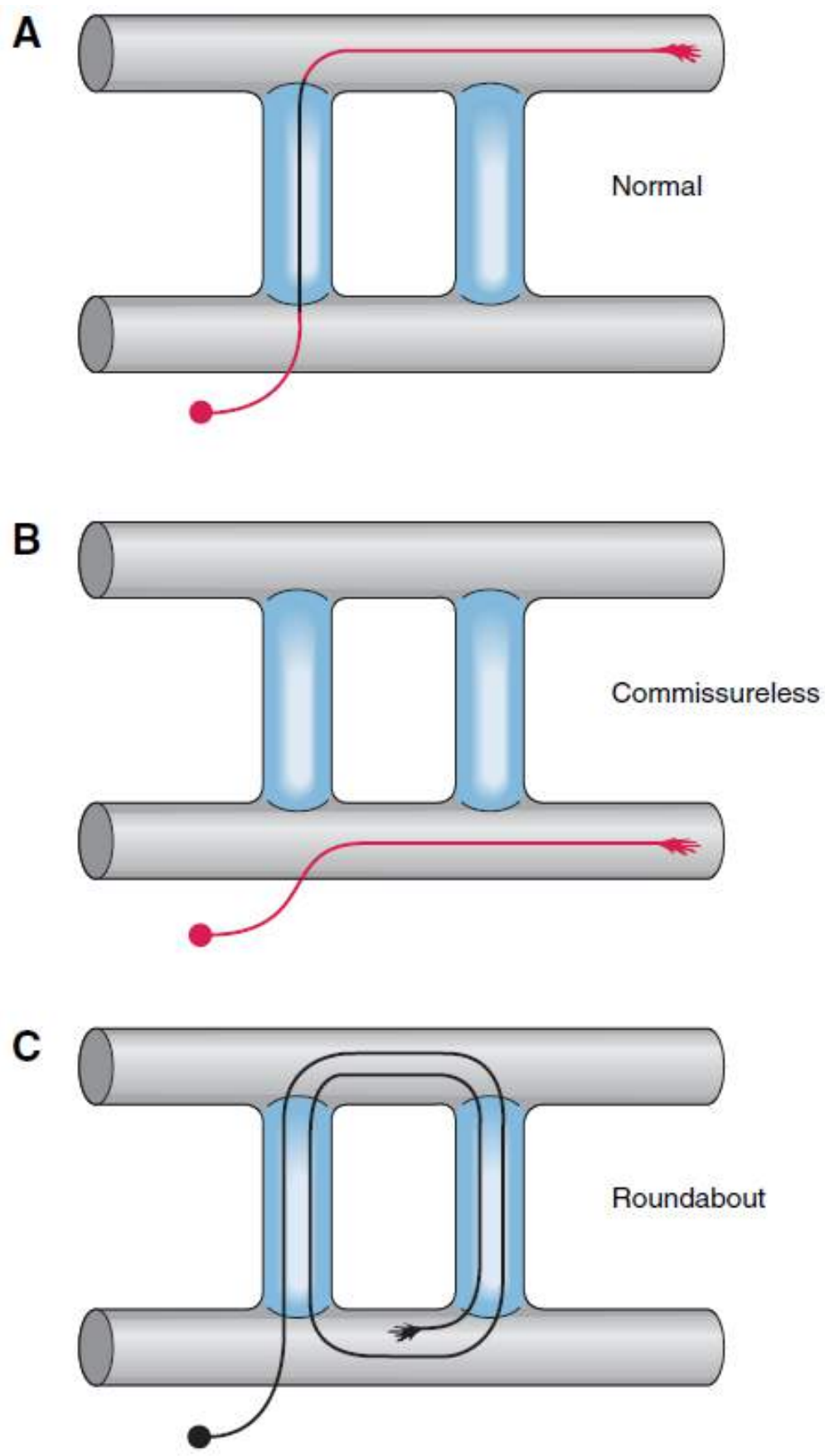

Figure 2.8: Midline crossing in Robo and Comm *Drosophila* mutants [74].

### 2.6.2 The Optic pathway

The optic pathway (from the retina to the tectum) is one of the most complete examples of guidance systems. It contains several kinds of guidance cues that guide the axons from the retina to their targets [74, 23].

The steps made by Retinal Ganglion Cells (RGC, named Ganglion because it is a sensorial cell) are the following (as described in [74, 23] and illustrated in Figures 2.10, 2.11 and 2.12):

1. First, retinal axons travel centrifugally from a peripheral location on the retina towards the optic nerve;
2. When encounter the optic nerve head, they found a high concentration of attractive netrin;

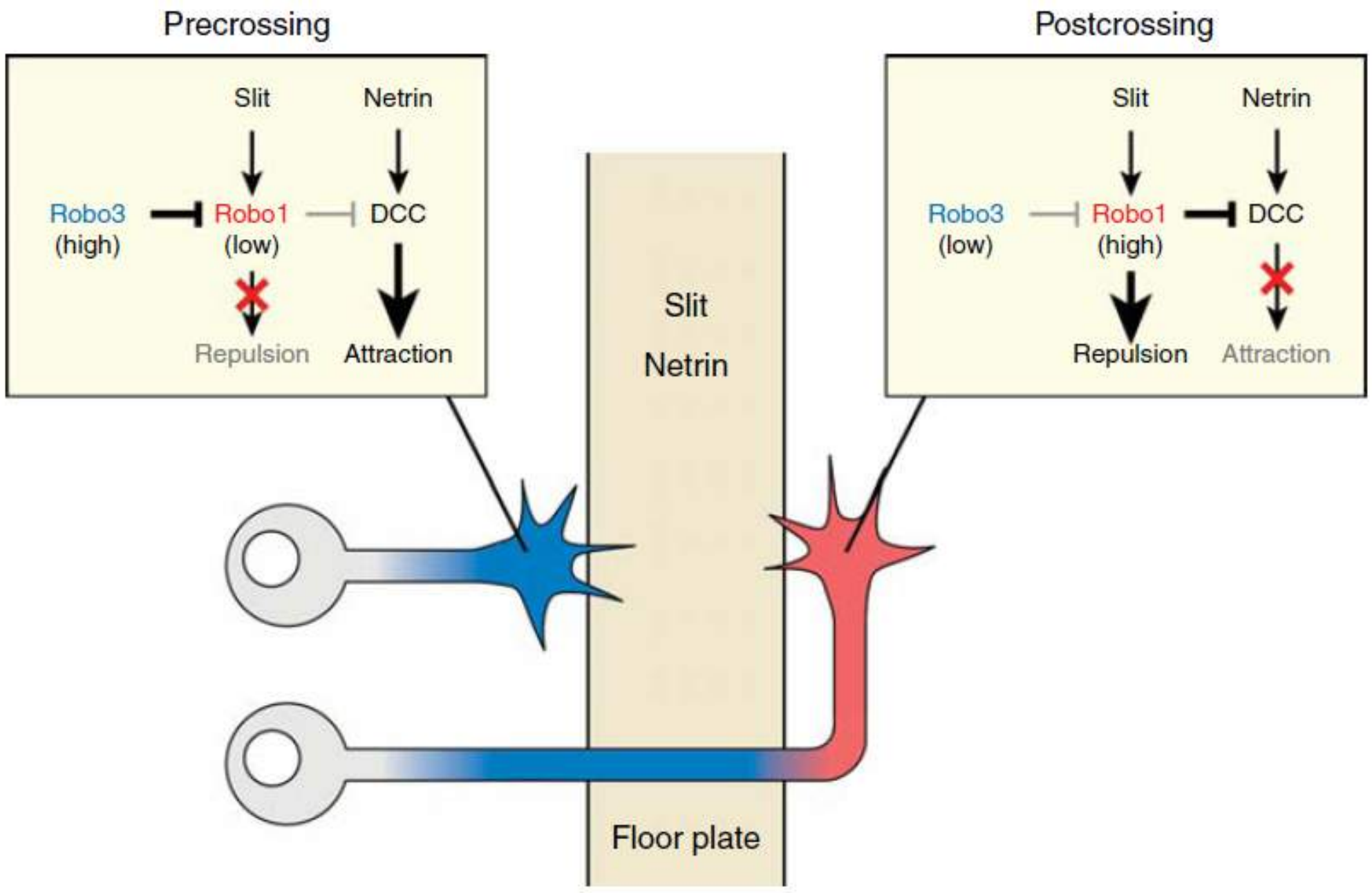

Figure 2.9: Midline crossing of Commissural axon in mice [22].

3. The combination of laminin and netrin is repulsive rather than attractive, which leads the retinal axons to turn away from the retina and grow in optic nerve where they travel until region near the chiasm;

4. Here they found a repulsive guidance molecule (Slit) and the morphogen Shh that also acts as a repellent. Thus, these cues help the retinal axons to grow further;

5. In the chiasm there is a high concentration of another repulsive guidance molecule called EphrinB. Depending on the receptors that the retinal axons express some cross the chiasm towards the optic tract, while others remain ipsilateral.

6. Once in the optic tract, retinal axons are repelled by the repulsive guidance cues (Sema3A and ECM heparan sulfate), allowing them to grow towards the tectum.

7. Finally, at the front of the tectum, the axons encounter a drop in the concentration of FGF (Fibroblast growth factor), which indicates that they have entered the target area. Then, in the target area an orthogonal gradient of EphrinA and EphrinB indicates the correct target region (see Figure 2.12).

This axonal navigation leads to the formation of the retinotectal maps which represent the connections between retina and tectum. This map can be activity-independent (see [34] for a recent review about activity-independent models for retinotectal map development) or activity-dependent [83, 95], but the present work we focus on the former mechanisms.

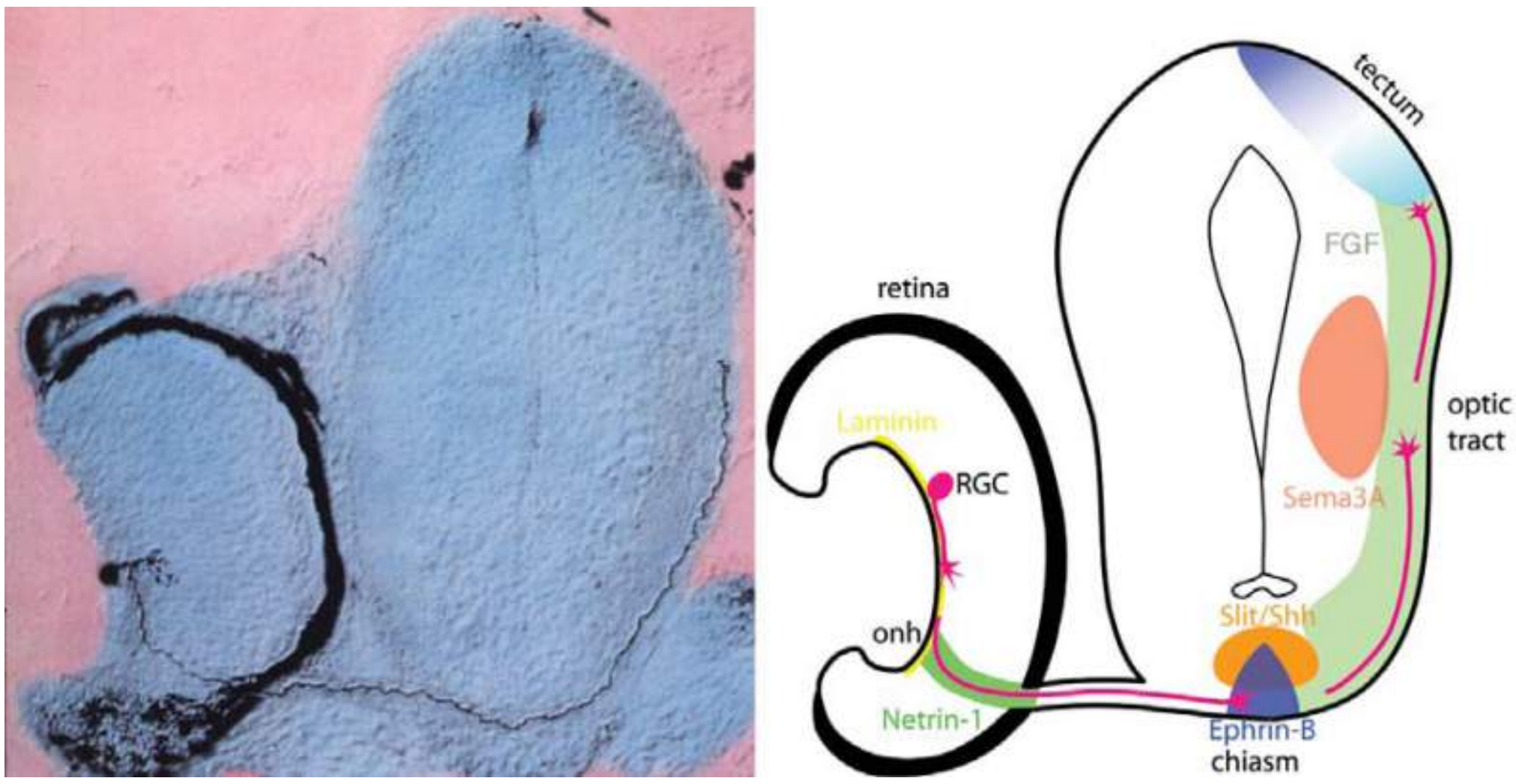

Figure 2.10: Two-dimensional representation of the guidance cues in the optic pathway [74].

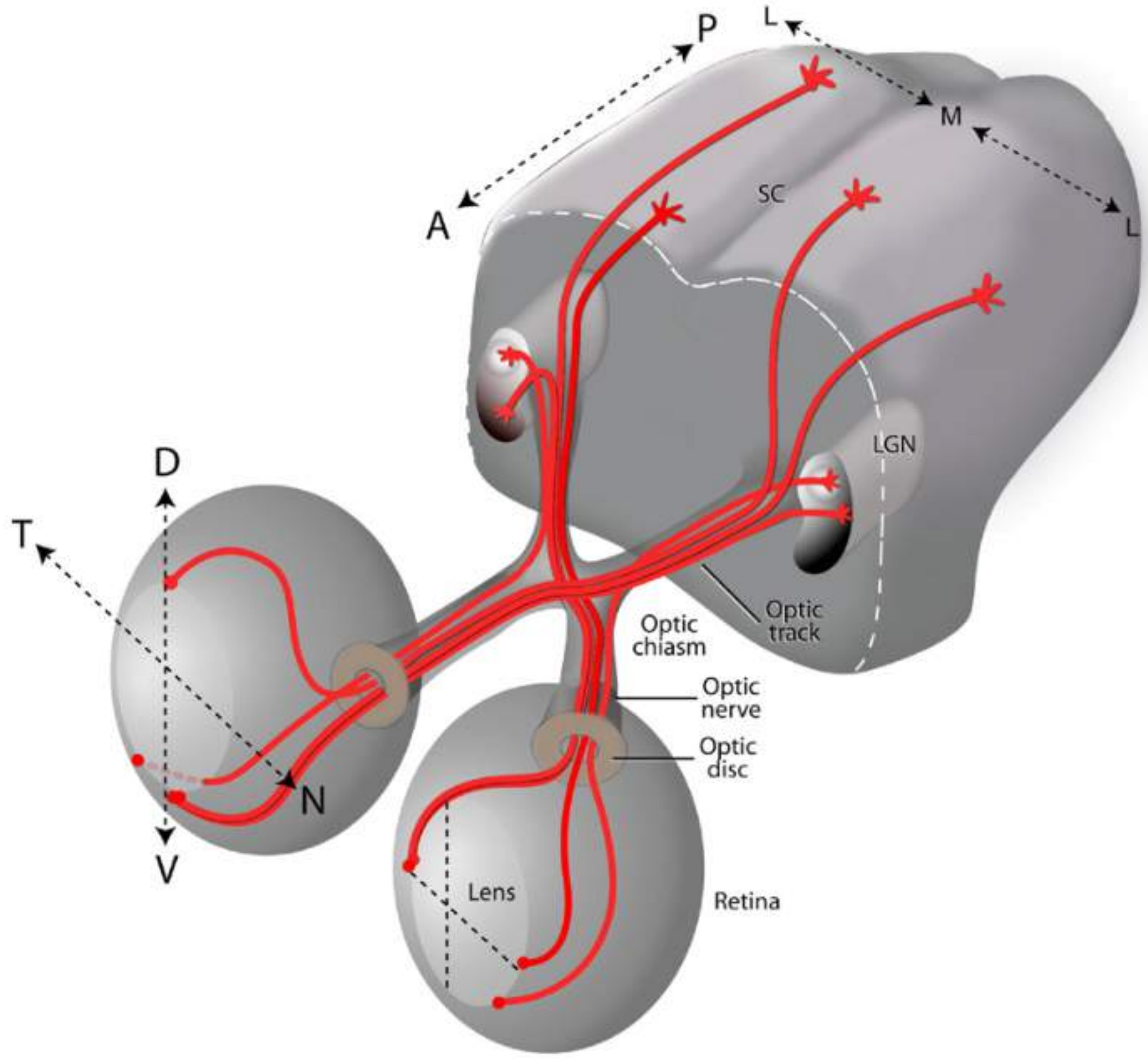

Figure 2.11: 3d model of the optic pathway [23].

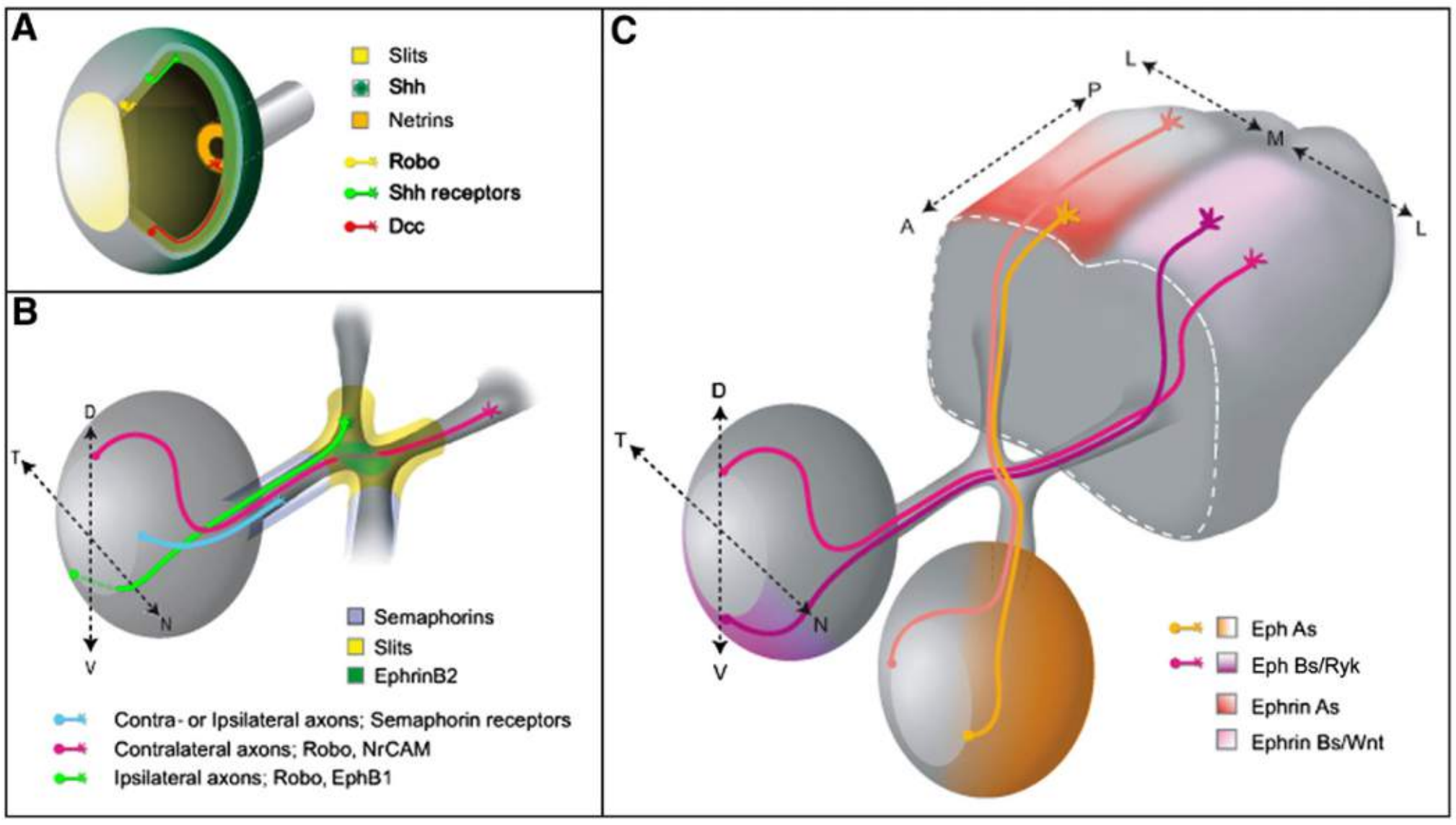

Figure 2.12: 3d model of the optic pathway (with guidance cues) [23].

# Chapter 3

# State of the art

There have been proposed several mathematical and computational models to study AG, which we introduce throughout this chapter. We also give an overview of the most important simulators used in this or similar fields.

## 3.1 Models

A literature review about the mathematical and computational models based on activity-independent principles is given in the next sections.

Comparing to experimental models computational and mathematical models have several advantages [34]. First, they required that assumptions are made explicit. Second, with these models it is easier to figure out the parameters range that are consistent with experimental work. Third, these models bring richer capabilities of predicting future experiments than qualitative models. The greatest drawback is the fact that they are an abstraction of the neurobiology.

Some of the mathematical models introduced here are accompanied with computational simulations. However, their starting point is always the mathematical models. On the other hand we introduce the computational models that start by modeling the problem using a computational approach, even with a mathematical basis.

For other review with a broader scope see [79], where are considered three types of models: axon guidance (that is divided in phenomenological, mechanistic and abstract), retinotectal and activity-dependent.

### 3.1.1 Mathematical

Since one decade ago several mathematical models have been proposed to study different aspects of axon guidance from a quantitative point of view. These works are reviewed in this section.

A mathematical framework for studying axon guidance was proposed by Krottje [54]. This framework allows the implementation of several models that are defined as concentration fields and finite-dimensional state vectors. These state vectors may represent growth cones, target neurons, other cells or impenetrable holes (see Figure 3.1). Several numerical methods are proposed to handle with time integration of this highly nonlinear system , which results from the combination of non-stiff ODEs (ordinary differential equations) and stiff diffusion equations. The last one is minored by modeling concentration fields using quasi-steady-state approximations, which turns the problem into a simplier one where is being only necessary to solve ODEs.

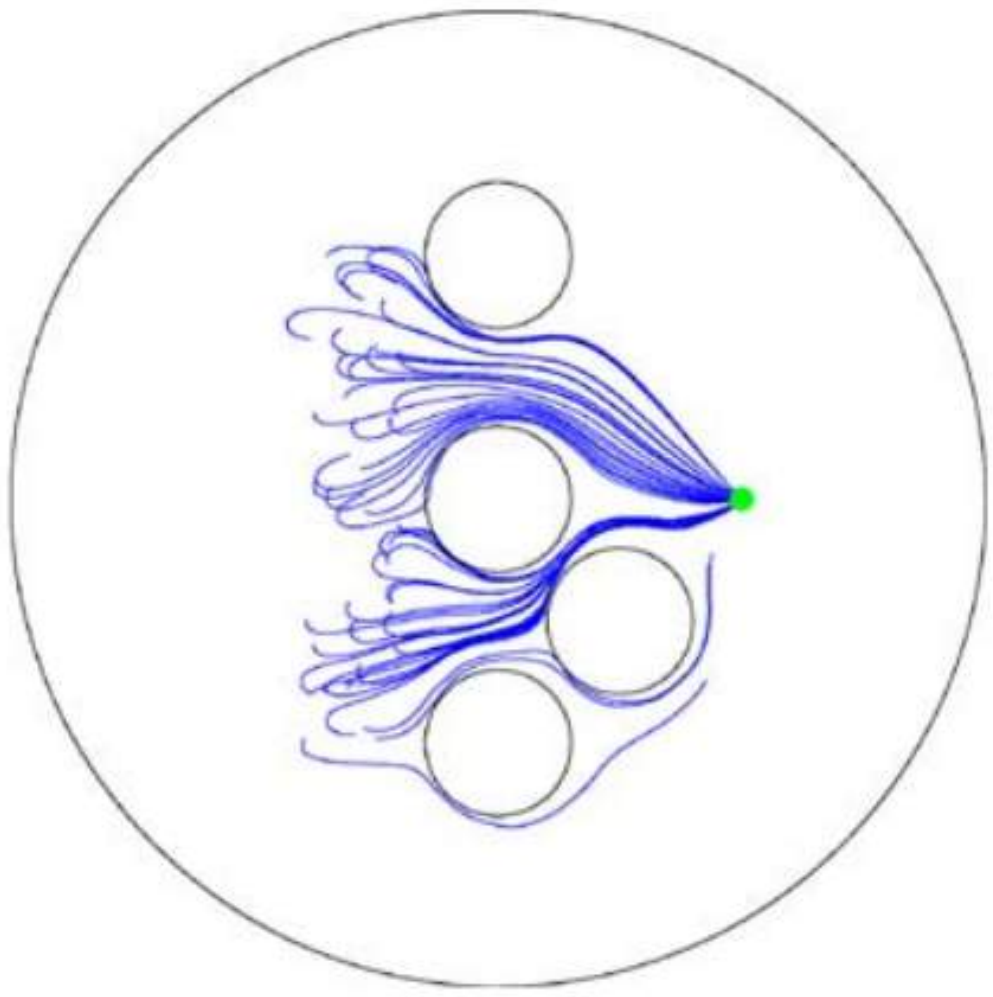

Figure 3.1: Two-dimensional representation of several axon paths with four holes [54].

Hentschel [40] studied the influence of diffusible and contact signals with attractive and repulsive effects on axon (de)fasciculation (see section 2.4.1) and AG towards a target. For this purpose two models were considered, model I, that includes only diffusible signals (long-range signals) and, model II, that includes contact interactions (short-range signals) for the axon-derived signals and diffusible signals from the target. The computer simulations showed that model I seems to be more effective because it: (i) forms well-structured fascicles, (ii) keeps topographical identity and a proper target innervation, (iii) is robust to parameters variations (to a certain extent) and (iv) captures the fasciculation and the defasciculation behavior. Moreover, the simulations also demonstrated that: (i) the diffusion of a target-derived attraction cue is an effective mechanism by which axons and axon fascicles can be guided to their target region; (ii) the outgrowth in too low or too high concentrations is zero; (iii) pathfinding (by individual axons) appears as an emergent property of the dynamics (random axon movements); (iv) contact attractive cues are enough to keep axons together (fasciculation), but random movements and repulsive cues must exist so that they can become together.

The mathematical model proposed by Aletti [3, 2] for the growth cone (GC) transduction chain takes into account the three most important GC components: gradient sensing, signal transduction and motion actuation. These components are described as a series of functional boxes that map inputs into outputs (see Figure 3.2). Each component is modeled using differential equations (to study the variability of each box) and have different time scales: 10s for gradient sensing, 200s signal transduction and 10m for motion actuation. These different time scales allows the axon to turn gradually, taking more time on the tasks that are more complex. As illustrated in Figure 3.2 the Sensing Device Box receives the extracellular concentration field $\Delta C$ and returns the directional vector $\tilde{P}$ that is received by the Signal Transduction Box, which returns the new $P_t$ considering the actual GC force. Finally, the Motor Actuator Box produces the new angle ($\alpha(t)$) of the GC in the instant $t$. Experiments with two dimensions in order to analyse the variability of the boxes were performed, the results obtained showed that AG appears to have an equilibrium between determinism and stochastic forces.

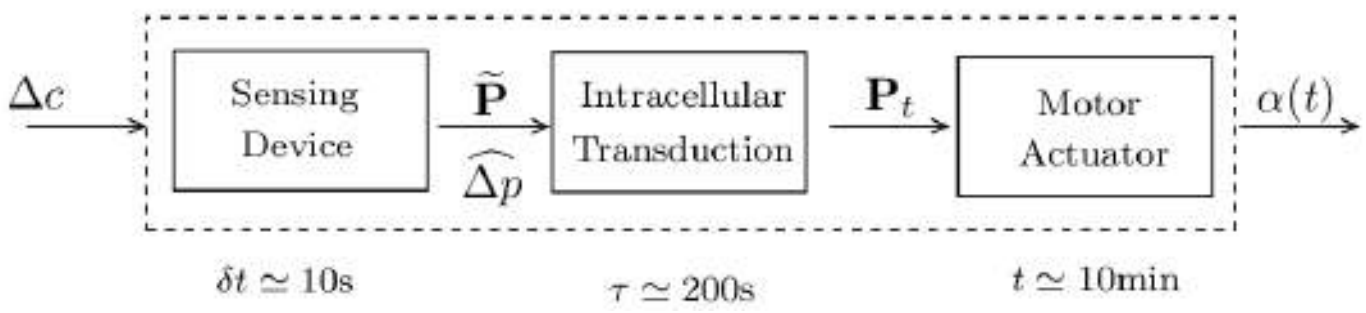

Figure 3.2: Functional subsystems of the GC transduction cascade introduced by Aletti [2].

Furthermore this work allows neuroscientists to compute sample sizes for detecting statistical significant differences of axonal response in different scenarios before performing experiments *in vivo*/*in vitro*. In [2] numerical simulations of single and multiple ligands were analysed. A biologically relevant numerical simulation was also done with commissural axons in the midline crossing focused on the possible silencing of DCC receptors by Robo receptors (see [82] for an analysis on this hypothesis).

Goodhill [31] investigated diffusible target gradients that consider the distance from the target and the time since the start of the diffusion. This spatiotemporal domain used by his model is modelled by Equation 3.1 that calculates the concentration $C(r,t)$ at a distance $r$ from the source and in time $t$ with a diffusion constant $D$ at a diffusion rate $q$. The $erfc$ is the complementary error function, that provides the probability of being outside the range $[-r, r]$.

$$C(r,t) = \frac{q}{4\pi Dr} erfc\left(\frac{r}{\sqrt{4Dt}}\right) \tag{3.1}$$

It is also used the parameter $p$ that determines the percentage change in the concentration across the GC width. This model was estimated based on the experimental literature, what includes the estimation of the diffusion constant ($10^{-6} cm^2/sec$ and $10^{-7} cm^2/sec$), the diffusion rate ($10^{-7} nM/sec$), growth cone diameter ($10\mu m$ and $20\mu m$), minimum ($10^{-1} nM$ and $10^{-2} nM$) and maximum ($100 nM$) concentration for gradient detection and the minimum change detectable by the GC (1% or 2%). The maximum guidance range found was 1 mm, for large times, which is supported by experimental work.

In an extension to this work and using the same approach was found a spatial limit on guidance of 1 cm for a substrate-bound gradient [30].

In other work Goodhill [32] proposes a simple theoretical model of axon lateralization after *Drosophila* midline crossing mediated by a Slit gradient. This model is based on a combinatorial Robo coding, where different combinations of the Robo family are studied (Robo, Robo2 and Robo3). The results of this model suggest that the most important differences between these Robo receptors are more quantitative than qualitative. The Slit gradient is modeled by the exponential Equation 3.2, where $C$ is the Slit concentration, $x$ the distance from the midline, and $\alpha$ the rate of decrease when moving away from the gradient. Two models are considered in order to generate a repulsive answer: (i) a linear model and (ii) a non linear model (Equation 3.3). The linear model is defined by $S(x) \times R$, where $R$ is a simple sum of the repulsion of the three Robo receptors constrained by experimental data. When the repulsive value reaches a certain threshold the axons began to be attracted by local cues. The non linear model considers both the receptor and the ligand concentrations ($[R]$ and $[L]$) and a $K_d$ dissociation constant that is similar to the threshold of the linear model, describing the affinity between the ligand and receptor.

$$S(x) = Ce^{-x/\alpha} \tag{3.2}$$

$$[RL] = \frac{[R][L]}{[L] + K_d} \tag{3.3}$$

More recently Mortimer et al. [66] introduced a bayesian model to predict the response of axons to molecular gradients based on the hypothesis that intrinsic receptor binding noise is one of the most important constraints on gradient detection. Using this model it is derived an equation that predicts the axon response to the gradient steepness and concentration. These predictions are then successfully validated using experimental assays. The results show the quantitative constraints for effective axonal guidance and the computational principles that may be present in the signal transduction pathways. Moreover, they found that the optimal behavior is based on the calculation of the sum of the positions of bound receptors, weighted by their distances to the GC center.

A simple model is introduced by Maskery et al. [61] which applies stochastic changes or external repulsive cues in the GC direction in order to demonstrate that these external cues are more efficient for axon guidance when the growth results from a balance between the stochastic and the deterministic effects. These results were obtained by measuring the resultant turning angles.

Aeschlimann [1] developed a two dimensional biophysical model of axonal pathfinding. This model reproduces the neurobiology of the GC in specific situations, such as in the detection of a diffusible gradient, or in the contact with various cells. The biophysical parameters of both the sensory and motor functions were estimated considering the information available from *in vivo* and *in vitro* experiments. Two methods for modeling the filopodia behavior were proposed, one for temporal detection and the other more related with the biology, which considers calcium levels. It was demonstrated that by using simple physical laws as well as some filopodia characteristics it is possible to obtain realistic behaviors in axonal pathfinding.

A more focused study was performed in [19] where is tested an assembly of neurons that releases a chemorepellant molecule and transfers this molecule to the GCs by axonal transport. This repellent gradient was modeled using an exponential function. The length of the axon was modeled with a linear function and the velocity of the axonal transport is constant. As a result, in this simple model the cones growth towards the regions of low chemorepellant concentration.

An unified mathematical and numerical simulation framework for morphological development of neurons is proposed in [36]. This framework includes four different stages: (1) neurite initiation, (2) neurite elongation, (3) axon pathfinding, and (4) neurite branching and dendritic shape formation. Graham and Ooyen state that this mathematical framework is the basis for a future computer simulator that could be used for studying neuronal development models.

Segev and Ben-Jacob [77, 78] introduced a 2D axon guidance model inspired by the growth of bacterial colonies and the aggregation of amoebae. In this model a single axon grows from each soma and its direction is influenced by extracellular cues (attractive or repulsive). In a first stage the soma emits a repulsive signal that repels the axon, then after a predefined axon length, the axon comunicates with the soma in order to start releasing an attractive signal. This behavior leads the axon to be attracted to the nearest soma. Using numerical experiments they analysed the network connectivity that appeared from different spatial distributions of somas.

### 3.1.2 Computational

The computational models presented in this section are either based on mathematical equations or purely computational approaches, but are always accompanied by simulations.

Goodhill et al. [33] proposes a probabilistic two-dimensional computational model for gradient sensing and movement of the GC mediated by filopodia. The GC direction results of the average of the filopodia directions that is calculated considering the ligand binding (see Figure 3.3 for a representation of the receptors distribution). New filopodia develops where this binding is high (for attraction) and where it is low (for repulsion). In order to obtain a realistic GC turning its new direction is equal to 0.97 times the previous direction, plus 0.03 times the new one. They demonstrated that simple mechanisms are enough to obtain realistic axonal paths for short-term and long-term responses. In this model the response to attractive and repulsive gradients seems to be asymmetric, it can predict the axon response to different gradient concentrations and steepness, the size of the intracellular amplificaion, and the differences required in the intracellular signaling for repulsive and attractive turning.

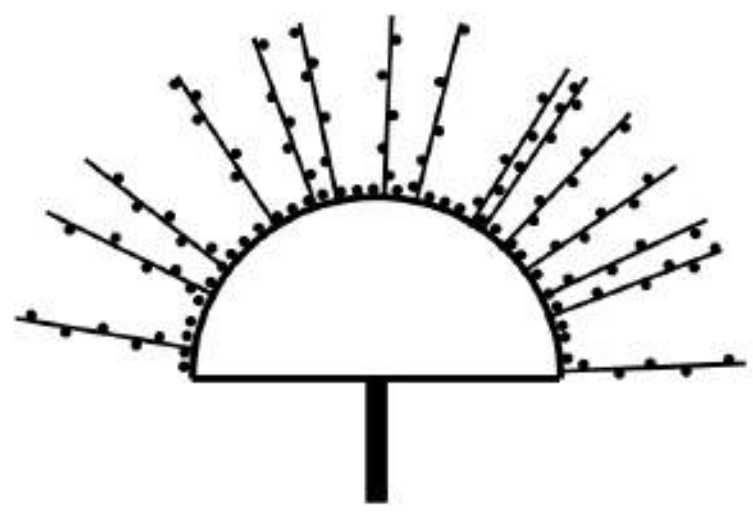

Figure 3.3: Representation of random receptor distribution (small circles) on the GC membrane and filopodia [33].

In [39] is introduced a model of dendritic development in neurons that considers cellular mechanisms based on experimental data. More specifically it includes the phosphorylation state of MAP2 protein. This model predicts how the branching patterns change some intracellular components (e.g. Calcium).

Merging experimental and computational methods [26] developed a new physiological computational model of the E18 hippocampal neurons that react to topographical and immobilized chemical cues. This model includes initiation, polarization (i.e. extension of neurites in all directions) and axon pathfinding based on substract cues. Their results show that the model successfully mimics and predicts the experimental polarization probability for some topographical feature sizes and a constant chemical cue spacing.

In an interesting approach, Feng et al. [24], proposed an open-L-systems-based framework for modeling, simulating and visualizing axon guidance. L-Systems are parallel rewriting systems that are defined as grammars and are useful to describe many systems, such as plant development , organism morphology or fractals [56, 71]. The open extention brings to L-systems the possibility of interacting with the environment. This framework is evaluated with simple simulations with one neuron and one or two guidance cues that are produces contact attraction, contact repulsion or chemoattraction (one for each simulation step).

A combination of L-systems and genetic algorithms was presented as an approach to evolve virtual neuron morphologies (using the EvOL-Neuron system) [91, 90]. Since the main aim of this work was to study the final morphologies and not how the axons perform their decisions, simple rules were evolved to describe the morphology generation process. Its fitness function is based on the comparison between virtual and real neurons.

A grammar language to study interaction in neurite models (precessor to axons and dendrites) during development is given in [38]. The system is specified with a graph grammar that led to simple realistic results for neurite development when compared with *in vivo* experiments.

Gathering inspiration from axon guidance, a new type of neural architectures implemented in circuits was proposed by Taba [84]. He introduced a novel technique for automatic connection rewiring between spiking neurons and analyzed its performance using a silicon implementation of a growth cone population model whose migration was driven and directed by patterned neural activity (i.e. activity-dependent).

Costa and Macedo proposed a two-dimensional simple multi-agent approach to model axon guidance [16]. This system modeled neurons and glial cells as agents and repulsive/attractive cues.

Several computational models have been proposed to study topographic development, mainly of the retinotectal maps [96, 34]. These models help to understand how the neural networks develop towards a specific topographic map taking into account the whole and the experimental data available. However no model has yet tried to include all the data available, as it is a very difficult task and hard to analyse [34].

## 3.2 Simulators

There are several simulators in different fields that could be relevant to this thesis. We consider the following fields: computational neuroscience, system biology and artificial life. During this section we give an overview of the current simulators for each field.

### 3.2.1 Computational neuroscience

In the field of computational neuroscience there are available and well established some simulators. Here we briefly describe their functions and main applications.

The most used neural simulators are NEURON[1] [41] and GENESIS[2] [10]. These two simulators focus mainly on the study of individual neurons or networks to analyse their neurophysiological properties (e.g. electrical signalling). These studies are facilitated by using near user-friendly graphical user interfaces for both model construction and results display. The first simulator is well-suited for empirically-based models due to its separation between biology and computational concerns and provides tools that are numerically sound and computationally efficient. While the second follows a "building block" approach, with modules that receive inputs, perform calculations and then produce outputs. Therefore, neuron models are built as a set of these basis modules, such as dendritic and ion channels components.

Topographica [3] [7, 8] is a simulator that helps to bridge across several levels of detail in topographic maps (see Figure 3.4). This simulator focus on the study of the effects of activity and organization in thousands of connections in topographic maps. It is designed to simulate the topographic maps of any cortical or subcortical, such as the ones related with vision, touch or audition. Currently, this tool has an easy interface with Python, C/C++, MATLAB, NEST, or NEURON.

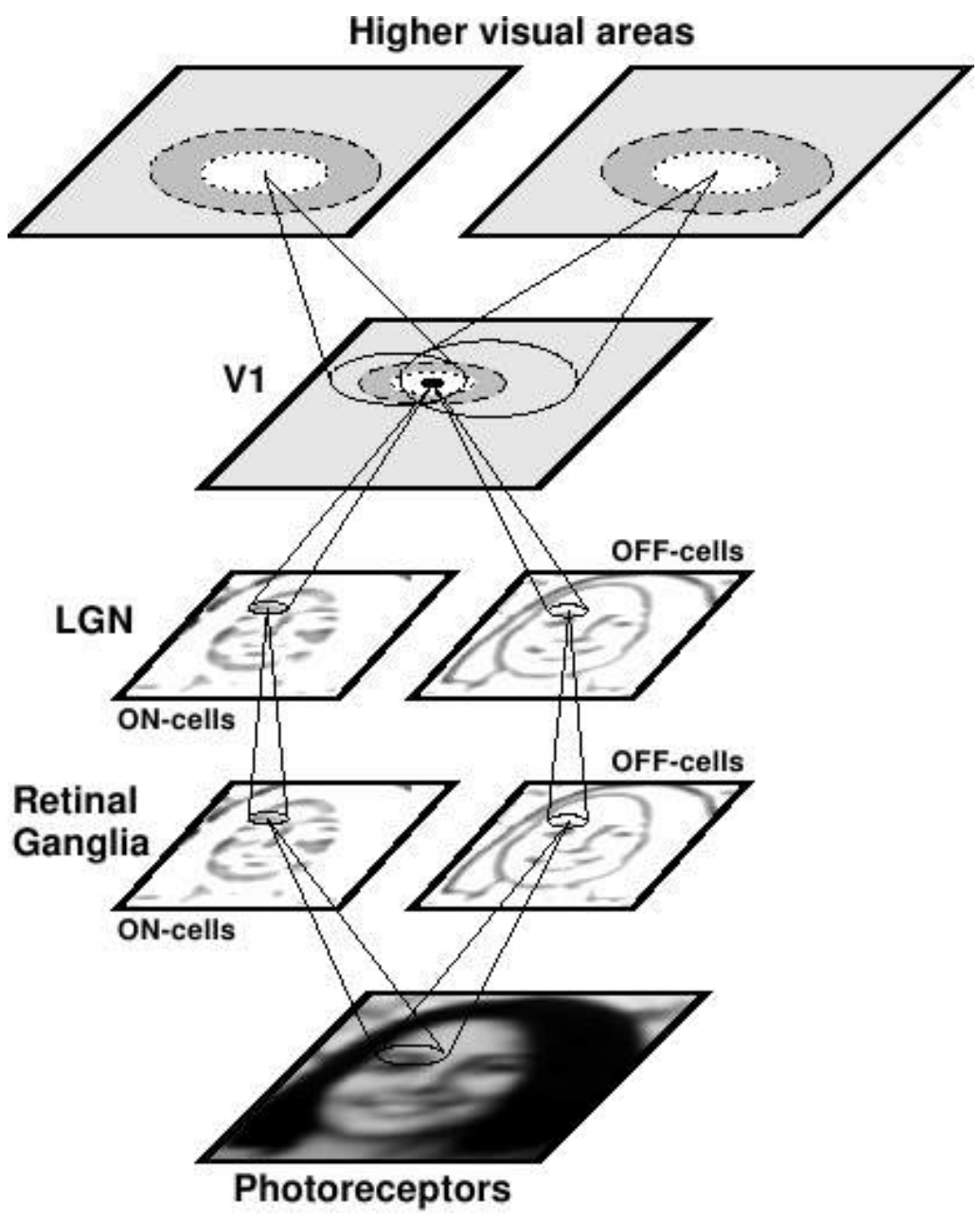

Figure 3.4: Representation of a sample Topographica model of the early visual system [7]. This model is composed by two groups of sheets. The first represents the eye with three sheets (one for the photoreceptors and the remaining two for the retinal ganglion cells). The second is the V1 area of the visual cortex and another one to represent higher visual areas. Sample connections between the different sheets are shown.

Other topographic simulator is NEST [4] [27] that is more focused on measuring topo-graphic maps and generating appropriate inputs for these maps.

[1]http://www.neuron.yale.edu
[2]http://genesis-sim.org
[3]http://topographica.org
[4]http://www.nest-initiative.org

Netmorph [5][51] is a general simulator capable of generating large-scale neural network morphologies and visualizing them in three-dimensions. The development is modeled from the prespective of growth cones in axonal or dentritic trees (with elongation, branching and turning) and its behavior is described with stochastic and phenomelogical mechanisms. Netmorph is a realistic and flexible tool that can be used to investigate questions regarding morphology and connectivity. It focus more on the final morphology and how this affects the sinaptic connectivity. The growth cone direction is simply defined by a probabilistic perturbation between a minimum and a maximum angles without considering extracellular information. Using a simple synapse formation rule based on proximity the simulations performed in Netmorph already lead to realistic morphologies and to small-world networks.

### 3.2.2 Systems biology

E-Cell [6][89, 85] is a cell simulator with a friendly graphical user interface that eases the process of results analysis (see Figureg 3.5). Besides its graphical capabilities with E-Cell it is possible to model, simulate and analyze complex, heterogeneous and multi-scale biochemical reaction systems (e.g. cell). All with numerical simulation algorithms, mathematical analysis methods, stochastic and deterministic algorithms.

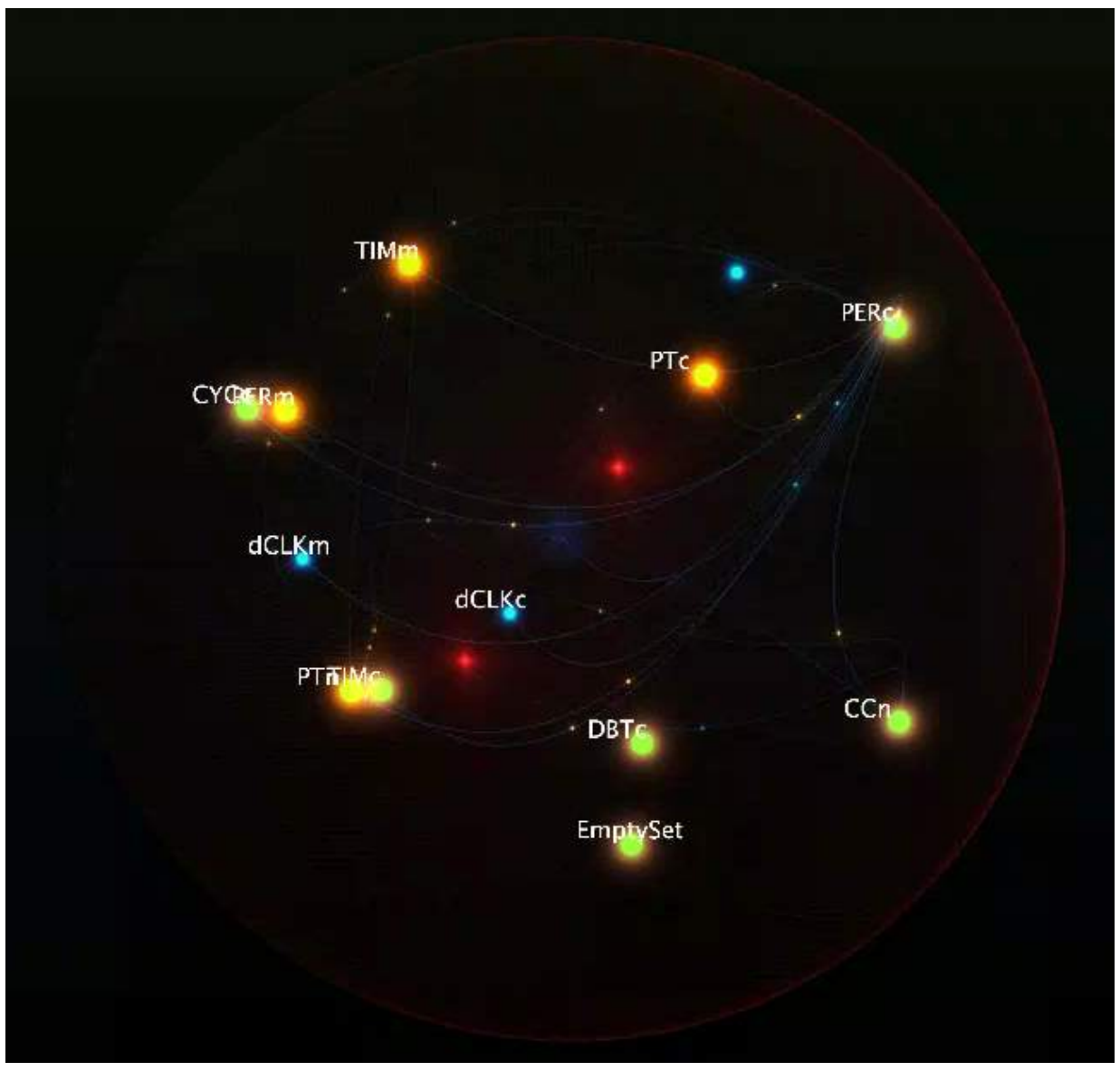

Figure 3.5: Visual representation of some intracellular pathways in E-Cell 3D simulator.

A more low-level tool is Avogadro[7] that is an advanced molecular editor that is used in areas such as, computational chemistry, molecular modeling, bioinformatics and materials science (see Figureg 3.6). It is a cross-platform software developed with the

[5]http://netmorph.org
[6]http://ecell3d.iab.keio.ac.jp
[7]http://avogadro.openmolecules.net

Qt framework [8] that was built to be of easy use for students, teachers and advanced researchers.

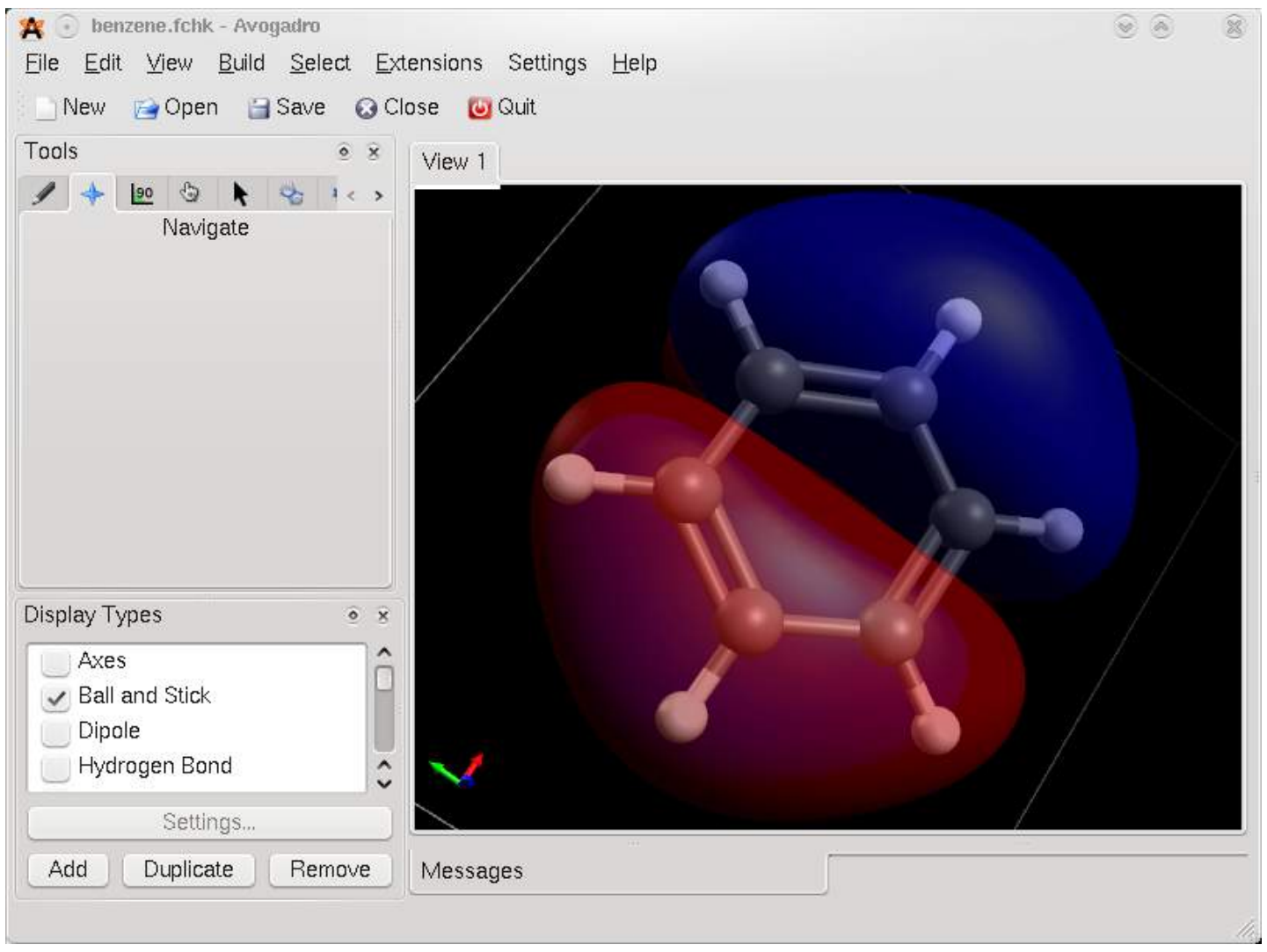

Figure 3.6: Avogadro screenshot with the three-dimensional structure of a chemical compound.

### 3.2.3 Artificial life

There are two interesting 3D simulators for artificial life: (i) Breve[9] [50] and (ii) Framsticks[10] [52]. The first is free, open-source that mases easy to build 3D simulations of multi-agent systems and artificial life by using the programming language Python (see Figure 3.7 for an example). It also includes physical simulation and collision detection and an OpenGL display engine. Its API as built-in several libraries about techniques such as genetic algorithms and creature morphology. Examples given are the simulation of 3D cellular automata or artificial creatures evolution. The second is a three-dimensional artificial creatures simulation project where each organism contains: physical structures, control structures and evolved based on specific fitness functions (e.g. speed).

## 3.3 Conclusions

To sum up, the last decaded has seen the emergence of several computational and mathematical models for axon guidance or neural development in general. Despite their obvious importance for studying several parameters they usually focus on a small subset of experimental data that are not enough to model an entire real system, such as midline or optic pathway. However, these studies specified very important parameters that are relevant both for axon guidance and for mathematical/computational models.

[8]http://www.qtsoftware.com
[9]http://www.spiderland.org
[10]http://www.framsticks.com

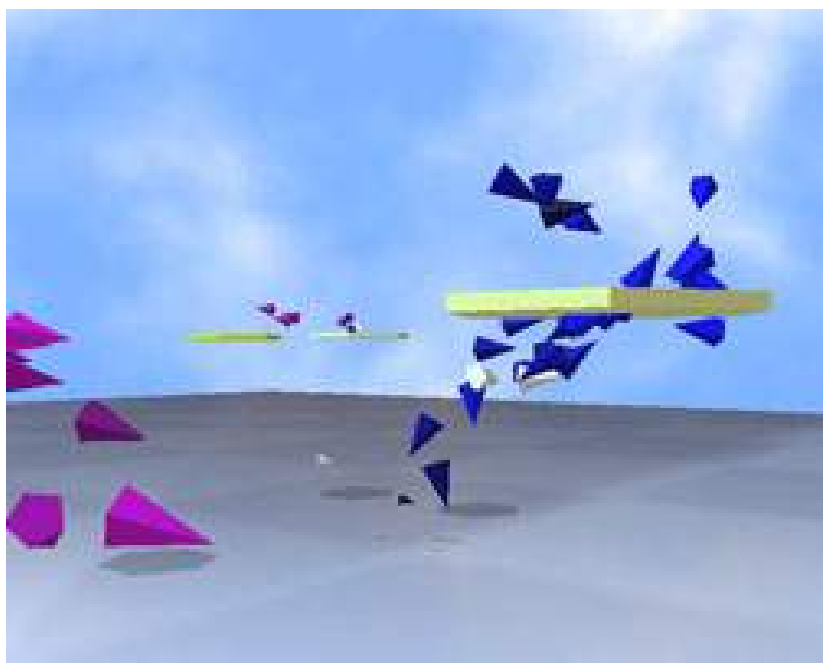

Figure 3.7: Breve screenshot of flocking agents that evolved strategies of capturing food sources [50].

The mathematical models based on dynamic equations, the multidimensionality of this problem is quite difficult to handle with them [84]. This is proved by the difficulties found when attempting to build a generic platform that brings together all the important issues of axon guidance.

L-System are rewriting grammars that are difficult do define and understand. More complicated is this process if all the relevant mechanisms and elements needed to obtain realistic behaviors are included(e.g. extracellular gradients or regulatory networks).

Currently, there are many simulators for different porposes in computational neuroscience and close fields. However none of them was built for studying specifically the axon guidance phenomenon.

## Chapter 4

# Computational model

This chapter introduces a computational model of axon guidance with simple elements and mechanisms that are biologically relevant. Our model (i) includes several elements, (ii) model these elements with simple interactions that together lead to realistic behaviors (emergence), and (iii) has a three-dimensional representation (euclidean). Furthermore, based on the neurobiological knowledge we introduce (v) a guidance model that defines how the growth cone should interpret the extracellular cues and a (vi) regulatory network between receptors and internal proteins.

The computational model proposed in this chapter was developed with the aim of being based on realistic rules and that could be implemented as a computational simulation so that could be applied in the study of biological systems (e.g. the midline and the optic pathway). By using such a model we are able to develop new hypotheses about specific systems that should guide future experimental work.

Our approach can be viewed as the multi-agent system described in [16]. The same basic principle is used here where there are several elements that interact between each other. However in this thesis we decided to keep the model more abstract; therefore we not defined the elements as agents, because it could be misleading. Consequently, we will not use any multi-agent platform or even the traditional communication languages used in these systems in order to obtain a better performance.

Furthermore, we focus on activity-independent mechanisms because they are more relevant for the axon guidance problem [34]. Activity-dependent mechanisms are more relevant for refinement of synaptic strengths, rather than for axon guidance. Considering the categories defined in [79] for axon guidance models, the model proposed here can be included in the abstraction models as it does not model explicitly the molecular details.

As our focus is axon guidance and not axon elongation we define the growth length as a constant $L$, which is used during this chapter.

During this chapter we present the elements and mechanisms that belong to our model (see in Figure 4.1 an overview of the model). All these mechanism are shown to be biologically relevant based on the simulations presented in the chapter , however depending on the specific problem some elements or mechanism may be irrelevant.

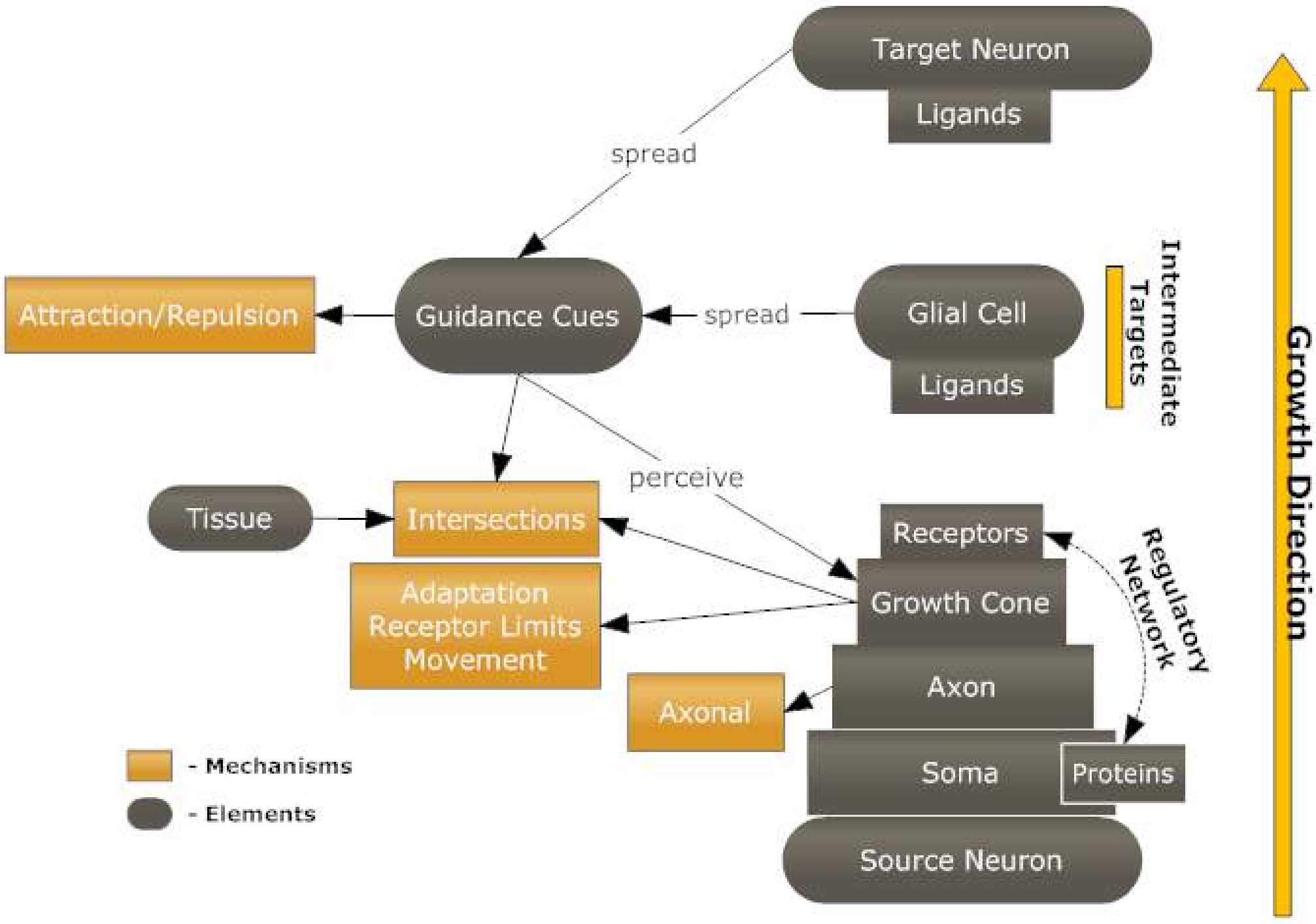

Figure 4.1: General structure of the Computational Model. This diagram illustrates the interaction between the elements and mechanisms.

## 4.1 Elements

In this section we present the elements included in our model and define how they are modeled. The chapter 6 contains several experiments where the relevance of these elements is evaluated from an axon guidance point of view. Each element that is defined in this section is implemented in the simulator as an object (Object-oriented Programming). Thus the reader can interpret all these elements as objects if this eases the model understanding.

### 4.1.1 Neuron

In our approach we define the neuron as a fixed element (i.e. already in its final position). Neuronal migration is a different problem that is not in our scope. In the following sections we introduce the components that belong to a neuron.

A neuron can be of two types, *source* or *target*. The source neuron is the neuron from which the axon starts growing, while the target neuron is the neuron until which the axon should growth and then establish a connection with it.

#### Soma

The soma is the main element of each neuron. It contains the nucleus where all the information about how the neuron should perceive and react to the the environment is stored. Therefore, most of the parameters described during this chapter that are related with the neuron are included in this component (e.g. types of receptors and guidance complexes). The most relevant parameters are presented in Table 4.1. The state parameter regulates if the neuron is active or not, when it is *on* the neuron performs its function (i.e. growth the axon if its a source neuron or releases ligands if its a target neuron). The proteins variable represents the list of intracellular proteins that the neuron contains (e.g. Comm), these proteins may play important roles by regulating receptors or other proteins.

Table 4.1: Soma parameters.

| Parameter | Definition |
|---|---|
| *State* | Defines if the neuron is On or Off. |
| *Position* | Defines neuron position (x,y,z). |
| *Proteins* | Set of internal proteins. |
| *Radius* | 0.15 $\mu$m (default value) |
| *Representation* | Three-dimensional representation (e.g. sphere). |

### Axon

The axon is the most relevant element of AG. We model the axon as a dynamic element that grows based on the decisions made by the GC. In Table 4.2 are given the parameters that we considered relevant for the axons. The initial angles define the initial direction that the GC has. These angles during the first simulation steps may change depending on the extracellular gradients. Thus, even if their values are far from the biological evidences they are corrected by the system dynamics (i.e. due to the extracellular gradients that attract or repel the GC). As our main concern is not the graphical representation of the elements; therefore we represent the axon as a set of $n$ lines, being $n$ the number of simulation steps. We do not consider branching in our approach, hence, each neuron contains a single axon. The branching process helps the axon to search the environment, however, is the decision capabilities of the GC that are essential by deciding the next step. The growth length parameter may be changed during the simulation due to the growth cone complexification mechanism.

Table 4.2: Axon parameters.

| Parameter | Definition |
|---|---|
| *Initial angles* | Defines the initial angles ($\alpha$ and $\beta$) of the axon in the membrane of its soma. |
| *Growth history* | List containing all the growth cone positions calculated during the simulation. |
| *Axon length* | Total axon length for the current simulation step. |
| *Growth length step* | Current axon growth length. |
| *Representation* | Three-dimensional representation (e.g. lines). |

Growth cone As explained in chapter 2 the GC belongs to the axon and it is the key element of AG. It perceives the extracellular cues and depending on them produces a response (i.e. a turning angle). In this model there is one GC on the tip of each axon and this GC contains a set of receptors that interact with extracellular ligands.

The fasciculation mechanism uses ligands released by the GC in order to attract other growth cones, leading to the formation of axon fascicles. Either the value of the GC radius or the receptor radius were chosen for a good visualisation, they do not affect the model calculations.

The graphical representation of the GC is a three dimensional green sphere. The GC receptors are defined by a small black sphere that rotates above the surface of the GC sphere (see chapter 6). This small sphere also represents the current GC direction. The GC parameters are described in Table 4.3.

Table 4.3: Growth cone parameters.

| Parameter | Definition |
|---|---|
| *Receptors* | List of receptors (see Section 4.1.4). |
| *Concentrations* | Array that saves the concentrations for each receptor. |
| *Location* | Three-dimensional location (x,y,z). |
| *Angles* | $\alpha$ and $\beta$ angles of the GC in a specific simulation step. |
| *Force* | Stores the current force being applied. |
| *Ligands* | List of ligands (see section 4.1.5). |
| *Radius* | 0.1 $\mu$m (default value). |
| $Radius_{receptors}$ | 0.02 $\mu$m (default value). |
| *Representation* | Three-dimensional representation (e.g. sphere). |

### 4.1.2 Glial cell

The glial cell is an abstract element that represents all types of glial cells (e.g. microglia and macroglia). Several studies have demonstrated that they perform an important role in axon guidance. Yet the influences of its different types remain to be clarified. This is why we decided to maintain the glial cells as a generic element that functions as an intermediate-target by releasing ligands gradients. For the parameters considered in glial cells check Table 4.4.

Table 4.4: Glial cell parameters.

| Parameter | Definition |
|---|---|
| *Radius* | 0.15 $\mu$m. |
| *Location* | Three-dimensional location (x,y,z). |
| *Ligands* | List of released ligands (see section 4.1.5). |
| *Representation* | Three-dimensional representation (e.g. sphere). |

### 4.1.3 Tissue

This element model vessels or tissue and is defined by a set of polygons. This tissue imposes a set of physical constraints that reduce the search space. In order to perform experiments, three-dimensional objects of specific axon guidance systems (e.g. midline and optic pathway) must be imported (see chapter 6 for an example). Hence, any tissue element must first be modelled in a 3D modeling software such as Blender[1] or Maya[2] and then imported into our model. The current model interprets these elements as cell adhesion, hence, when the GC collides with it rotates so that could grow paralelly to it.

[1] http://www.blender.org
[2] http://www.autodesk.com/maya

### 4.1.4 Receptors

The receptors play a crucial role in our model, being the sensors of the GCs (see their parameters in Table 4.5). They are activated by ligands and may be down-regulated or up-regulated by internal proteins or other receptors (see section 4.1.7). The removal and production rate are only activated when the adaptation mechanism is enable. The same happens with receptor limits mechanism for the activation and saturation. The receptors are only removed (i.e. down-regulated) when they are activated in order to apply the adaptation principle (desensitization), see section 4.2.4 for more information. Therefore this rate depends on the activation, stronger activation leads to an higher removal rate. On the other hand the production rate does not depends on the activation, but only on the concentration level, we assume that the neuron produces new receptors at a constant *production rate* until it reachs its maximum concentration (resensitization).

Table 4.5: Receptor parameters.

| Parameter | Definition |
|---|---|
| *Receptor type* | e.g: UNC5, DCC and Robo. |
| *Receptor distribution* | Defines how the receptors are distributed in the membrane. See uniform function in Algorithm 1. |
| *Removal rate* | Rate (per simulation step) at which the receptors are removed from the membrane. Should be faster than the production rate. One of the biological processes that is involved in this removal is endocytosis. |
| *Production rate* | Rate (per simulation step) at which the receptors are produced. This production stops when the maximum receptor concentration is reached (resensitization). |
| *Activation* | Threshold of minimum activity above which the GC starts reacting. |
| *Saturation* | Ligand concentration above which the receptors response (attraction or repulsion) is not increased. |

We assume an uniform distribution of the receptors along the GC membrane. This leads to the calculation of spherical coordinates (define the GC orientation) and force given by the Algorithm 1. This algorithm defines how the receptors are affected by the gradients by calculating the new direction of the GC based on the extracellular gradients and on the previous direction.

**Input**: Growth Cone $gc$, Guidance Complex $c$, Gradient $g$
**Output**: Double $force$, Angle $\alpha_n$, Angle $\beta_n$

$A1, A2 = intersection(g_{center}, gc_{center}, gc_{center}, gc_{radius})$; ▷ see Algorithm 2

**if** $\sqrt{g_{center}^2 + A1^2} < \sqrt{g_{center}^2 + A2^2}$ **then**
  | **if** $gc_{behavior} == repulsive$ **then** $P = A2$ **else** $P = A1$
**else**
  | **if** $gc_{behavior} == repulsive$ **then** $P = A1$ **else** $P = A2$
**end**

$P_x = P_x - gc_x$
$P_y = P_y - gc_y$
$P_z = P_z - gc_z$

$\alpha = atan2(\sqrt{P_z^2 + P_x^2}, P_y)$
$\beta = atan2(P_x, P_z)$

$updateAngles(\alpha, \beta)$; ▷ see Algorithm 3

$force = diffusion(\sqrt{g_{center}^2 + gc_{receptor_{center}}^2}, c_{range}) \times c_{force}$

**Algorithm 1**: Calculation of $\beta$, $\alpha$ and $force$ for uniform function. First are determinated the line-sphere intersection points, then we consider the close intersection point as the attractive reference and the further as the repulsive reference. After, based on this new point that represents the maximum turning point we calculate the $\alpha$ (alpha) and $\beta$ (beta) angles, that are then updated considering the current growth cone angles, using Algorithm 3. Finally, the current $force$ is calculated considering a specific $diffusion$ function (e.g. exponential) and the guidance complex $range$ and $force$.

**Input**: Point $P1$, Point $P2$, Point $S$, Double $r$
**Output**: Point $A1$, Point $A2$

$a = (P2_x - P1_x)^2 + (P2_y - P1_y)^2 + (P2_z - P1_z)^2$

$b = 2((P2_x - P1_x)(P1_x - S_x) + (P2_y - P1_y)(P1_y - S_y) + (P2_z - P1_z)(P1_z - S_z))$

$c = S_x^2 + S_y^2 + S_z^2 + P1_x^2 + P1_y^2 + P1_z^2 - 2(S_x P1_x + S_y P1_y + S_z P1_z) - r^2$

$i = (P2_x - P1_x)^2 + (P2_y - P1_y)^2 + (P2_z - P1_z)^2$

**if** $i == 0$ **then** ▷ `Only one intersection`
  $\mu = \frac{-b}{2a}$
  $A1_x = L1_x + \mu(L2_x - L1_x))$
  $A1_y = L1_y + \mu(L2_y - L1_y))$
  $A1_z = L1_z + \mu(L2_z - L1_z))$
**else**
  **if** $i > 0$ **then** ▷ `Two intersections`
    $\mu = \frac{-b + \sqrt{b^2 - 4ac}}{2a}$
    $A1_x = L1_x + \mu(L2_x - L1_x))$
    $A1_y = L1_y + \mu(L2_y - L1_y))$
    $A1_z = L1_z + \mu(L2_z - L1_z))$
    $\mu = \frac{-b - \sqrt{b^2 - 4ac}}{2a}$
    $A2_x = L1_x + \mu(L2_x - L1_x))$
    $A2_y = L1_y + \mu(L2_y - L1_y))$
    $A2_z = L1_z + \mu(L2_z - L1_z))$
  **end**
**end**

**Algorithm 2**: Calculates the intersection points between a line (defined by points $P1$ and $P2$) and a sphere with center $S$ and radius $r$. The $i$ represents the number of intersections and, $A1$ and $A2$ the intersection points.

**Input**: Angle $\alpha$, Angle $\beta$, Growth Cone $gc$
**Output**: Angle $\alpha_n$, Angle $\beta_n$

$\alpha_n = \alpha - gc_\alpha$
$c = |\beta - gc_\beta|$
$b_n = min(c, 2\pi - c)$
**if** $c == b_n$ **then**
  **if** $\beta < gc_{beta}$ **then** $\beta_n = -(b_n)$
**else**
  **if** $\beta > gc_{beta}$ **then** $\beta_n = -(b_n)$
**end**

**Algorithm 3**: Calculation of the $\beta_n$ and $\alpha_n$ angles that are the new angles based on the current growth cone angles. This function allows the GC to turn smoothly even when the angles reach its limits (i.e. $2\pi$). Thus the new angles result from the difference between the current values to the previous values.

### 4.1.5 Ligands

Ligands bind to specific receptors and are modeled as diffusion functions (see parameters in Table 4.6). In this model we consider that ligands always activate specific receptors.

What we define as ligands range is defined in other works as gradient steepness [33, 2]. The idea is the same, this range influences the decay of the diffusion function. The range term is used in order to keep it closer to axon guidance knowledge where it is known that different ligands have different ranges. This range is directly related with the ligand disintegration by the extracellular environment.

Table 4.6: Ligand parameters.

| Parameter | Definition |
|---|---|
| *Ligand type* | E.g: Sema4D and Netrin. |
| *Diffusion function* | Exponential, exponential-square, inverse and constant functions (see Table 4.7). Normalized between 0 and 1. |
| *Range* | Range of detection by receptors. |
| *Diffusion rate* | Defines the velocity of diffusion into extracellular environment (0-100%). Faster diffusion leads to a faster effect on the GCs. |

Table 4.7: Ligand diffusion functions, where $d$ is the distance from the ligand to the receptor and $r$ the ligand range.

| Function | Equation | Definition |
|---|---|---|
| *Exponential* | $e^{-\frac{d}{r}}$ | With this function the effect of the gradient decreases exponentially with the distance. The range controls the decay of the exponential function, for larger values it drops gradually, while for shorter values it decreases sharply. |
| *Square-Exponential* | $e^{-\sqrt{\frac{d}{r}}}$ | This has the same behavior has the previous one, but here even far distances have a strong impact. |
| *Inverse* | $\frac{r}{d}$ | In this function the gradient effect decreases sharply as the distance is inversely proportional to the range. |
| *Constant* | $r$ | The $r$ range must be defined between 0 and 1. |

### 4.1.6 Guidance cues

In this section we present how the guidance cues are included in the model. Each guidance cue is represented by a guidance complex that links one ligand and the respective receptor (see Table 4.8). The complex range for the guidance cue should be the same defined by the ligand and the behavior can either attractive or repulsive. Altought the guidance complexes are based on neuroscience knowledge of a specific system, they can be easily changed in the model to test new possibilities.

In Table 4.8 the *force* column is undefined because it depends on the axon guidance system that is being considered (normalized between 0 and 1). This value defines the strength of the interaction between the pair ligand-receptor, from a chemical point of view it englobes mechanisms such as ligand affinity and the effects on the intracellular pathways after receptor activation.

Table 4.8: Examples of guidance cues known from experimental studies (see section 2.3 for details).

| Ligand | Receptor | Behavior | Range | Force |
|---|---|---|---|---|
| Netrins | DCC-UNC5 | Repulsion | Chemo (long) | - |
| Netrins | DCC | Attraction | Chemo | - |
| Netrins | UNC5 | Repulsion | Chemo (short) | - |
| Ephrins | Eph | Repulsion | Contact | - |
| Slits | Robo | Repulsion | Chemo | - |
| Sema4D | PlexinsB1 | Repulsion | Contact | - |
| Cadherin (CAM) | Cadherin (CAM) | Attraction | Contact | - |
| Laminin (ECM) | Integrin (CAM) | Attraction | Contact | - |

### 4.1.7 Regulatory network

During chapter 2 were introduced some examples of regulation between receptors and internal proteins. For instance, when the Robo receptor is activated the response produced by the DCC receptor declines (inhibitory regulation). The regulatory links defined in the network lead to increases or decreases in the receptors or proteins concentrations.

Table 4.9 presents some examples of regulation and Algorithm 4 defines how this network affects the receptors and proteins concentrations. This network is a key feature in our model because it brings to the GC a greater number of possible navigation decisions. Receptors activation may regulate others, changing the GC response to extracellular stimuli.

The guidance model proposed in section 4.3 can be tested with and without this regulatory network between the regulatory elements (i.e. the receptors and internal proteins).

It is important to note that regulation at different levels (i.e. a receptor that regulates other which in turn regulates another one) emerges during the simulation and not in a single step. This is realistic because biological pathways are not processed in a single step, but as a dynamic and continues process.

Table 4.9: Examples of regulation.

| Source | Regulation | Target |
|---|---|---|
| UNC5 (receptor) | inhibits | DCC (receptor) |
| Robo (receptor) | inhibits | DCC (receptor) |
| Comm (protein) | inhibits | Robo (receptor) |

## 4.2 Mechanisms

In this section all the mechanisms contemplated by our model are explained. Some of these mechanisms can be turned off during the simulation, what allows to study specific aspects, such as the importance of axonal transport for axon guidance. The remaining

**Input**: Concentrations $[Q]$, Regulatory Network $rn$, Growth Cone $gc$
**Output**: Concentrations $[Q]$

**for** $rn\ i = 1$ **to** $n$ **do**
  $q = [Q](rn_i)$
  **for** $rn\ j = 1$ **to** $n$ **do**
    **if** $rn_{i,j} \neq 0$ **then**
      $q = q + q \times rn_{i,j} \times gc_{force(rn_j)}$

**Algorithm 4**: Algorithm for the Regulatory Network on the GC. The element $i$ on the regulatory array $rn$ with concentration $q$ is regulated by the element $j$. Meanwhile, the concentrations array $[Q]$ is updated with the new concentrations (step not shown). This process runs a single time for each GC.

ones are essential to the simulation, thus, can not be disabled (see in chapter 6 which can be disabled).

### 4.2.1 Axonal

The axonal mechanisms refer to the mechanisms that are directly related with the axon (including the GC).

#### Turning

The turning of the GC is not immediate, when perceiving a new information from the extracellular environment this new information do not reflects immediately in the GC direction. This is partially due to the polarization effect described in section 2.4.4. In our model this smooth turning process emerges with the system dynamics due to the fuzzy value of the guidance complex force, the ligand diffusion and the axonal transport mechanism (see section 4.2.1).

Other possible approach could have been to define the new angle equal to the previous angle plus a small part (3%) of the new value (similar to the work of Goodhill in [33]), however, in our model this is already obtained by the own system dynamics, i.e. without this principle the GC turns gradually.

#### Fasciculation

The fasciculation mechanism is modeled by adding an attractive guidance cue that specifically explores this behavior (see an example in section 6.1.3). This new complex should then be added in the GCs (as receptors and ligands). In this way, the axons will be attracted to each other and navigate together, forming fascicles. However, the attraction force should be studied in order to find a good balance between fasciculation and the attraction from the remaining cues.

#### Transport

The axonal transport is important when modeling axon guidance. The turning of the GC based on the new information from the extracellular environment and the transport of new elements to the GC (e.g. receptors) are two examples where axonal transport is relevant. In both situations should be present a factor that is influenced by the length of the axon and consequently the time of the axonal transport. This transport may be made from the GC to the soma (retrogade) and vice-versa (anterograde). In order to obtain a simple model and due to the lack of studies relating this two types of transports with axon guidance, we decided to model a generic type of axonal transport as defined in Equation 4.1 that is only influenced by the axon length. In the equation, $L$ represents the length of each growth step and $N$ the current number of simulation steps.

$$force = force \times \left( \left( 1 + \frac{L}{\sqrt{N}} \right)^{N} \right)^{-1} \quad (4.1)$$

**Growth Cone Complexification**

Growth cone motility increases when the GC faces complex extracellular environments (i.e. with several guidance cues). The idea is that if more than one receptor is activated in a single simulation step, the growth length decreases. The Equation 4.2 calculates the new growth length $s$ in simulation step $N$ based on the predefined growth length $L$ and on the number of activated $receptors$ in that specific simulation step.

$$s(N) = \frac{L}{1 + 0.1 Receptors_{activated(N)}} \quad (4.2)$$

### 4.2.2 Attraction and Repulsion

The attraction and repulsion forces are caused by extracellular cues that depend mainly on three factors:

1. The effect of the activated receptors (attractive or repulsive)
2. The force of this effect (between 0 and 1)
3. The distribution of the receptors on the membrane (uniform function)

The Algorithm 5 describes how we model the attraction/repulsion mechanism, by using a guidance model (Section 4.3) and the receptors distribution function presented in section 4.1.4.

### 4.2.3 Movement

Here we define how the growth cone moves on the three-dimensional space based on the angles calculated by Algorithm 5 that are defined as spherical coordinates. The movement equations are defined in Equation 4.3 where the spherical coordinates are converted into cartesian coordinates. The first coordinate system is represented by $\beta$ and $\alpha$ angles, and $L$ (growth length) that are converted in $X$,$Y$,$Z$ cartesian coordinates.

$$\begin{cases} X(t) = X(t-1) + cos(\beta) sin(\alpha) L \\ Y(t) = Y(t-1) + cos(\alpha) L \\ Z(t) = Z(t-1) + sin(\beta) sin(\alpha) L \end{cases} \quad (4.3)$$

**Input**: Growth Cone $gc$, Guidance Complexes $c$
**Output**: Angle $\beta_n$, Angle $\alpha_n$, Double $force$

**for** $c$ $i = 1$ **to** $n$ **do**
    **if** $distance(c(i)_{ligand}, c(i)_{receptor}) \geq range$ **then**
        $out(i)_{\beta,\alpha,force} = DistFunction(gc, c(i), c(i)_{ligand})$; ▷ See Algorithm 1
    $out(i)$=Mechanisms($out$,$c(i)_{receptor}$)
$out$=RegulatoryNetwork($out_{[Q]}$,$gc_{regulatoryNetwork}$,$gc$); ▷ See Algorithm 4
**for** $out$ $i = 1$ **to** $n$ **do**
    $\alpha_{current} = \alpha_{current} + out(i)_{\alpha} \times out(i)_{force}$
    $\beta_{current} = \beta_{current} + out(i)_{\beta} \times out(i)_{force}$
    $force_{current} = force_{current} + out(i)_{force}$

**Algorithm 5**: General algorithm for attraction/repulsion behavior. First, the values of $\beta$, $\alpha$ and $force$ are calculated for each guidance complex that belong to the GC, using the uniform distribution function defined in Algorithm 1 and the several Mechanisms defined during section 4.2. After, the regulatory network is applied in order to change concentrations $[Q]$ in structure $out$. Finally, the new angles and forces of each receptor are used to calculate the new values based on the previous ones.

### 4.2.4 Adaptation

The adaptation principle is modeled using the parameters, removal and production rate mentioned in section 4.1.4. This process is only activated when the receptors become active, what allows the GC to enter in a desensitization process (at a specific removal rate). After being stimulated by a specific cue it starts to recover the sensitivity usually at a slower rate, that is defined by the production rate (i.e. resensitization process). When this mechanism is disable the removal and production rate do not affect the GC behavior.

### 4.2.5 Receptor limits

In neurobiology has been observed that sometimes the GCs do not react to low levels. On the other hand it as also been observed that above a threshold the GCs become saturated and do not increase their responses.

In this model the activation and saturation of receptors (section 4.1.4) perform this role, and this receptor limits mechanism control when they affect or not the GC behavior.

### 4.2.6 Intersections

We propose two different intersection algorithms to deal with the presence of tissue (three-dimensional objects). The first decreases the effect of a gradient when there are tissue layers between the receptor and the source (see section 6.1.3 for an example). This is a difficult and important problem because in the reality the ligands collide with the tissues and other elements that block their passage to other spaces. In order to model this problem precisely it would be necessary to develop a map of intersections of the gradient with the surrounding elements. But as we are only concerned with the decisions made by the GC we defined an algorithm that simply calculates the number of intersections of a line drawn from the current GC position to all its gradient sources (see Algorithm 6 for more details). A function available in the OpenSceneGraph is used to compute the intersections[3].

**Input**: 3DModel $m$, Point $s$, Point $e$
**Output**: Double $d$

$i = computeIntersections(m, s, e)$; ▷ `OpenSceneGraph function`

$d = distance_{gradient,receptor} \times i$

**Algorithm 6**: Algorithm for calculation of the number of intersections between the receptor and the gradient source. First, the number of intersections $i$ is calculated between a start point $s$ (i.e. the receptor) and a end point $e$ (i.e. gradient source) in a three-dimensional model $m$ that may represent the midline. Then, the final distance $d$ between $s$ and $e$ is obtained by multipling the original euclidian distance by $i$ in order to increase the distance and consequently decrease the effect of that specific gradient.

The second intersection algorithm detects collisions with the three-dimensional models in order to model the cell adhesion behavior (see Algorithm 7). This algorithm will avoid the GC to pass through the tissues. In the neurobiology it is known that the physical constraints of the tissue helps to guide the axons by reducing its search space.

## 4.3 Direct mapping guidance model

After having defined all the elements and their parameters must be decided how the growth cone will convert all the extracellular information into a turning angle.

The current neurobiological knowledge does not allow to propose a complete model that can deal with the intracellular pathways and molecular concentrations (e.g. Calcium, Sodium, Fosfatum, cAMP and IP3). Therefore, throughout this chapter we have been proposing a model that focus on a direct mapping between the receptor activation into a turning movement. This direct mapping is already possible to implement because there are enough neuroscience knowledge that can support most interactions between the receptors and the extracellular environment. The only intracellular element that we consider are the proteins that may regulate receptors or other proteins. This approach is based on direct mapping between the response returned by the receptors into a attractive/repulsive force.

## 4.4 Evaluation

The evaluation of a computational model is very important because it allows to measure how good the abstraction is.

Comparing the results obtained by our model with current neuroscience knowledge is a possible approach (similar to the target system validation defined in [92]). From this comparison we defined several metrics:

[3]http://www.openscenegraph.org/projects/osg/browser/OpenSceneGraph/trunk/src/osgSim/LineOfSight.cpp#L152

**Input**: 3DModel $m$, Point $s$, Angle $\alpha$, Angle $\beta$, Growth length $L$
**Output**: Angle $\alpha$, Angle $\beta$

$e = move(\alpha, \beta, L)$ ; ▹ Equation 4.3
$e_1 = move(\alpha, \beta, LK)$ ; ▹ Equation 4.3
$e_2 = move(\alpha + \frac{\alpha}{100}, \beta + \frac{\beta}{100}, LK_2)$ ; ▹ Equation 4.3
$i, P1 = computeIntersections(m, s, e_1)$; ▹ OpenSceneGraph function
**if** $i > 0$ **then**
    $i, P2 = computeIntersections(m, e, e_2)$; ▹ OpenSceneGraph function
    **if** $i > 0$ **then**
        **if** $distance(s, p1) < distance(s, p2)$ **then**
            $\alpha = atan2\left(\sqrt{(P2_z - P1_z)^2 + (P2_x - P1_x)^2}, P2_y - P1_y\right)$
            $\beta = atan2\left(P2_x - P1_x, P2_z - P1_z\right)$
        **else**
            $\alpha = atan2\left(\sqrt{(P1_z - P2_z)^2 + (P1_x - P2_x)^2}, P1_y - P2_y\right)$
            $\beta = atan2\left(P1_x - P2_x, P1_z - P2_z\right)$

**Algorithm 7**: Algorithm for calculation of the new $\alpha$ and $\beta$ angles when the GC collides with the three-dimensional model. The point $e$ corresponds to the next position of the GC. While the point $e_1$ represents the next GC position and is calculated by increasing $L$ (growth length) $K$ times, being $K$ a constant (e.g. 10). The point $e_2$ follows the same principle, but with a different constant ($K_2$, with $K_2 > K$). $P1$ and $P2$ are the two intersection points that allow us to determine define a line between them and then determine its direction. After, the direction of this new line is calculated and and the resultant angles will define the new direction of the GC. With this algorithm the new angles make the GC to grow parallely to the 3d model in a specific collision zone, functioning as cell adhesion.

1. **Decisions on critical points** ▹ In all AG systems there are several points that are critical in the pathfinding. The direction that the axon follows when crossing these points is one of our metrics. Using this metric we can make structural comparisons with the biological systems. Possible values: *down*, *up*, *left*, *right*, *continues* (when axonal navigation do not changes).

2. **Receptors and proteins** ▹ One of the most important ways of analysing such a model is by measuring the differences in the activities and concentrations of the receptors and internal proteins during the simulation. Then, compare their relative differences with what is known from experimental works.

3. **Axon length/number of simulations** ▹ The final axon length and the total number of simulations are interesting metrics to measure the effect of the components in a model. For instance, they are useful to analyse the relevance of the *axonal transport* mechanism.

4. **Final topographic map** ▹ This metric defines with which target neurons the source neurons established connections and allows us to analise the system from a network prespective. May help to understand how a variation on the initial state can prevented the axon from finding its target.

5. **Visual** ▹ This is the simplest metric but probably the best to understand what is happening from a macroscopic point of view. This metric is based on a visual comparison (in static images or in video) between the simulation results and the visual results captured by experimental works, using techniques such as imunoflourescence.

All the metrics listed above are considered in our experiments based on biological systems (see chapter 6).

Other possible way of evaluating our model is by comparing it with other works, however, this is difficult as do not exist benchmarks, the several approaches tend to be quite different and they focus on different problems.

# Chapter 5

# yArbor, an axon guidance simulator

Simulators currently available in close fields could have been adapted to our needs, but they were developed for different purposes what would reveal difficulties during the implementation and a low control of all the system. Therefore, we decided to develop our own simulator so that we can control the system completely and implement the model with less effort.

Performance is a critical point when dealing with complex system; therefore, the simulator was developed in C++ a rich programming language that can achieve good performances.

The programming approach applied is object-oriented with the design pattern model-view-controller (MVC) pattern in mind. With this model was possible to isolated the user interface and the data from the computational model. With this approach the graphics can be easily disabled in order to improve the computation performance.

The simulator was developed in Qt[1] a cross-platform application and user interface framework that reduces the effort of developing a software of this dimension.

The fact that the software Avogadro (see section 3.2.2) is implemented using the Qt framework reinforced our trust in this platform.

In terms of software complexity our simulator has a total of 36 classes, some are directly related with the elements and mechanisms, others with auxiliar features.

This simulator allows the study of axon guidance in an unique way because several experiments can be performed and its parameters can be varied easily. It has a user-friendly graphical user interface (GUI) so that anyone with some knowledge about the problem could use it.

The QSettings class[2] of the Qt framework was used to save the simulations in the yArbor format with the extension yrb. The GUI allows the user to easily open, save or create simulation files (see for example Figure 5.3). The OpenSceneGraph[3] (OSG) toolkit was used to create and import three-dimensional objects. OSG is an open source high performance 3D graphics toolkit, well established and widely used by developers in fields such as visual simulation, games, virtual reality, scientific visualization and modelling.

[1] http://www.qtsoftware.com
[2] http://doc.trolltech.com/4.5/qsettings.html
[3] http://www.openscenegraph.org

Both the Qt framework and OSG support the different operative systems; therefore will be possible to deploy the simulator for Linux, OSX and Windows.

In this chapter we briefly explain the five modules of our simulator (see Figure 5.1).

Figure 5.2 illustrates the simulation structure of yArbor. The computational model captures information from the data and graphics modules and invokes in all elements the next simulation step. These elements apply the mechanisms, which leads to an update in the state of the computational model. The results of each update are then drawn in three-dimensions and the respective results (plots, axonal length and number of simulation steps) are updated.

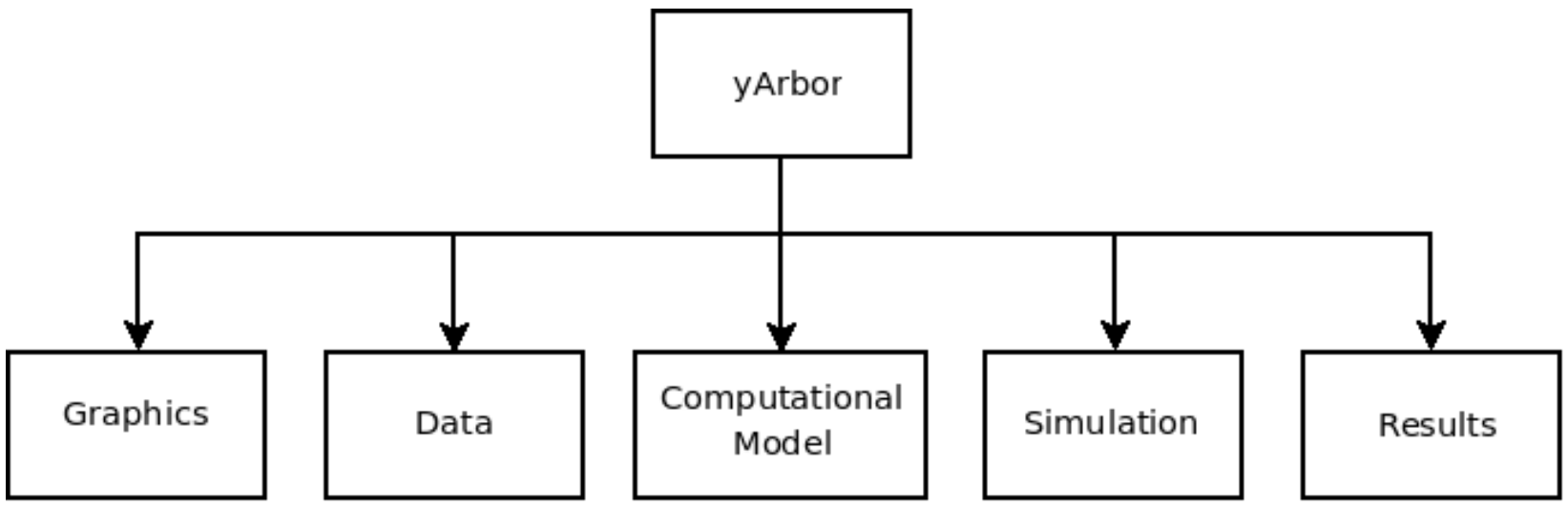

Figure 5.1: yArbor modules.

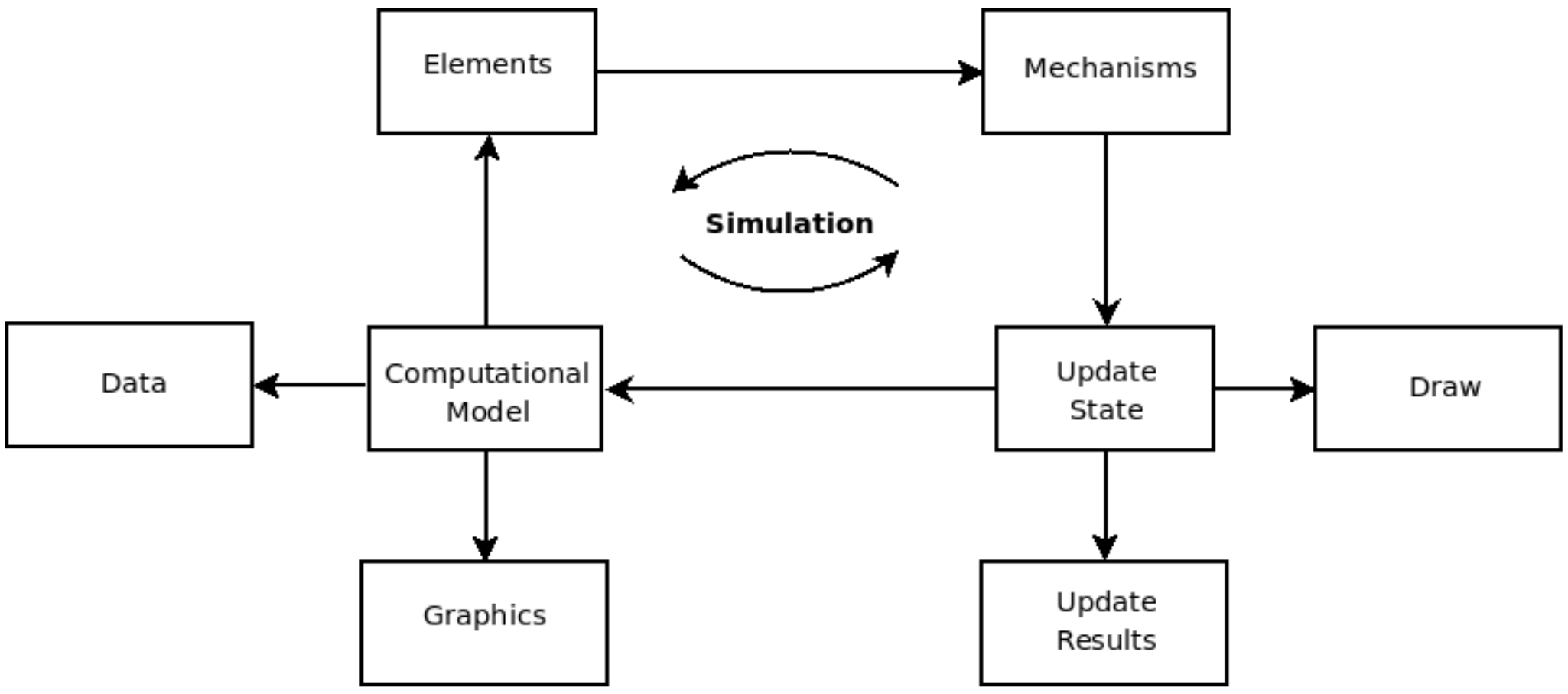

Figure 5.2: yArbor simulation cycle. The computational model captures information from the data and graphics modules and invokes in all elements the next simulation step. These elements apply the mechanisms, which leads to an update in the state of the computational model. The results of each update are then drawn in three-dimensions and the respective results (plots, axonal length and number of simulation steps) are updated.

The name *yArbor* results from the combination of *y* that is a tribute to *Santiago de Ramón y Cajal* a neuroscientist pioneer in this and many other topics. The word *Arbor* means tree in *latim* and it is a metaphor between the perfection and beauty found in *arbors* and the development of neuronal *arbors*.

## 5.1 Data

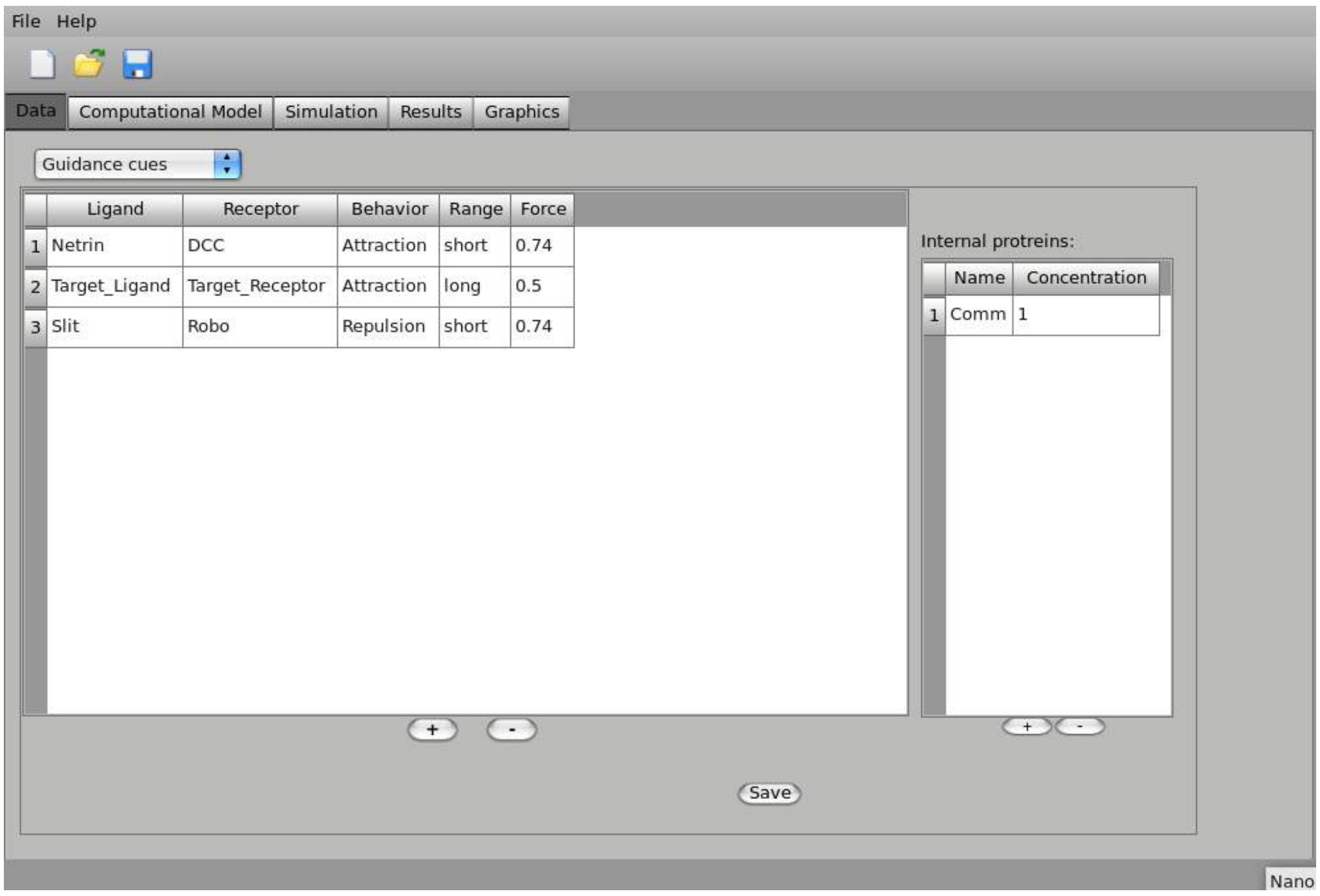

Figure 5.3: GUI of guidance cues in the data module. The screenshot was taken using the Drosophila Midline simulation.

The data module is where all neuroscience knowledge for a specific system must be added. More specifically, this module contains the following sub-modules:

- Guidance cues ▹ Figure 5.3 represents this sub-module, where is possible to add guidance complexes and proteins.
- Receptors and ligands ▹ Where can be added/removed/edit the receptors and ligands.
- Topographic Map ▹ Defined by a bidimensional matrix where it is possible to define the source-target neuron connections.
- Regulatory Network ▹ Defined by a square matrix with values between -1 and 1.

This module brings a great flexibility in the management of neuroscience data (e.g. guidance complexes).

## 5.2 Computational model

The computational model presented in chapter 4 is the main basis for the simulator. In this module is possible to add elements and configure their initial state. By using the mouse over the widget on the right side of Figure 5.4 it is possible to rotate, scale and translate the three-dimensional scene.

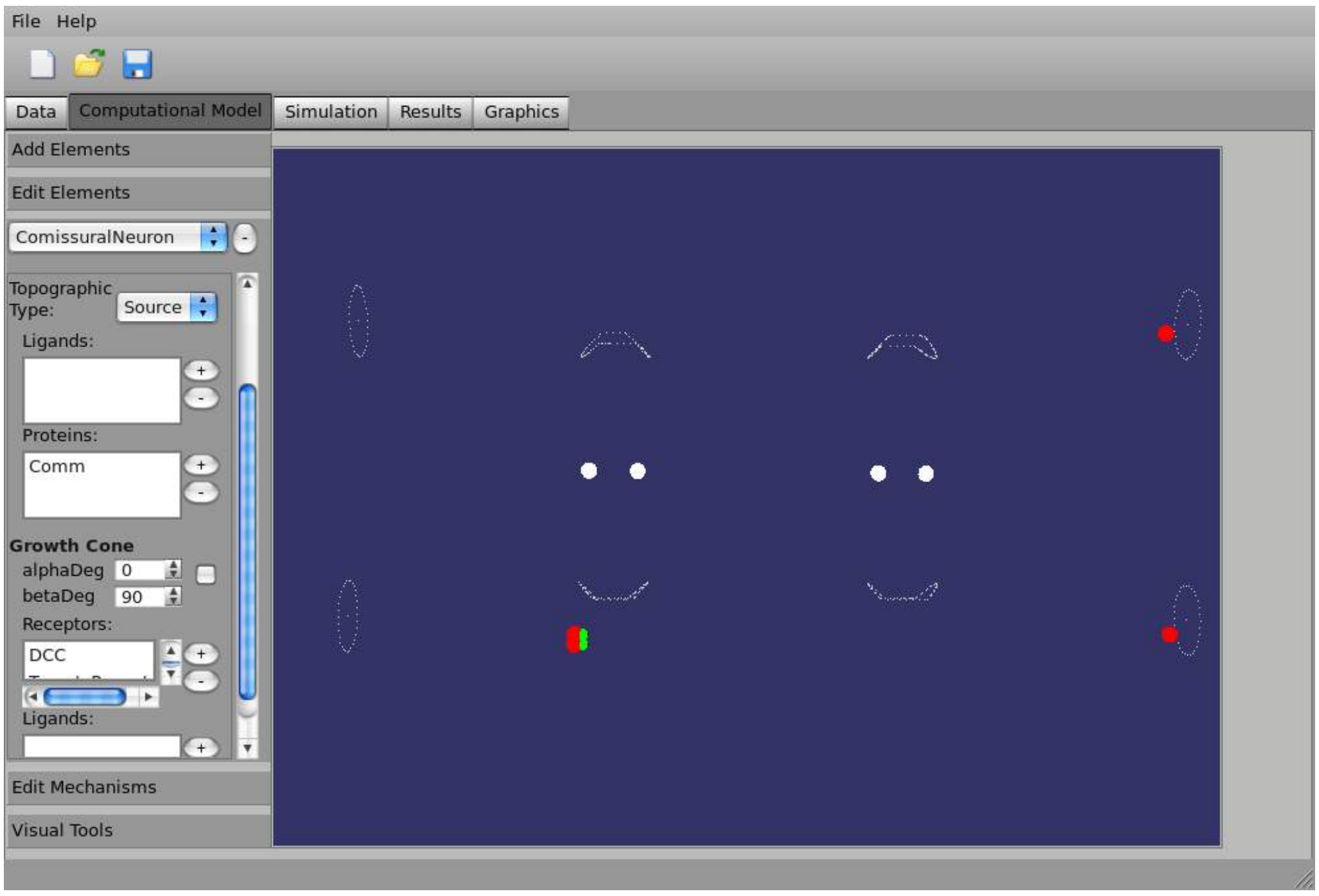

Figure 5.4: GUI of a source neuron edition in the computational model module. The screenshot was taken using the Drosophila Midline simulation.

## 5.3 Simulation

The simulation module allows the user to control the simulation, by simulation *n* steps (see a screenshot in Figure 5.5). This module interacts with the results module in order to update the simulation results. Note that the changes made in the GUI during simulation do not affect the initial state of the computational model defined in the computational model module, but could affect the simulation current state.

## 5.4 Results

This module allows to plot the concentration and activities of receptors and proteins during the simulation (in tab *simulation*), Figure 5.6). It also shows the total axonal length and number of simulation steps (in tab *others*). In the GUI the user can add vertical lines to mark a specific simulation step, for instance to mark the simulation step when the axon passed a critical point. The user can also save the plot for posterior analysis.

## 5.5 Graphics

Three-dimensional objects representing tissue or other neural obstacles can be added in this module (see Figure 5.7). These objects can be created in modeling software such as Blender[4] and then be imported in this module. Due to the use of the OSG graphics toolkit, the system supports the most used 3d formats, such as: 3D Studio, COLLADA or Alias Wavefront.

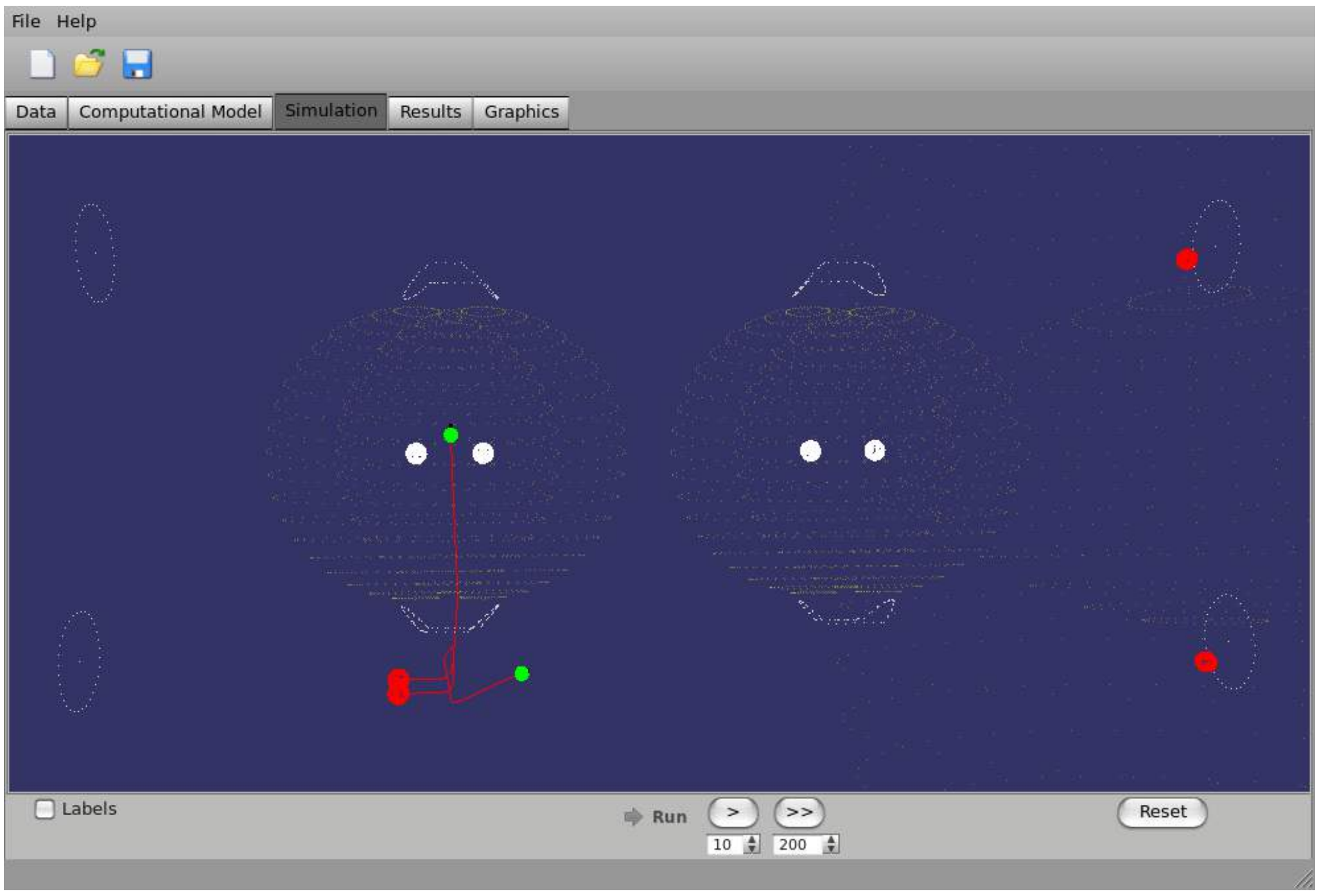

Figure 5.5: Screenshot of the simulation module.

[4] http://www.blender.org

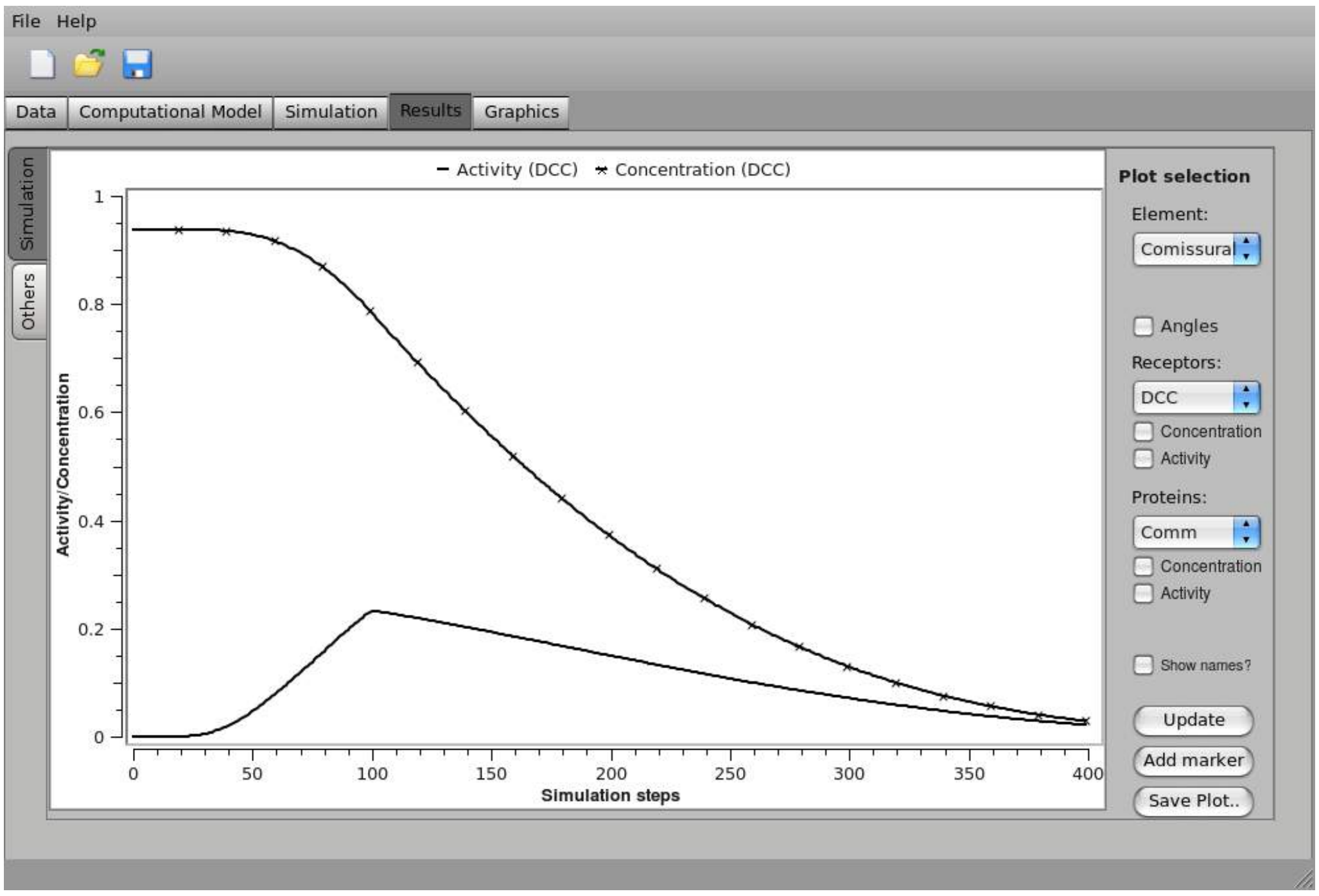

Figure 5.6: GUI the results module, with a plot for activation and concentration levels of the Robo receptor. The screenshot was taken during the Drosophila Midline simulation.

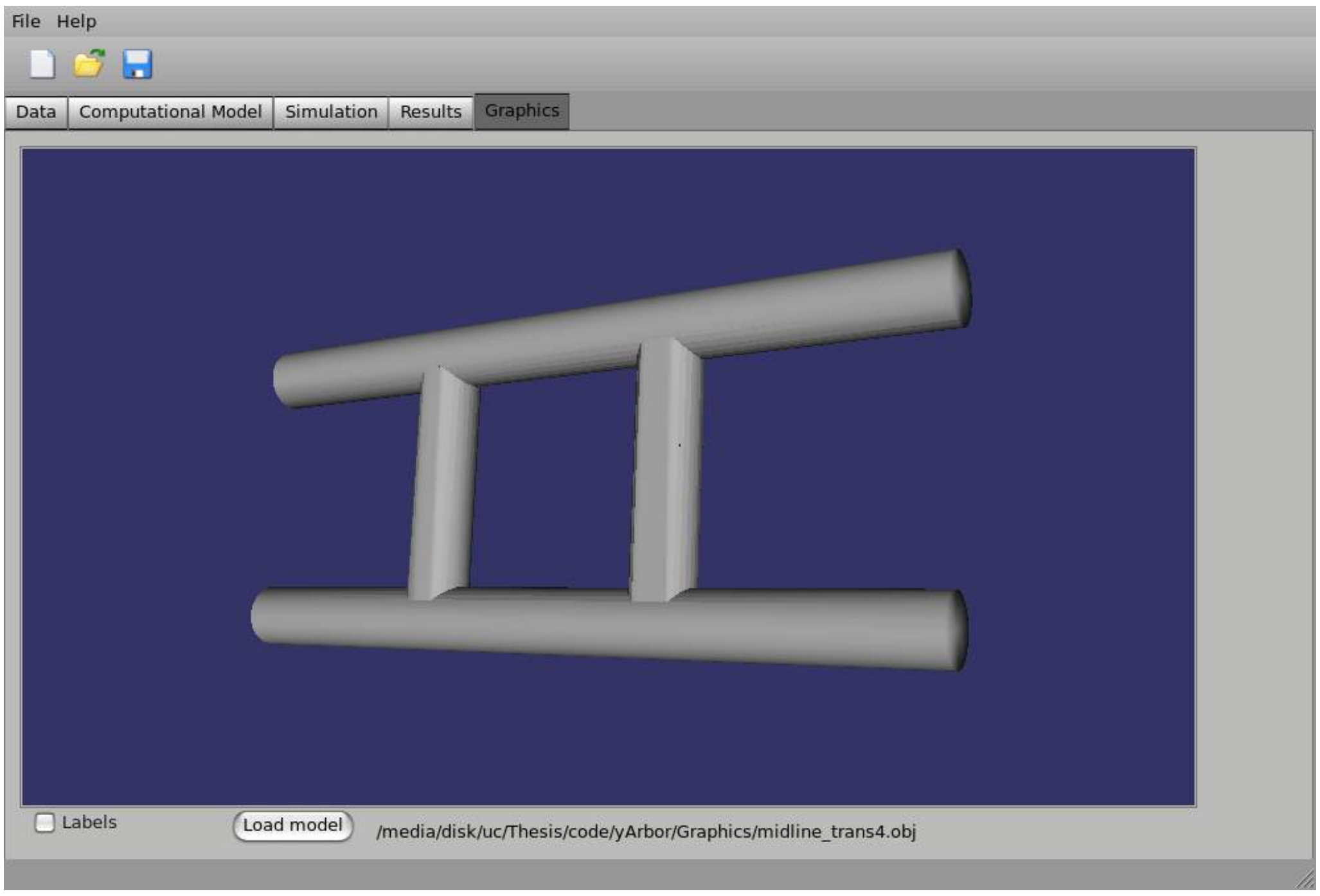

Figure 5.7: Simple GUI for the graphics module where 3d models can be imported (e.g. three-dimensional midline).

# Chapter 6

# Experimentation

This chapter presents the experiments made in order to evaluate our model. First, simple studies were performed to analyse how the model behaves for each element independently of all the others. Second, experiments with the *Drosophila* midline crossing model, a well characterized axon guidance system, were done. These simulations consider three different neurobiological scenarios (normal, comm mutant and robo mutant) and the evaluation process takes into account (i) the structure (decision points, axon length and target innervation), (ii) the activity and concentration of the GC receptors and proteins , (iii) the growth angles and (iv) the resultant visual output.

It is important to note that all the experiments were performed with a growth step length of 0.005 $\mu$m. This value empirically revealed to be a good choice. The option for microns as the reference unit was arbitrary; what is more important is the relative variations during and between simulations. The plots given throughout this chapter represent the values obtained during all the simulation; these values can be radians (for the alpha and beta angles) or level of activity/concentration for receptors and proteins. During the experiments presented in this chapter the $K$ and $K2$ intersection constants are 10 and 15, respectively.

## 6.1 Simple studies

In order to evaluate each part of the model it was necessary to perform independent studies of each element and mechanism. This section introduces these studies and discuss the relevance of each component for the the simulations. Several other experiments could be added to this section, but it was decided to include only the most relevant ones, that could allow the comprehension of the importance of each component included in our computational model.

### 6.1.1 Single source-target pair

In this experiment a simple source neuron (with its axon and GC), the correspondent target neuron and an attractive guidance complex were simulated. The GC contains the receptor of this guidance complex and the target neuron diffuses the ligand; this will lead the GC to be attracted by the target neuron, as can be seen in Figure 6.1.

Figure 6.2 demonstrates the angles variation (alpha and beta) and the increase in the receptor (ReceptorX) activity, while the receptor concentration keeps constant at 0.9.

The ligand is defined by an exponential function with a range of 1.5 and the guidance complex has a force of 0.6.

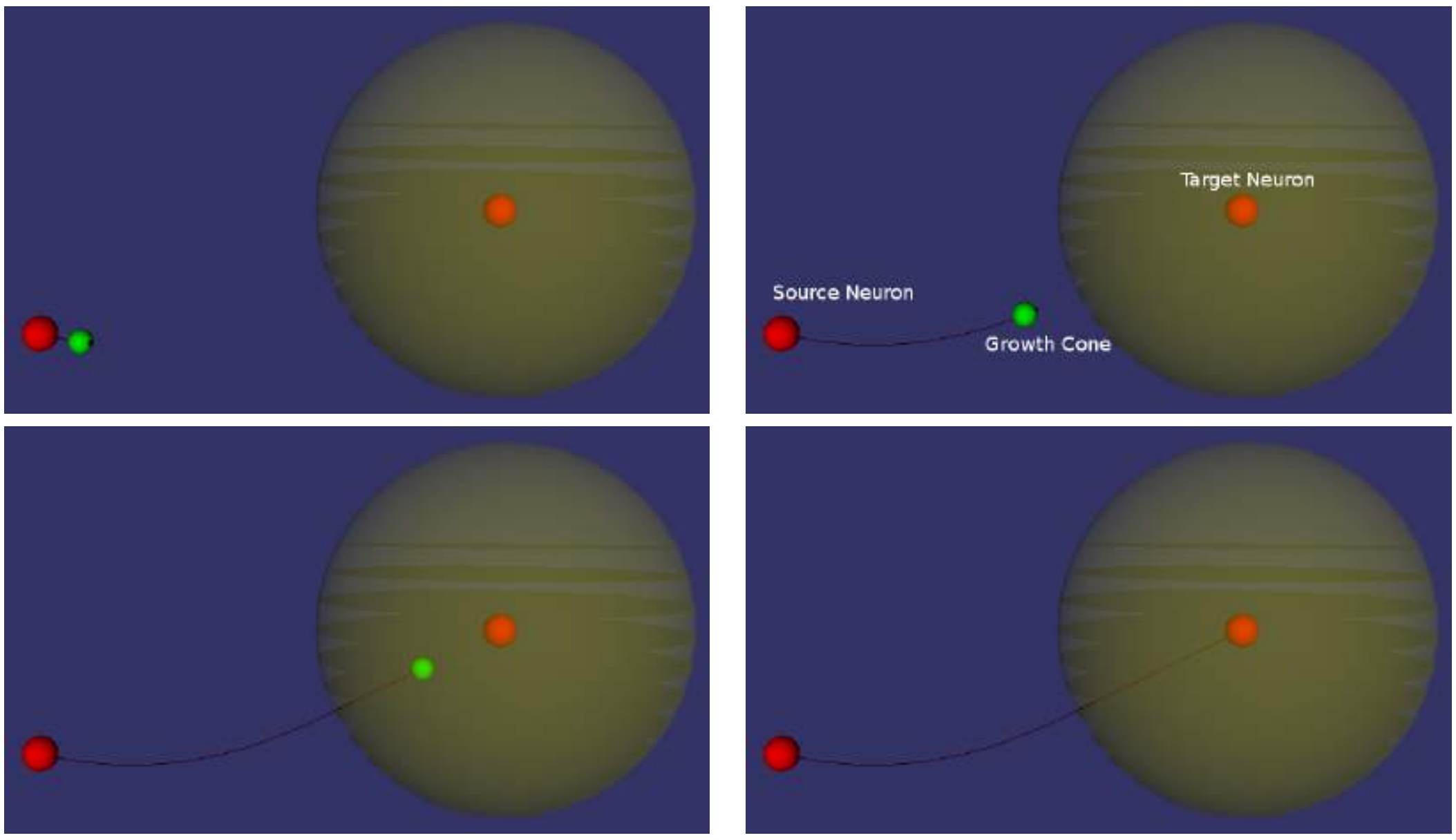

Figure 6.1: Sequence of steps from a simple simulation with a single source-target pair and one guidance complex. The GC of the source neuron grows toward the target neuron.

### 6.1.2 Simulation with manifold elements

In this experiment two source neurons, three attractive target neurons, four repulsive glial cells and two guidance complexes (one attractive with 0.6 force and the other repulsive with 0.3 force) were simulated (see Figure 6.3). The source neurons can connect to any target neuron because their topographic map was not predefined. The axon path is affected by both the attractive and repulsive forces. Near the targets these forces become similar, hence the variations observed in Figure 6.4 close to the 1500$^{th}$ simulation step.

### 6.1.3 Mechanisms

As it is shown in this section the mechanisms considered in our model appear to be relevant for axon guidance modeling. Each mechanism explores different prespectives of the problem and some seem to be more important than the others.

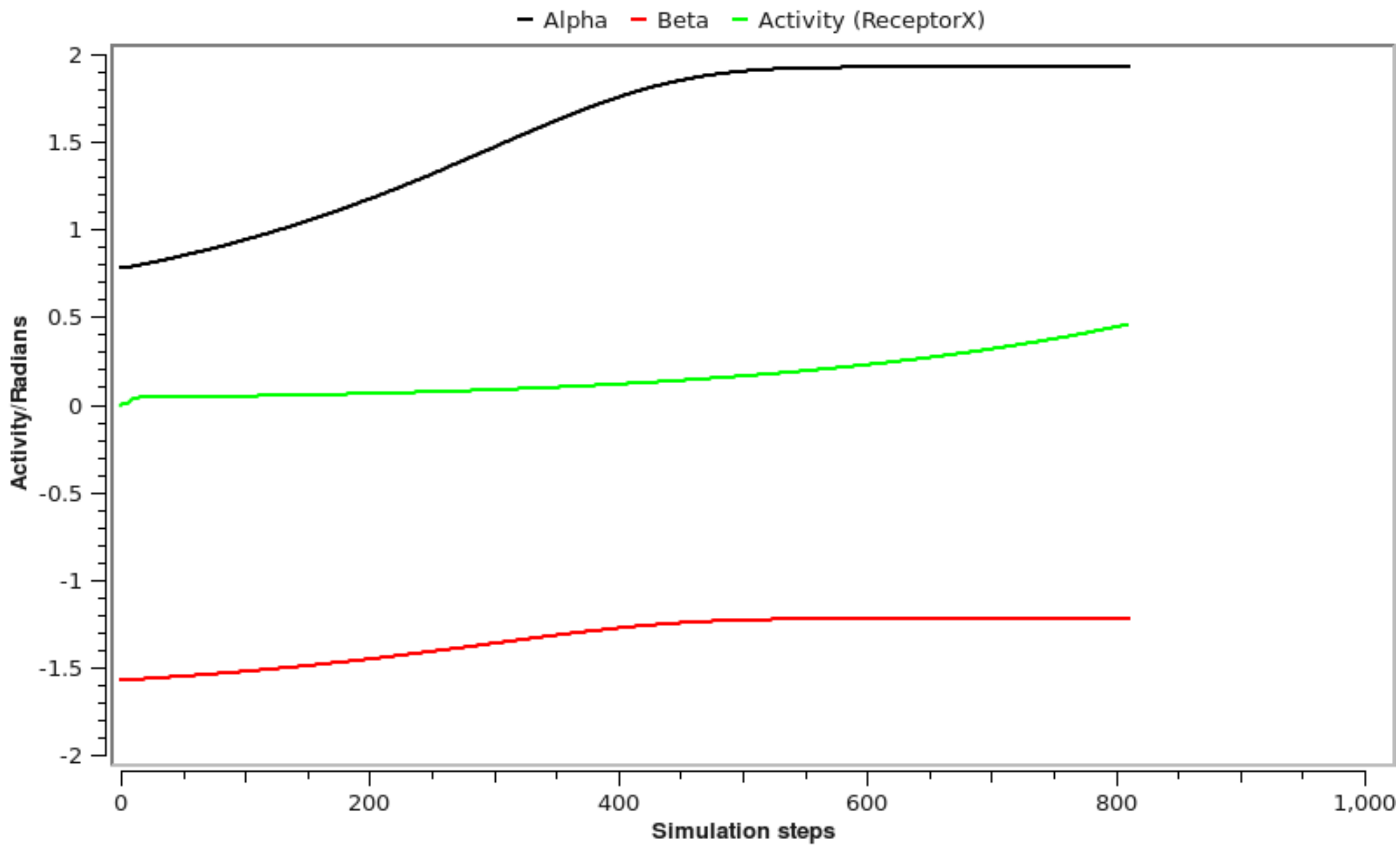

Figure 6.2: Receptor activity (ReceptorX) and angles variation of the simple simulation with a single source-target pair. Beta and alpha represent the rotation of the GC in spherical coordinates during the simulation.

### Adaptation

As mentioned during this thesis, adaptation is one of the most important mechanisms. It allows the GC either to adapt to attractive sources that represent intermediate targets (e.g. glial cells) or to ignore the repulsive ones by removing/adding receptors from/to the GC membrane. This mechanism is demonstrated in this section with a simple experiment containing one source-target pair and a glial cell (see Figure 6.5). Figure 6.6 shows a sharp drop in the concentration of receptor ReceptorX2 (desensitization), while it is activated by the ligand released by the attractive glial cell. Then this allows the GC to grow toward its target, as demonstrated by the increase in the ReceptorX activity.

### Fasciculation

The fasciculation experiment was defined by adding a new ligand that is released by the pioneer GC and a new axon with the corresponding receptor, both linked by a new attractive guidance complex. This mechanism produces shorter axons (see Figure 6.7), a mechanism that seems to be relevant for the development of neural networks that are more efficient in terms of information propagation.

### Axonal Transport

The axonal transport affects the speed of the GC reaction, the longer is the axon the slower is its reaction to the extracellular cues. This is reflected in Figure 6.8, where the axonal transport is enable, leading to the development of a longer axon.

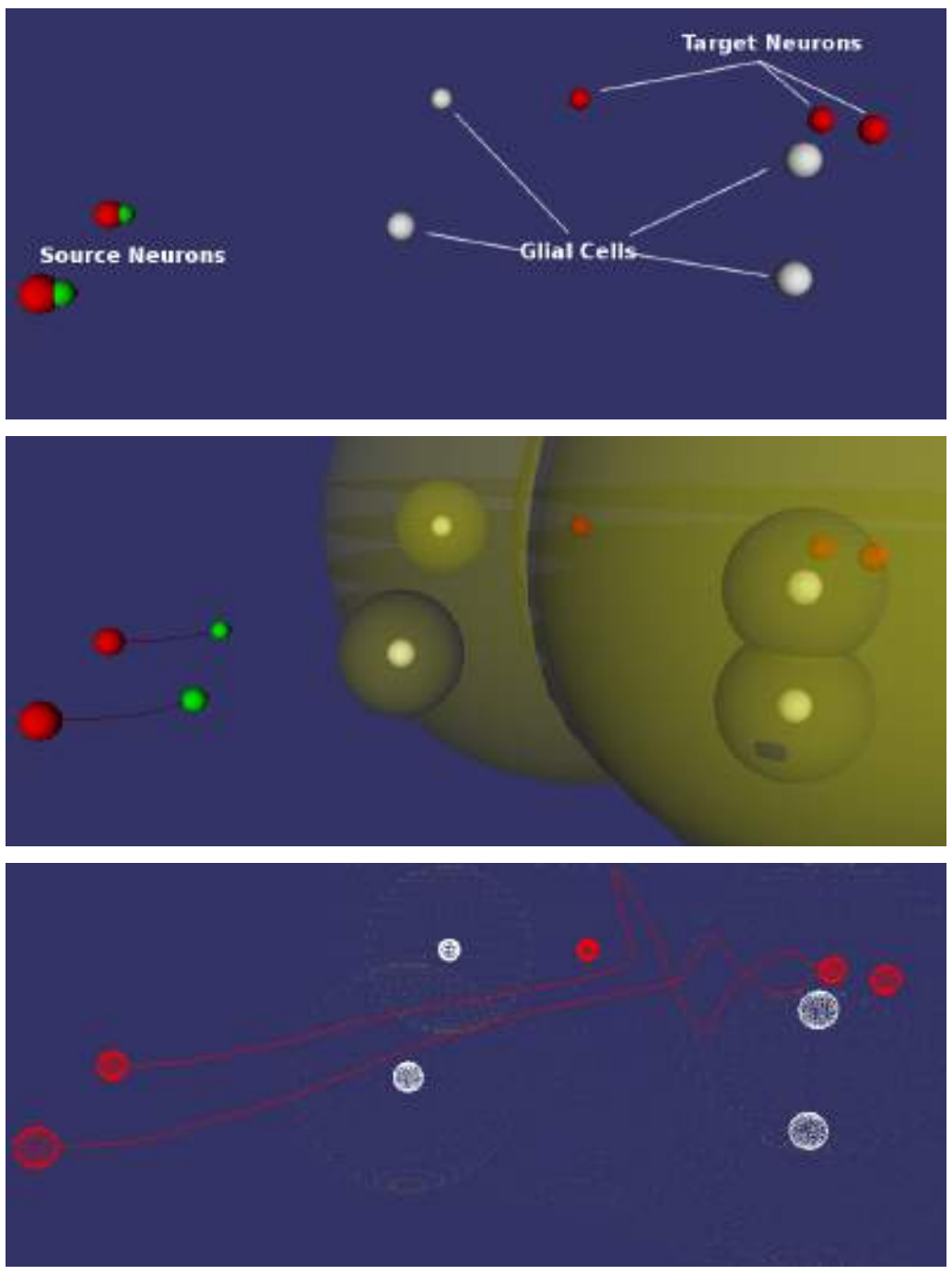

Figure 6.3: Sequence of simulation with manifold elements. The experiment includes two source neurons (red spheres on the left), three target neurons (red spheres on the right), four glial cells (white spheres on the center) and two guidance complexes. The GCs of the source neurons are repelled by the gradients of the glial cells and attracted by the gradients of the target neurons

#### Receptor limits

Unlike the previous mechanism this one is not affected by the axon length but by the proximity to the ligand source. Figure 6.9 illustrates a simple simulation where the minimum activation level of the receptors is 1% and the saturation level 5%. As a result, the receptors only start reacting to the extracellular cues when the activation reaches at least 1% and its response is not increased after being stimulated more than 5%.

#### Growth cone complexification

The growth cone complexification decreases the axon speed when the GC faces complex extracellular environments (i.e. with several guidance cues). Considering this simulation we can reinforce what is known from neurobiology, that states that the decrease of the GC speed in more complex extracellular environments allows the GC to decide more accurately. With the three gradients illustrated in Figure 6.10 the final axon length was 3.62 $\mu$m with this mechanism enabled and 3.66 $\mu$m when it was disabled.

#### Intersections

As described in chapter 4 we consider two types of intersections with the neural tissue. The first deals with the need to decrease the activation level when tissue is present between the GC and a gradient (for an example, see Figure 6.11). The second detects collisions with the tissue (e.g. midline) and lead the axons to growth paralelly to this tissue (functioning as cell adhesion, see Figure 6.12).

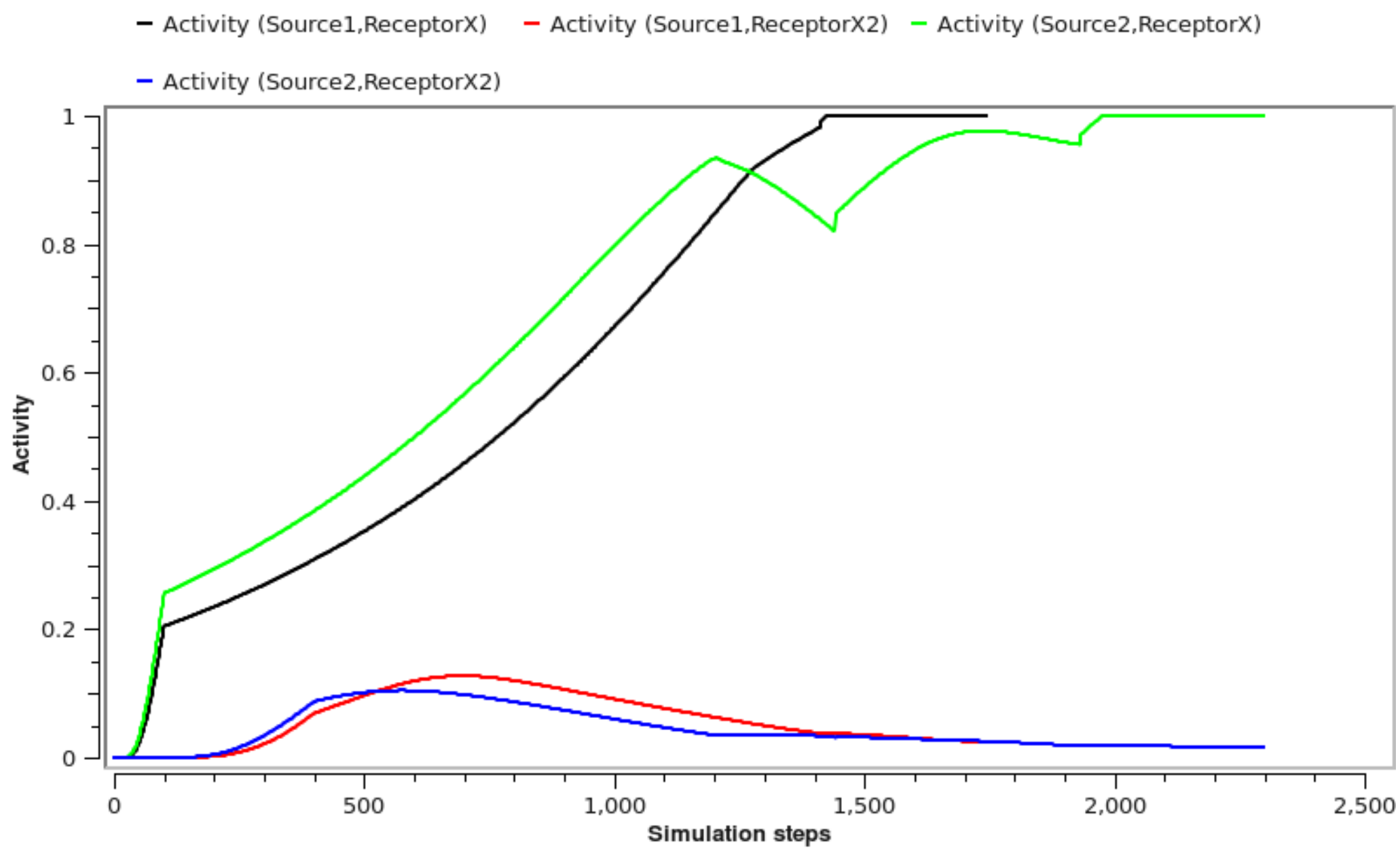

Figure 6.4: Activities of the receptors on both source neurons (Source1 and Source2) in a simulation with manifold elements.

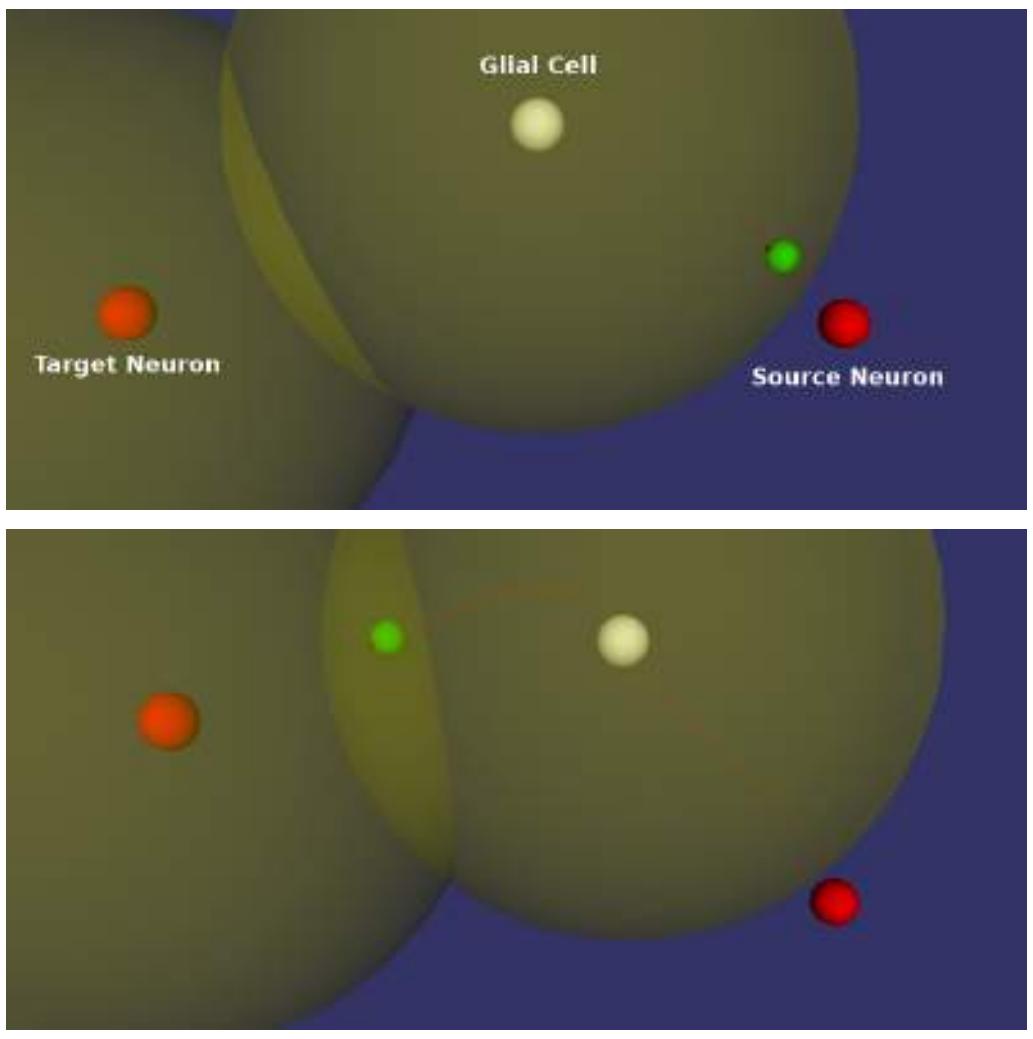

Figure 6.5: Sequence of simulation demonstrating the adaptation mechanism. The adaptation process allows the axon to adapt to the attractive gradient released by the glial cell (white sphere). This occurs due to the desensitization of the receptor that is stimulated by the ligand released by the glial cell and lead the GC to grow towards its target due. The GC passes through the glial cell because the detection of intersections between elements was not considered in our model.

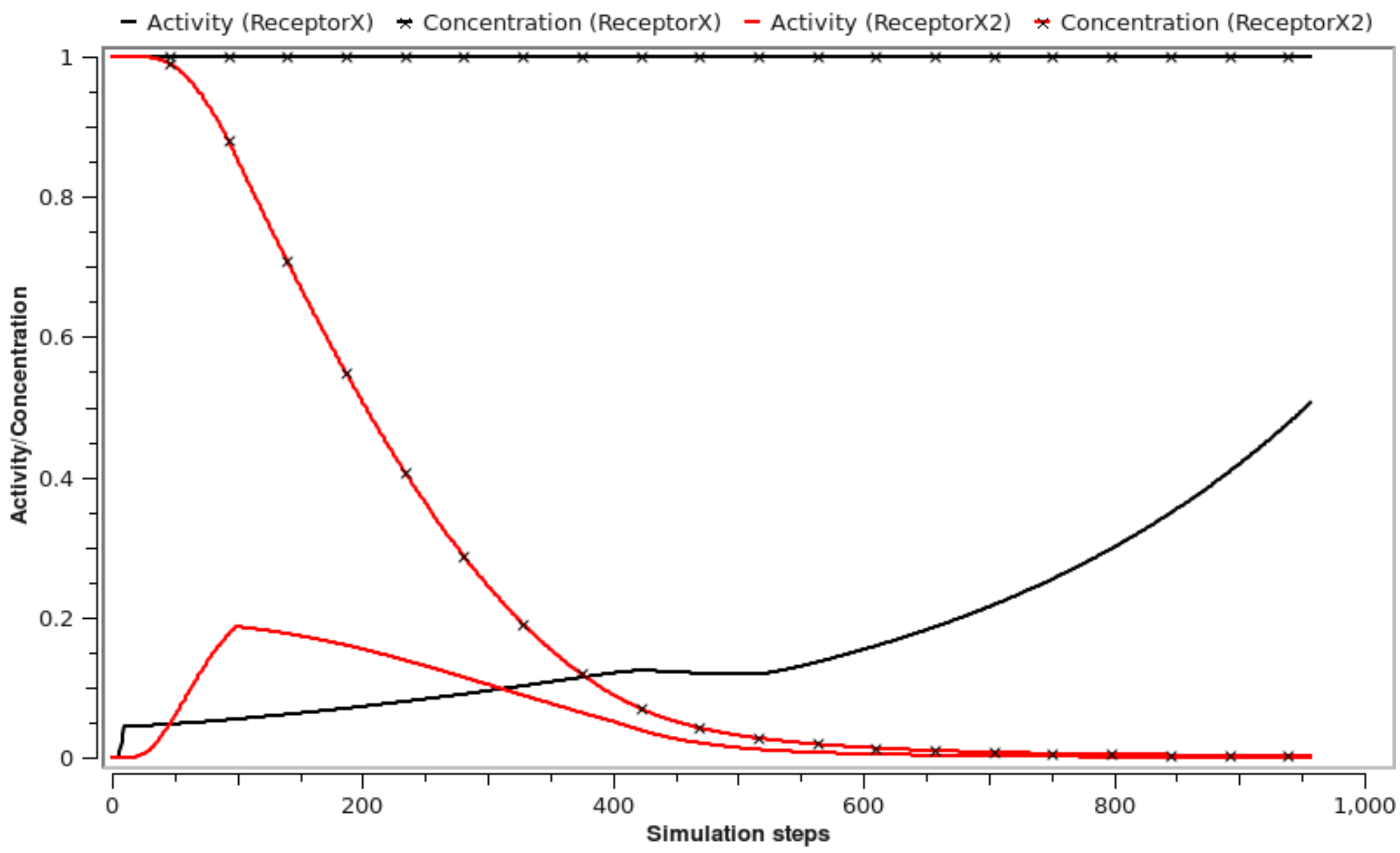

Figure 6.6: Activity and concentration of the source neuron during a simple simulation of the adaptation mechanism. The glial cell releases a ligand that stimulates the ReceptorX2 and the target neuron a ligand that activates the ReceptorX. The removal rate of the ReceptorX2 is 2% and the production rate 0%. This leads to a fast desensitization of the ReceptorX2.

### 6.1.4 Discussion

In the last sections we presented simple simulations that aimed to evaluate the importance of the elements and mechanisms independently of each other. Even with simple simulations we could to achieve interesting results concerning their benefits and about their impact on axon guidance. This is consistent with the computational model proposed by Goodhill [33] where he showed that simple mechanisms are sufficient to obtain realistic paths.

## 6.2 The *Drosophila* Midline

Unlike the previous experiments in this section we present a study based on a biological system that includes most elements and mechanisms. We propose a computational model based for the *Drosophila midline*, one of the most studied system. However it is still a system with many uncertainties and mysteries that our model could help to solve.

This system was introduced in section 2.6.1. Here we present a computational model of axon guidance for the *Drosophila* midline crossing that was obtained using our generic computational model and discuss the results obtained by the simulations.

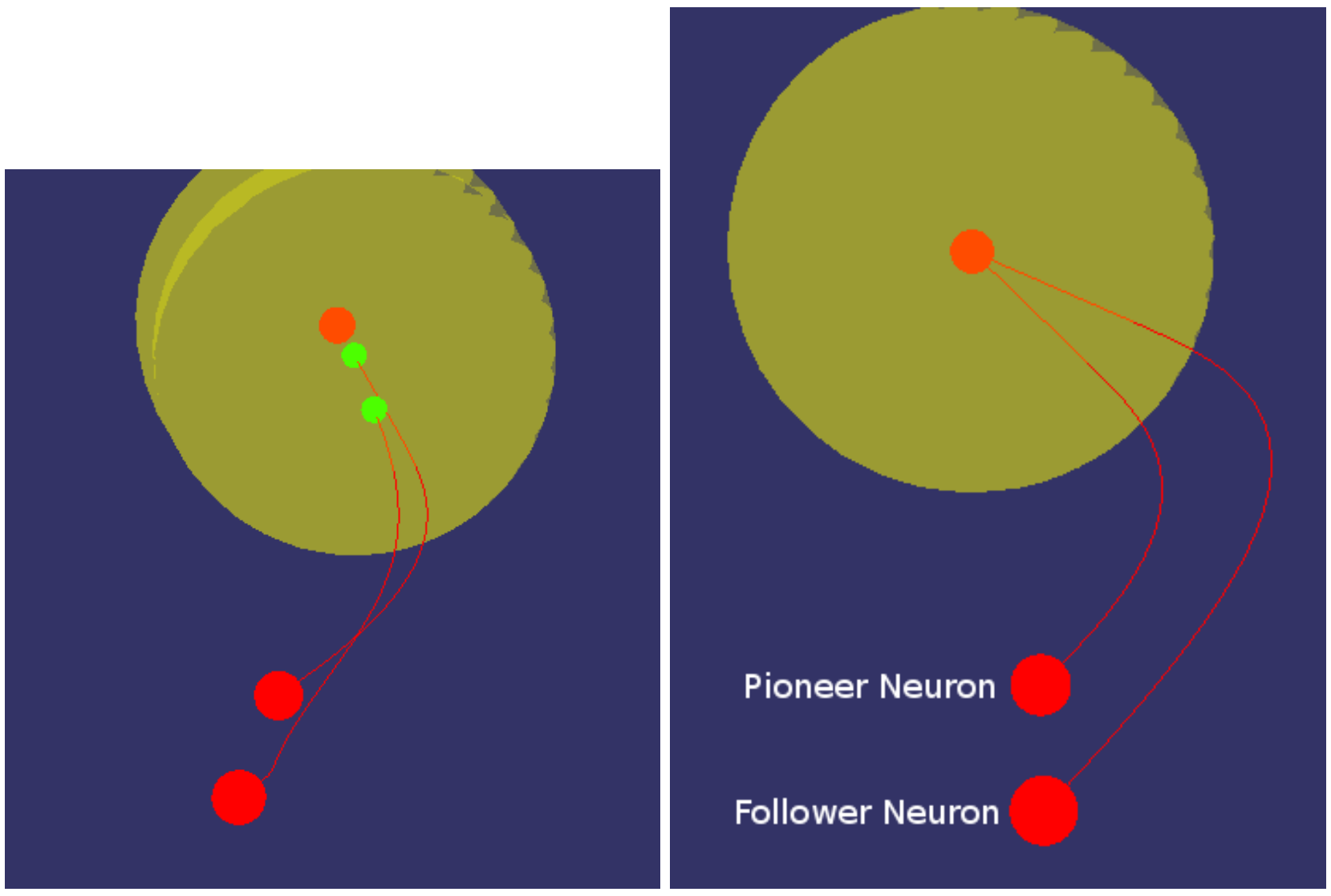

(a) With fasciculation (follower axon ended with 4.85 $\mu$m).

(b) Without fasciculation (follower axon ended ended with 5.54 $\mu$m).

Figure 6.7: Simple simulation with the fasciculation mechanism.

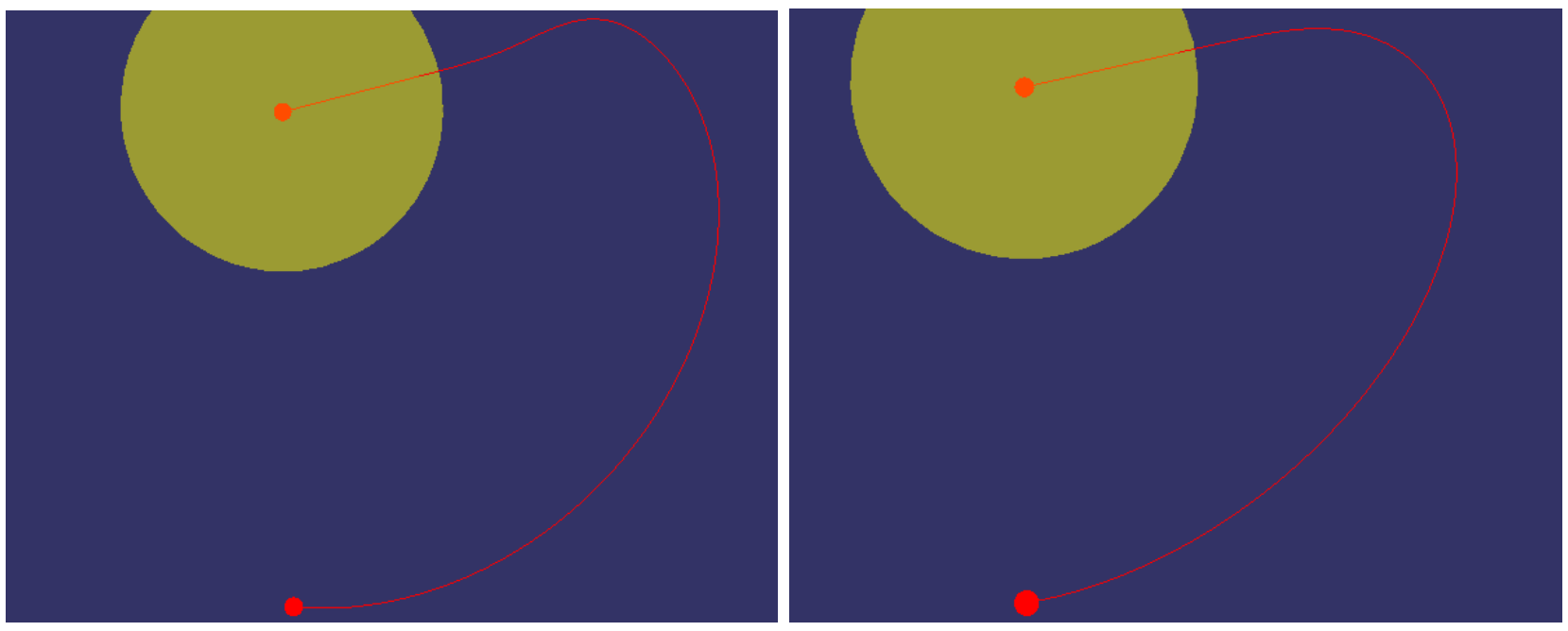

(a) With axonal transport (axon ended with 18.34 $\mu$m and 3667 simulation steps).

(b) Without axonal transport (axon ended with 15.29 $\mu$m and 3058 simulation steps).

Figure 6.8: Simulation with a single source-target neuron to studys the axonal transport mechanism.

### 6.2.1 Midline model

In this section we explain how the midline model was defined considering our computational model. All the elements introduced have a direct or at least indirect neurobiological grounding. A three-dimensional model that represents the midline was developed (see Figure 6.13(a) and 6.13(b)) based on the representation given in section 2.6.1, Figure 2.8. This model is a possible representation of the Midline that considers two possible crossing points. Using this simple three-dimensional model, it is possible to: (i) study the most important decision points and (ii) introduce a physical barrier similar to the real

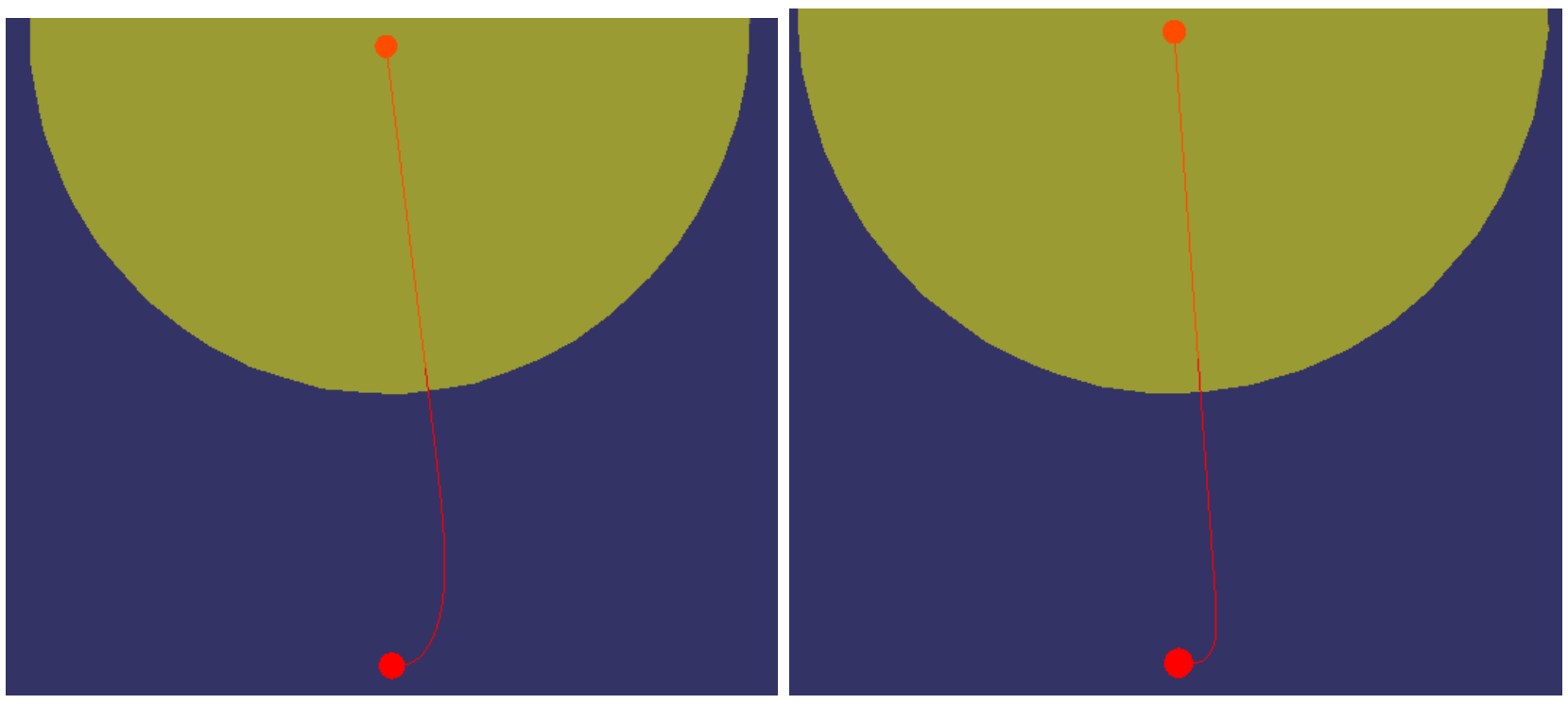

(a) With receptor limits (axon ended with 8.12 $\mu$m and 1623 simulation steps)

(b) Without receptor limits (axon ended with 7.87 $\mu$m and 1573 simulation steps)

Figure 6.9: Simulation with a single source-target neuron to studys the receptor limits mechanism.

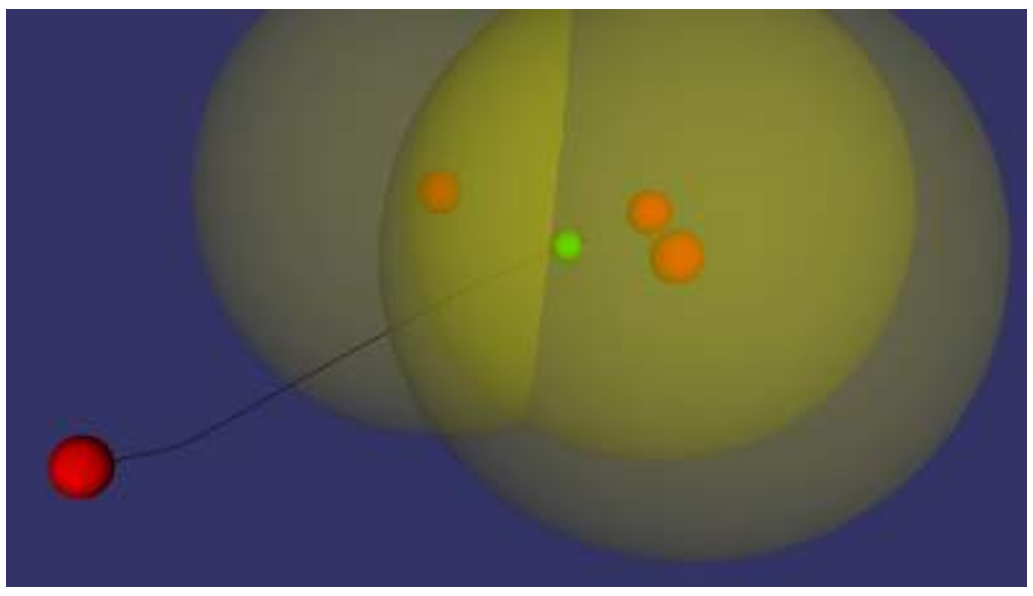

Figure 6.10: Simulation with the growth cone complexification mechanism (with three attractive cues on the center). When the GC starts to perceive the three gradients its growth speed decreases, allowing more accurate decisions.

midline tissue.

Figure 6.13 shows the general structure of the midline model and all the parameters considered are given in Tables 6.1-6.8.

The parameters presented throughout this section gave realistic results. This parameters were obtained by starting with values close to what is known from neuroscience and then several variations were introduced in order to obtain realistic results. This fine-tuning process was performed towards the most realistic paths (i.e. similar to the observations made *in vivo*) for three different situations known from the neuroscience:

1. Normal pathfinding, where the commissural axon cross only once and then connects to its target.
2. Comm mutant pathfinding, where both types of axons never cross the midline.
3. Robo mutant pathfinding, where both types of axons cross and recross the midline.

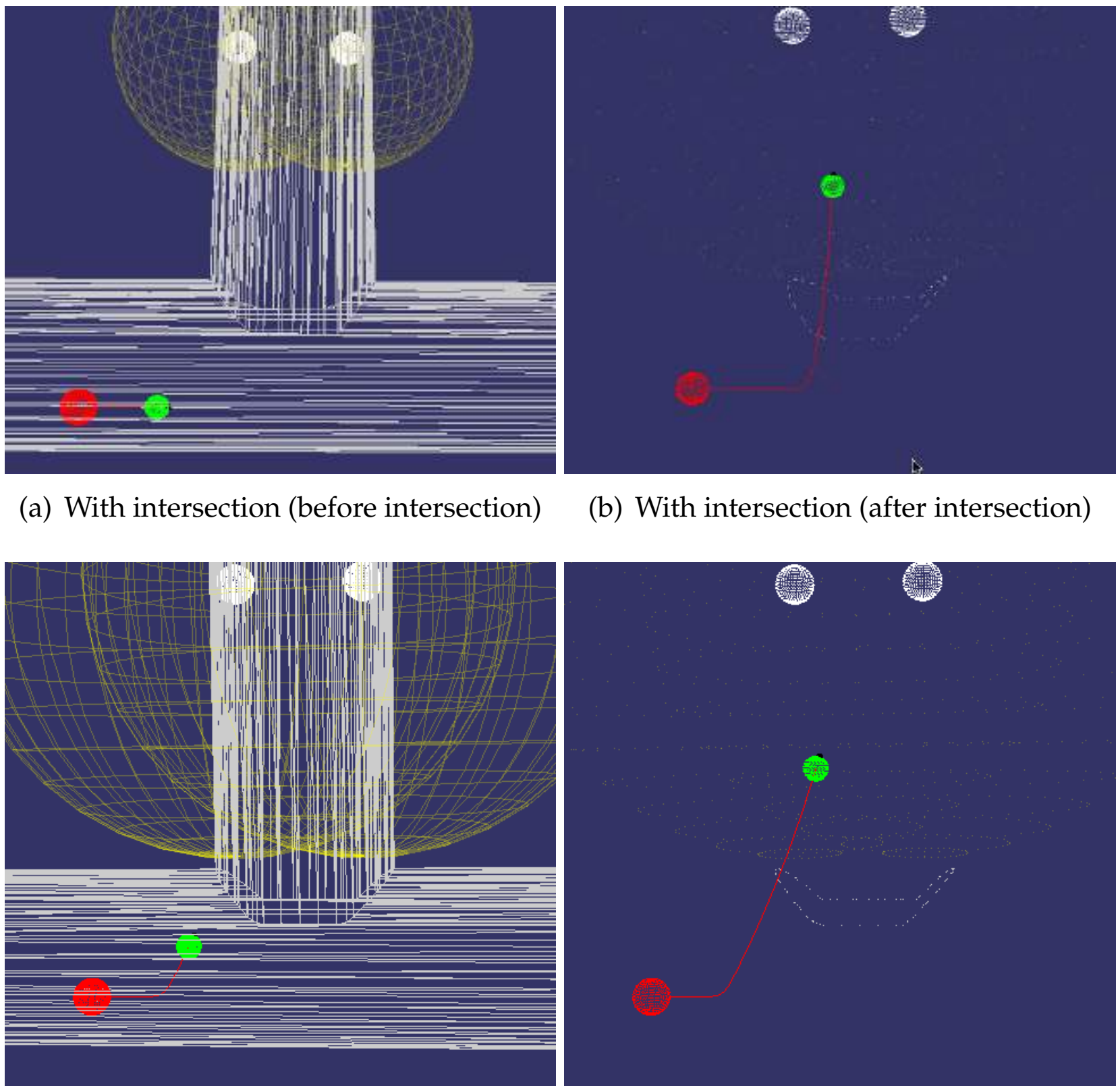

(a) With intersection (before intersection) (b) With intersection (after intersection)

(c) Without intersection (first figure) (d) Without intersection (second figure)

Figure 6.11: Intersection type 1. When the intersection is disabled the GC turns to soon. With greater diffusion rates in the gradients the GC would have passed through the tissue. With this intersection mechanism the GC behavior is more realistic because the ligands produce a weaker stimulation when there are obstacles (e.g. tissue) between them and the receptors.

During this fine-tuning process were found several situations where the axon paths were realistic, revealing that the system tolerates slight variations.

### Neurons

Our model considers the two types of neurons known in this system, the ipsilateral neurons that do not cross the midline and the commissural neurons that cross the midline (in a normal pathfinding).

Both are source neurons (see Table 6.1 for their parameters), hence each one contains an axon and a growth cone that should connect to specific targets, forming a simple topographic map (see Table 6.2).

The source neurons represent sensorial neurons that take sensorial stimuli to the brain, their position in the midline is based on the observations made *in vivo*.

The target neurons can be seen as the *Drosophila* brain that receives connections from the sensorial neurons. Their positions in our model could be further away (i.e.

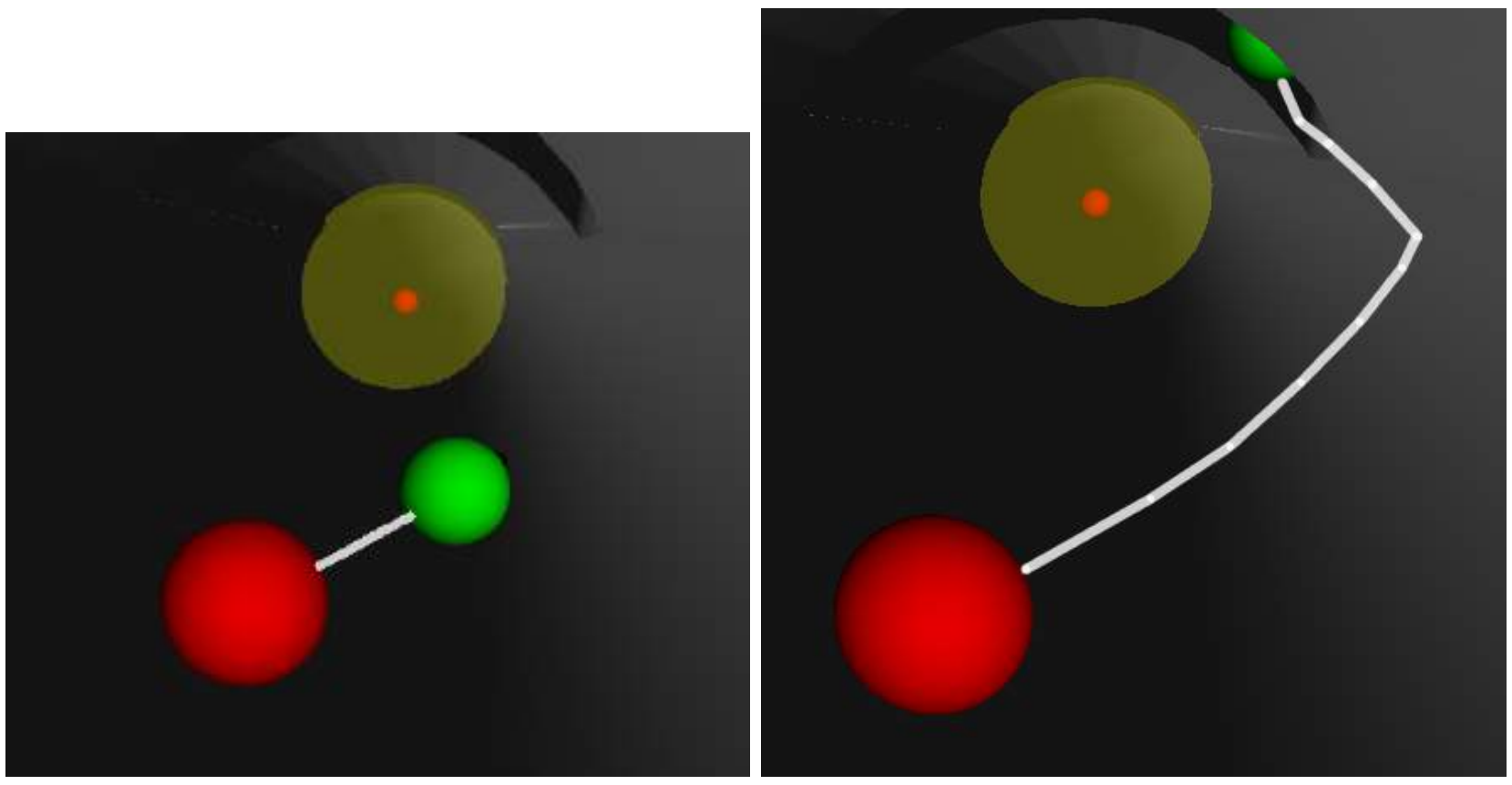

(a) Before intersection (b) After intersection

Figure 6.12: Intersection type 2. In 6.12(b) it illustrated the GC colision with the gray object (that could represent tissue) and its growth along with it.

closer to the real brain position), but what is important is that they function as the real target to the sensorial neurons (the *Drosophila* brain). Moreover with a predefined target we were able to compare different simulations by measuring the final axon length and the total number of simulations needed to reach the target.

### Glial Cells

Glial cells are important neuron partners during axon guidance. They seem to be particularly important when axons cross the midline, where they are both sources of attractive and repulsive forces. In this model several positions and concentrations of glial cells were tested, but the pattern presented in Figure 6.13(c) by the white spheres was empirically found to be the most appropriated (see their parameteres in Table 6.3).

This quantity and position of glial cells is at the same time central, which may suggest that there is a higher concentration of glial cells in the central region of the midline or/and that the glial cells diffuses more guidance molecules in this region.

### Guidance cues

The guidance cues are released into the extracellular environment by the glial cells and the target neurons, as can be seen by the spheres envolving these elements (Figure 6.13(c)). The ligands considered are: Netrin, Slit and a specific Target ligand (see Table 6.4 for all their parameters). The first two are known to be involved in the midline crossing, while the last one is more related with target innervation or target detection, which are relevant topics to understand how the axon identifies its target. This target cue provides information to the axon about the direction that it should take (towards the target), otherwise it would be affected only by the netrin and slit gradients and would never reach its target. These ligands bind to the receptors on the growth cone membrane of the ipsilateral and commissural neurons. The receptors that were included are: DCC, Robo and Target receptor (see Table 6.5 for all their parameters). As for the ligands, the first

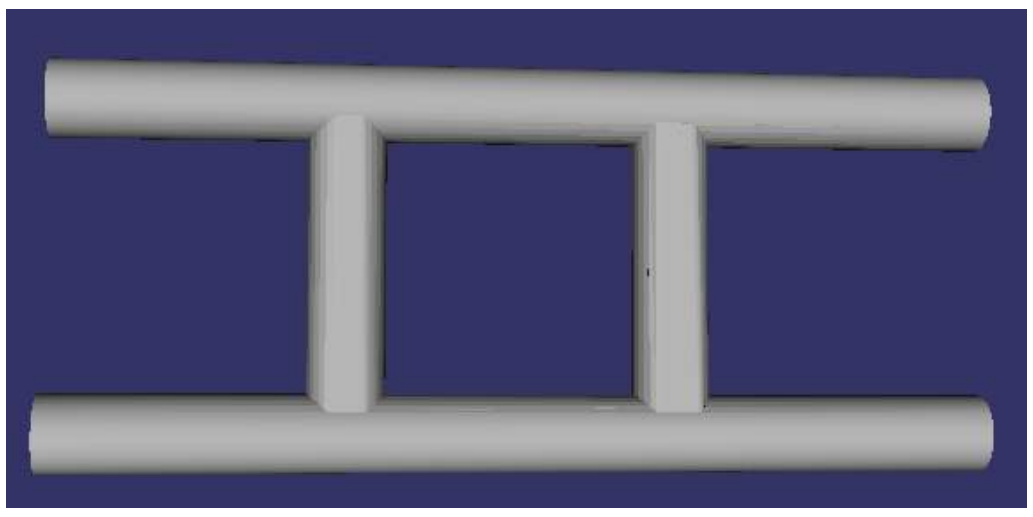

(a) Three-dimensional midline model

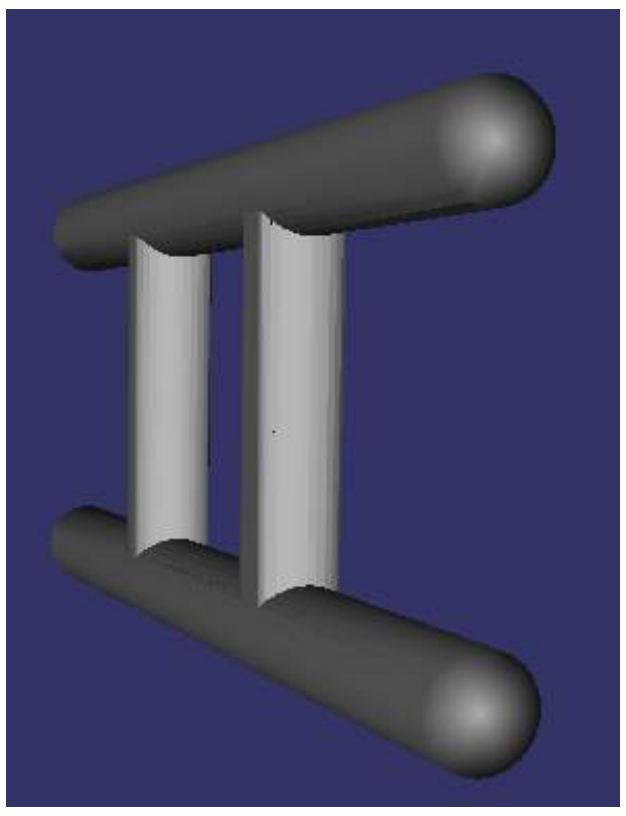

(b) Different prespective

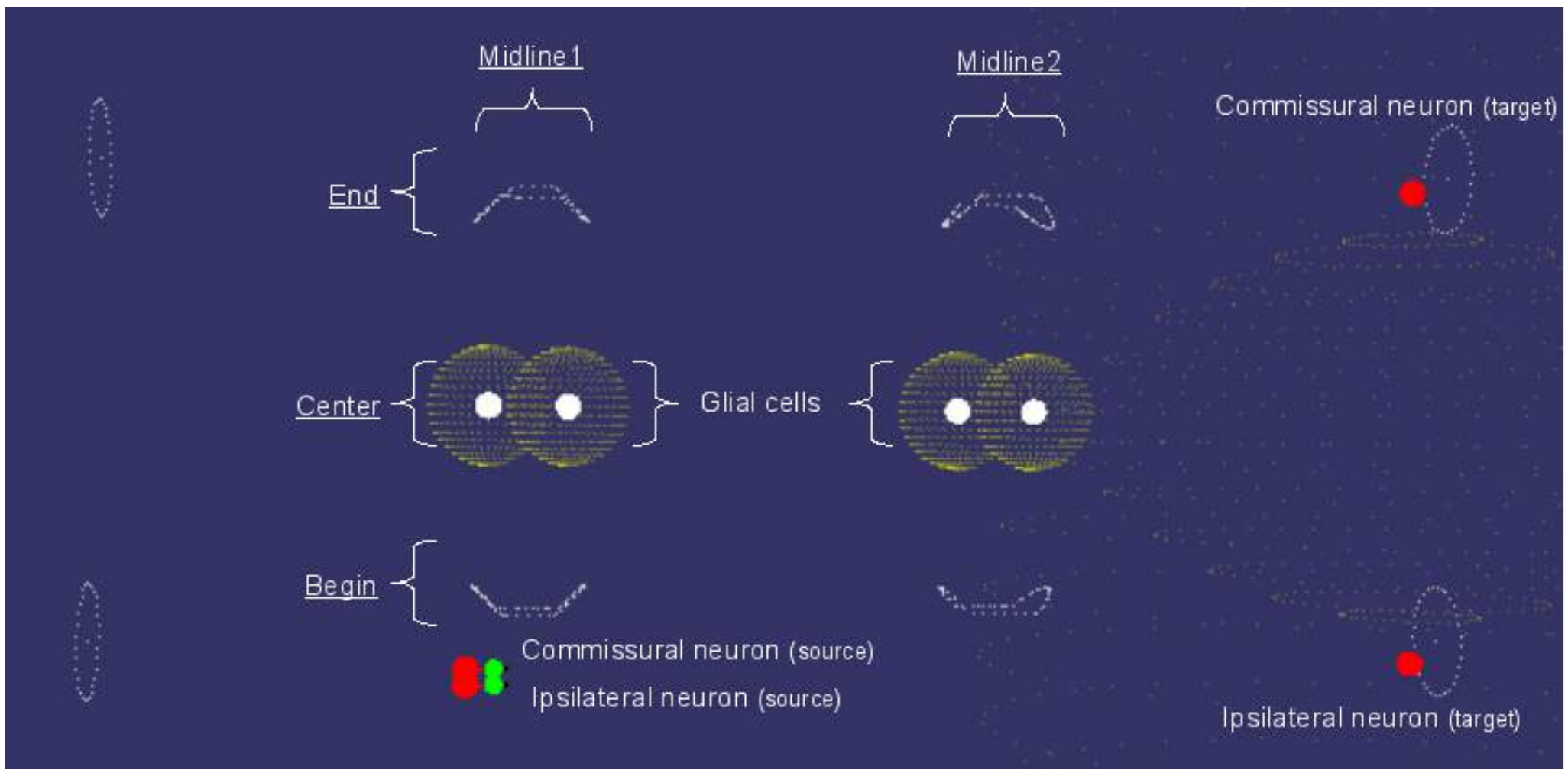

(c) Graphical representation of all the midline elements. The critical decision regions considered during the evaluation process are underlined. The white points represent the limits and the intersections of the midline model illustrated in 6.13(a) and 6.13(b) two are based on direct experimental evidences and the last one was added in order to allow the axon to perceive its target. All these ligand-receptor pairs led us to include three guidance cues in our model (see Table 6.6). The parameters of each ligand and receptor were found empirically. The ligand range and the receptors removal/production rate play an important role in this system, with the system being sensible to both, because they define when the GC becomes insensitive to the attractive Netrin and starts reacting to the repulsive Robo. The diffusion is an incremental process (i.e. the ligands range starts in zero and reaches its final range with a certain speed), starting when the simulation begins.

Figure 6.13: Midline model.

Table 6.1: Neurons parameters.

| | Comm Source | Ipsi Source | Comm Target | Ipsi Target |
|---|---|---|---|---|
| **Position** | 2.5,-0.35,0.45 | 2.5,-0.35,0.25 | 2.5,12,6.7 | 2.5,12,0.5 |
| **Topographic type** | Source | Source | Target | Target |
| **Proteins** | Comm | | | |
| **Receptors** | DCC,Robo,Target | DCC,Robo,Target | | |
| **Ligands** | | | Target | Target |
| **Alpha,Beta** | 0°,90° | 0°,90° | | |

Table 6.2: Midline topographic map.

| Source | Target |
|---|---|
| Commissural Neuron | Commissural Target Neuron |
| Ipsilateral Neuron | Ipsilateral Target Neuron |

Table 6.3: Glial cells parameters.

| Name | Position (x,y,z) | Ligands |
|---|---|---|
| **Midline 1.1** | 2.5,0.9,3.75 | Netrin,Slit |
| **Midline 1.2** | 2.5,-0.1,3.75 | Netrin,Slit |
| **Midline 2.1** | 2.5,6.9,3.75 | Netrin,Slit |
| **Midline 2.2** | 2.5,5.9,3.75 | Netrin,Slit |

Brankatschk and Dickson [11] state that the Netrin acts as a short range cue in the midline, which is consistent with our model (as can be seen in Table 6.4).

Table 6.4: Parameters of the midline model ligands.

| Name | Diffusion | Rate | Range |
|---|---|---|---|
| *Slit* | Exponential | 1% | Short (2.5) |
| *Netrin* | Exponential | 1% | Short (2.5) |
| *Target* | Exponential | 10% | Long (5.5) |

Table 6.5: Parameters of the midline model receptors.

| Name | Distribution | Removal | Production | Min | Sat | Concentration |
|---|---|---|---|---|---|---|
| *DCC* | Uniform | 6.16% | 4.25% | 0% | 100% | 0.9 |
| *Robo* | Uniform | 0% | 2.6% | 0% | 100% | 0.1 |
| *Target* | Uniform | 0% | 0% | 0% | 100% | 1 |

Table 6.6: Parameters of the midline model guidance cues.

| Ligand | Receptor | Behavior | Range | Force |
|---|---|---|---|---|
| Netrin | DCC | Attraction | Short (2.5) | 0.74 |
| Slit | Robo | Repulsion | Short (2.5) | 0.74 |
| Target | Target | Attraction | Long (5.5) | 0.5 |

### Proteins and Regulatory Network

It is known that the Comm protein plays a very important role in the *Drosophila* midline crossing; its presence blocks the expression of the Robo receptor allowing the Commissural Neuron to enter the midline. In our model this protein starts with the maximum concentration (see Table 6.7) and it decreases due to a regulatory link with the DCC receptor. A regulatory network between some receptors and proteins allows the model to capture what seems to be one of the most important dynamics in axon guidance, and in cell biology in general, that is the regulation of certain elements by others when these are stimulated.

As can be observed in the Table 6.8, we defined two inhibitory links: (i) one between the Comm protein and the Robo receptor and (ii) other between the DCC receptor and the Comm protein. The first link is well established, while for the second some evidences from neurobiology start to emerge (see section 6.2.6). It is important to note that this regulation affects the concentration of the regulated element and not directly its activity.

Table 6.7: Midline model protein.

| **Name** | **Concentration** |
|---|---|
| Comm | 1 |

Table 6.8: Regulatory network of the midline neurons.

| | **DCC** | **Robo** | **Comm** | **Target**$_{Receptor}$ |
|---|---|---|---|---|
| **DCC** | 0 | 0 | 0 | 0 |
| **Robo** | 0 | 0 | -0.05 | 0 |
| **Comm** | -0.03 | 0 | 0 | 0 |
| **Target**$_{Receptor}$ | 0 | 0 | 0 | 0 |

#### Mechanisms

Experiments were performed in order to identify which are the most relevant mechanisms for this model (see in Table 6.9 the selected mechanisms). The midline model presented so far does not consider fasciculation and receptor limits, but their possible role is discussed in section 6.2.5.

Table 6.9: Midline mechanisms.

| **Name** | **Enabled?** |
|---|---|
| Axonal transportation | Yes |
| Adaptation | Yes |
| GC complexification | Yes |
| Intersections | Yes |
| Receptors limits | No |
| Fasciculation | No |

### 6.2.2 Normal pathfinding

The model presented during the last sections lead to results consistent with the available experimental evidences.

Structural simulation results presented in Table 6.10 show that our computational model did the same choices in the decision regions that are observed in biology (see section 2.6.1). As expected, the commissural axon is longer and needs more simulations than the ipsilateral axon to reach its target.

Figures 6.14, 6.15 and 6.16 show the values of receptors, proteins and angles throughout the simulation for either the commissural or the ipsilateral neurons. During the simulation in the decision regions, where the midline crosssing should be avoided, the growth cones pass through some sudden perturbations.

Table 6.10: Structural results for the midline model in a normal situation. The directions expressed in the decisions regions (Midline1 (begin), Midline1 (enter), etc.) mean: Continues - Continues in the same direction, Up - Grows up, Right - Grows by turning right

| | **Commissural Neuron** | **Ipsilateral Neuron** |
|---|---|---|
| Midline1 (begin) | Up | Continues |
| Midline1 (center) | Up | |
| Midline1 (end) | Right | |
| Midline2 (begin) | | Continues |
| Midline2 (center) | | |
| Midline2 (end) | Continues | |
| Target | Yes | Yes |
| Axon Length ($\mu$m) | 11.104 | 8.325 |
| Total simulation steps | 2753 | 1832 |

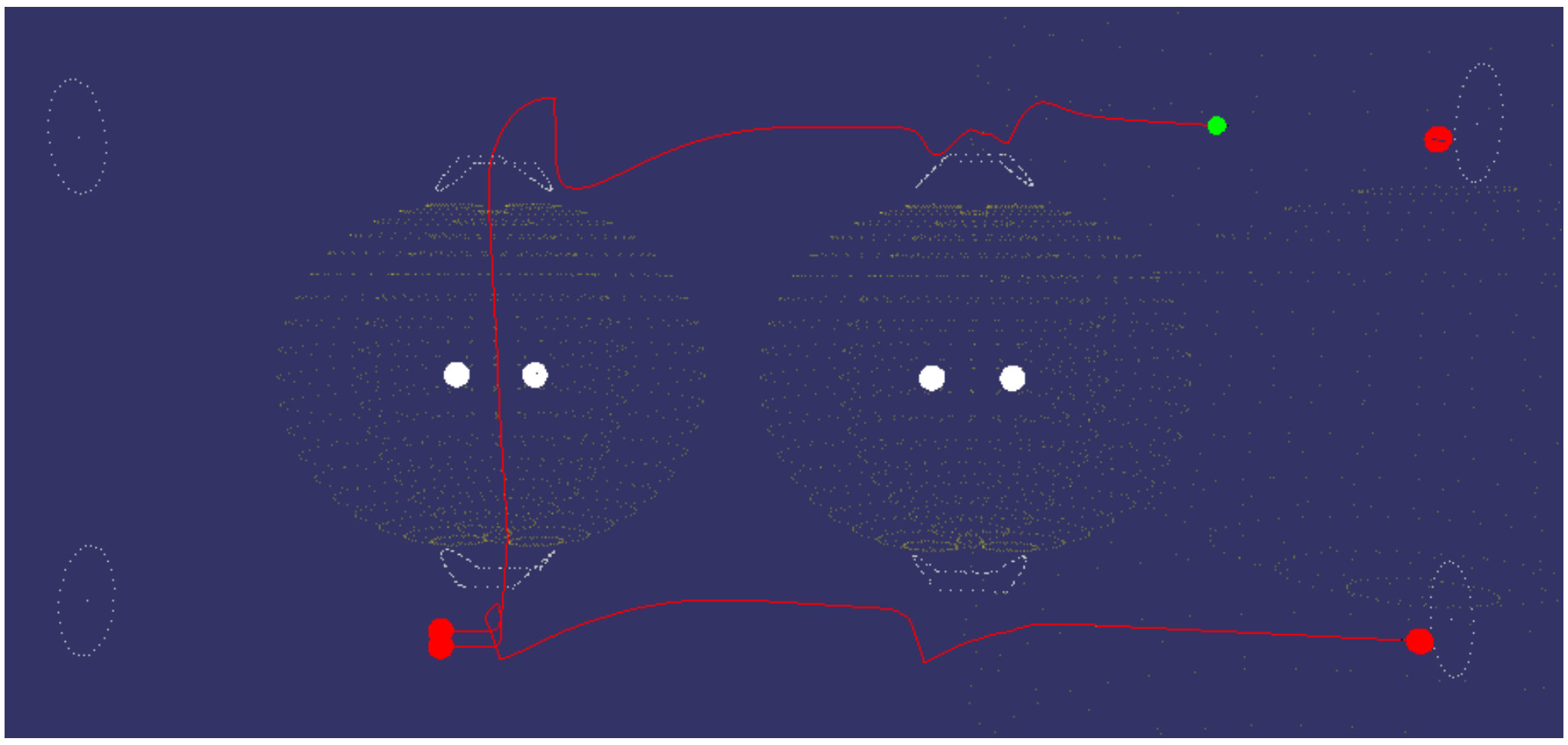

Figure 6.14: Simulation step from the normal midline simulation.

### 6.2.3 Comm mutant pathfinding

Comm mutant neurons avoid the midline; therefore both commissural and ipsilateral do not cross it. When this was replicated in our model by removing the same behavior Comm protein we obtained the same behaviors observed experimentally. (see Table 6.11). As can be observed in Figure 6.17 the Comm mutant simulation is consistent with the available experimental evidences, since the commisural axon does not cross the midline, reinforcing the realism of our model.

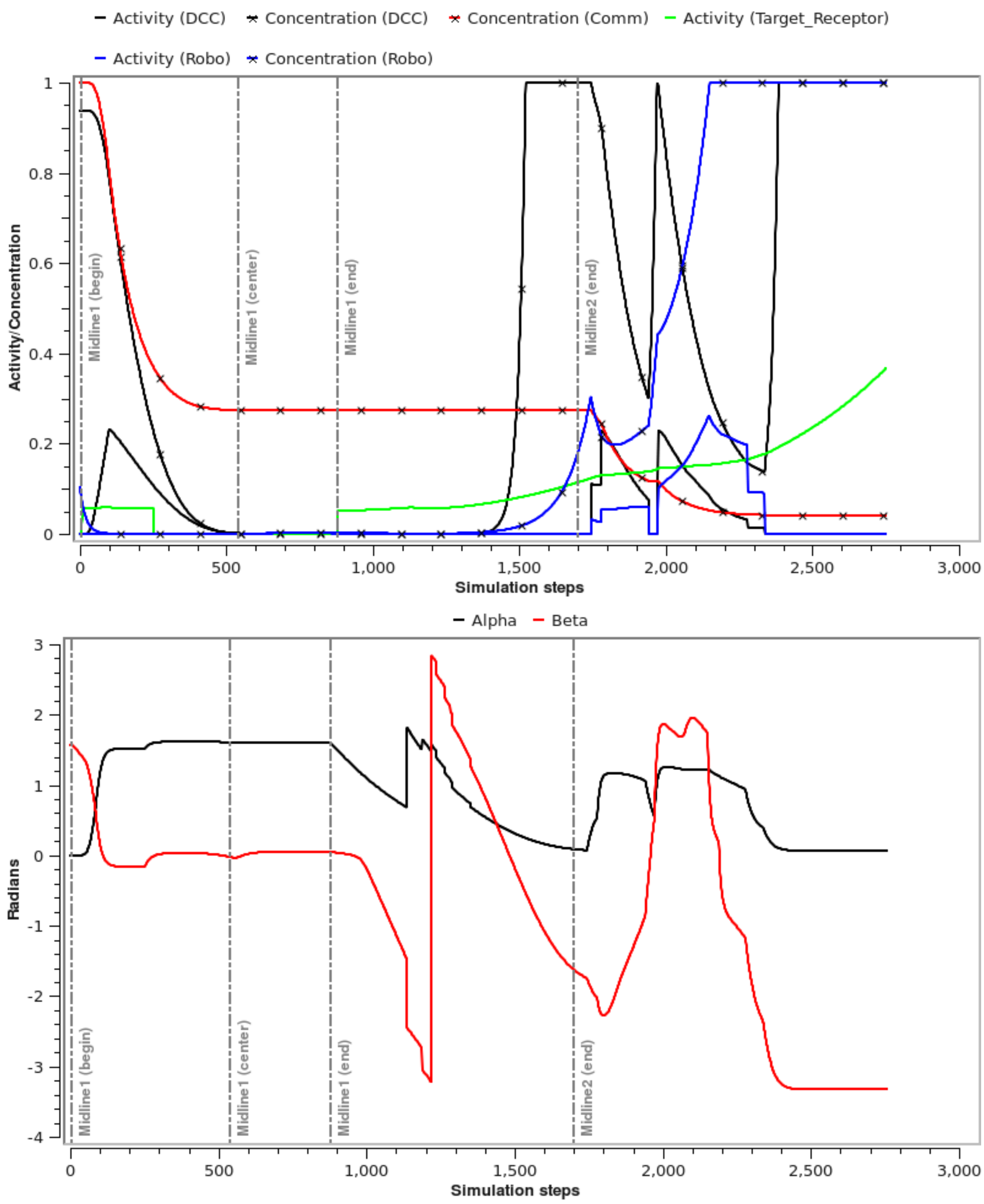

Figure 6.15: Normal midline simulation Activities (Robo, DCC and Target Receptor), concentrations (Robo, DCC and Comm) and angles (Alpha and Beta) of the Commissural neuron. The activity level of the Comm protein is equal to its concentration, hence it was not plotted. The concentration of the target receptor was not plotted because it maintains constant at 1.

Figure 6.18 shows the variation of receptors and angles during the simulations for both commissural and ipsilateral neurons. The ipsilateral values are not plotted because they are equal to the normal pathfinding.

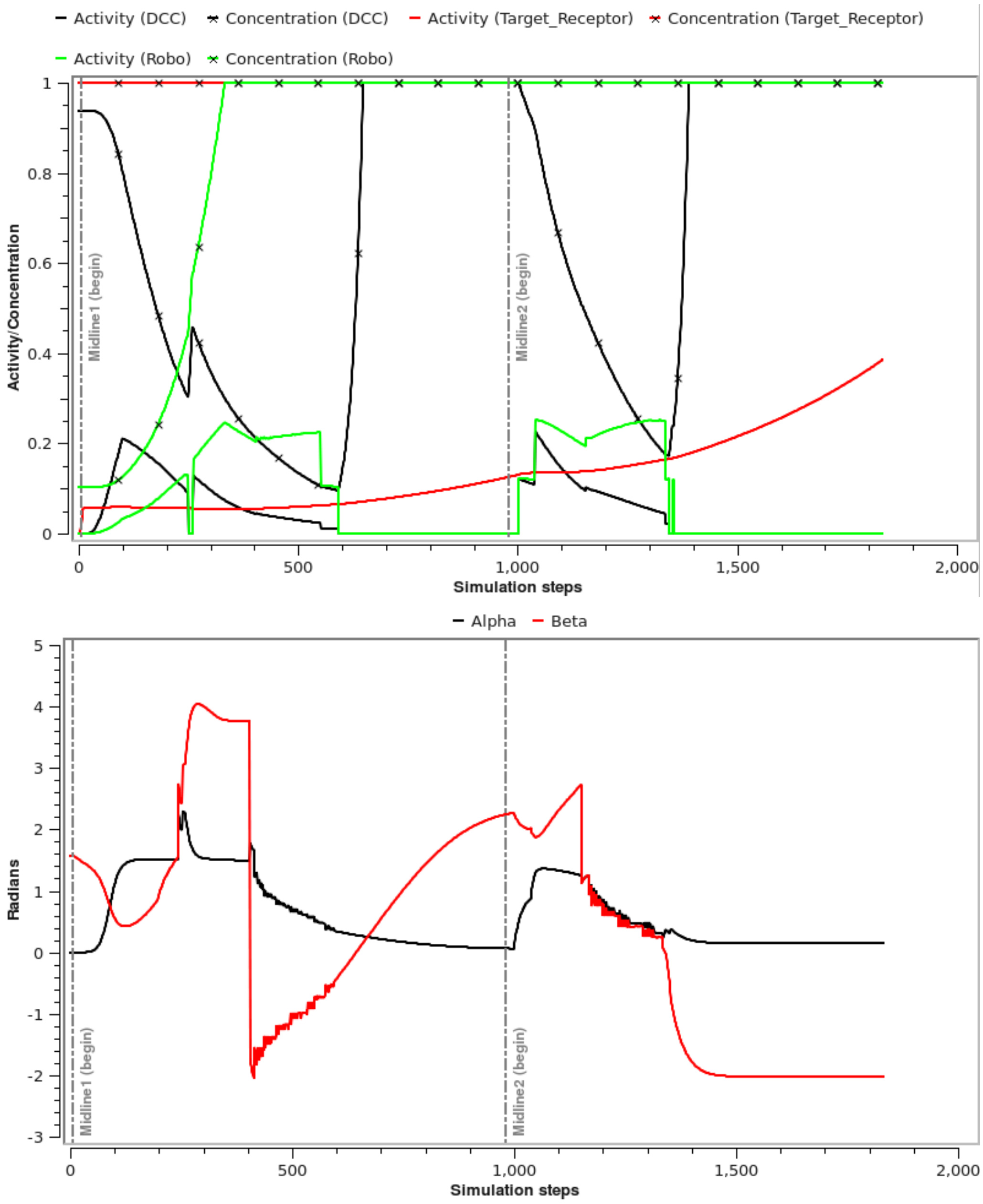

Figure 6.16: Normal midline simulation: Activities (Robo, DCC and Target Receptor), concentrations (Robo, DCC and Target Receptor) and angles (Alpha and Beta) of Ipsilateral neuron. The concentration of the target receptor was not plotted because it maintains constant at 1.

Finally, it is important to note that the only difference between this simulation and the normal pathfinding is the lack of the Comm protein, similarly to the biological Comm mutant.

Table 6.11: Structural results of midline model without the Comm protein. The directions expressed in the decisions points mean: Continues - Continues in the same direction

| | Commissural Neuron | Ipsilateral Neuron |
|---|---|---|
| Midline1 (begin) | Continues | Continues |
| Midline1 (center) | | |
| Midline1 (end) | | |
| Midline2 (begin) | Continues | Continues |
| Midline2 (center) | | |
| Midline2 (end) | | |
| Target | No | Yes |
| Axon Length ($\mu$m) | | 8.32077 |
| Total simulation steps | | 1878 |

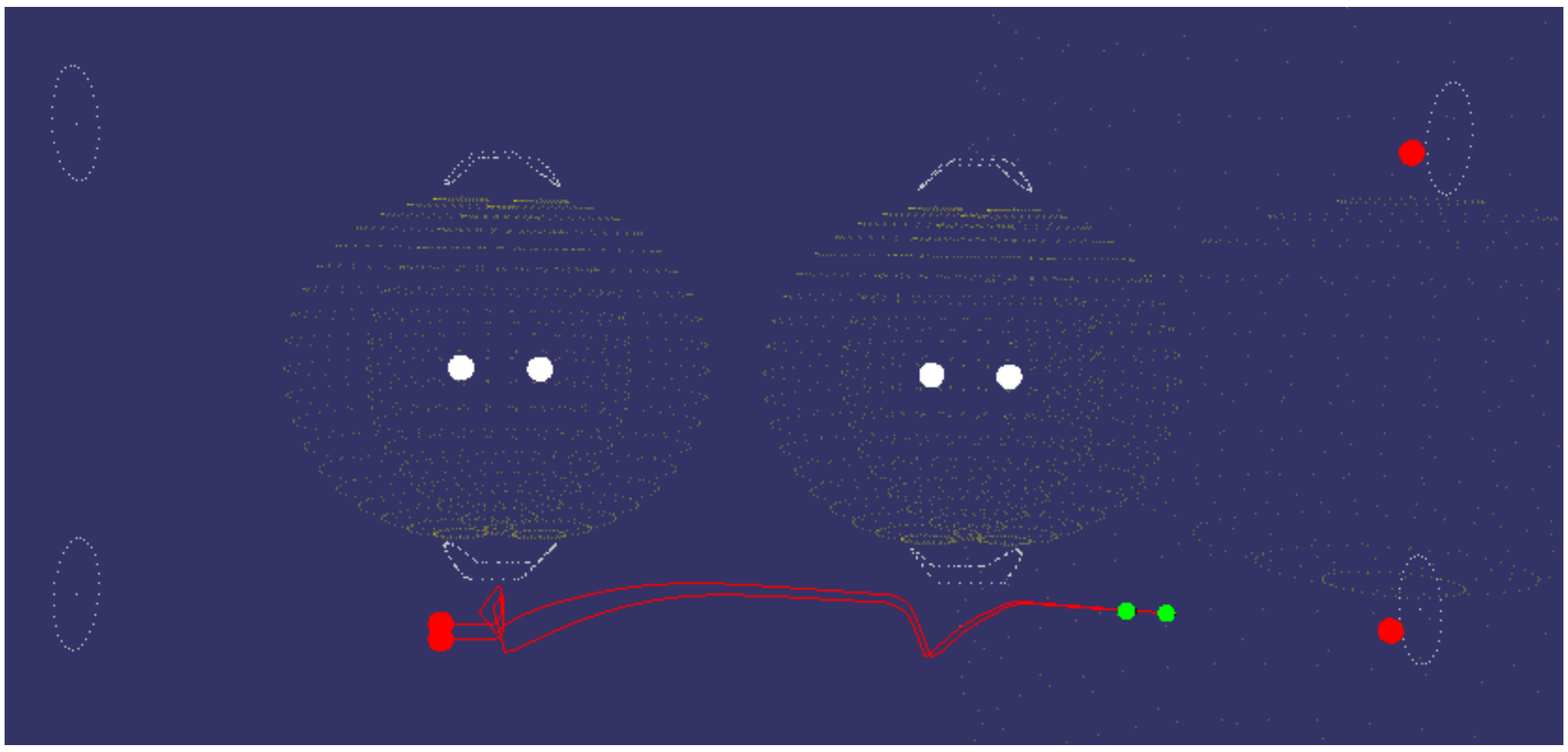

Figure 6.17: Simulation step of the Comm mutant midline.

### 6.2.4 Robo mutant pathfinding

Another situation that is well studied from *in vivo* experiments is the robo mutant. In this mutant both axons cross and recross the midline. Similarly to the previous mutant our model is consistant with the experimental evidences. However, unlike the graphical representation in section 2.6.1 where the axon keeps crossing in circles, in our model the axons only recross once, i.e. after recrossing both axons continue growing towards the target (see Table 6.12).

But as stated by [76] only some axons recross in circles. Furthermore, our three-dimensional model can easily be extented in order to include more midlines, which could lead to more cross and recross behaviors, in agreement with the most frequent behavior for this mutant.

Figures 6.20 and 6.21 show the variation of receptors and angles during the simulations for both commissural and ipsilateral neurons in Robo mutants. It is important to note that the only difference between this simulation and the normal pathfinding is the lack of the Robo receptor, like the biological Robo mutant.

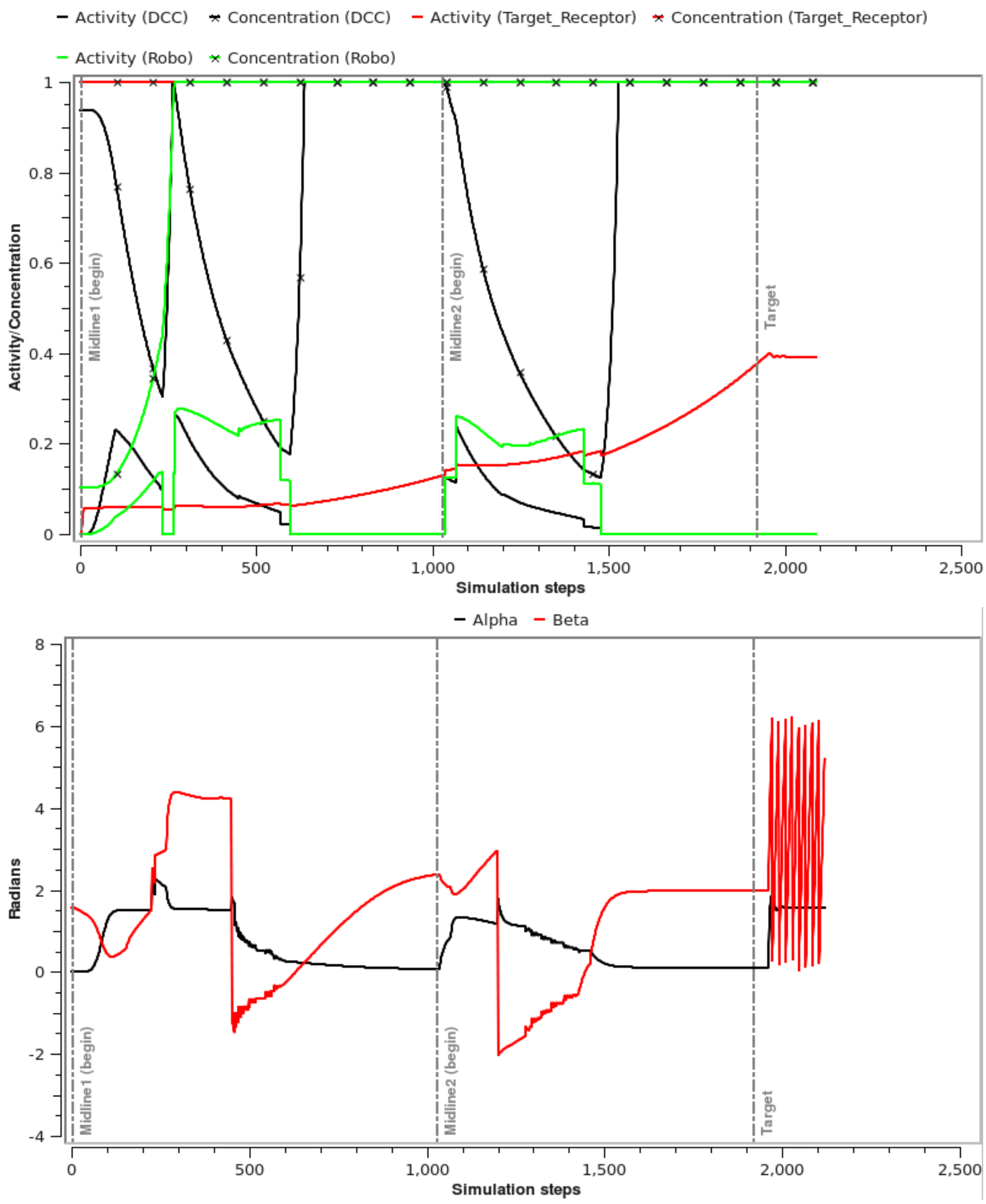

Figure 6.18: Comm mutant simulation: Activities (Robo, DCC and Target Receptor), concentrations (Robo, DCC and Target Receptor) and angles (Alpha and Beta) of the commissural neuron. The concentration of the target receptor was not plotted because it maintains constant at 1.

### 6.2.5 Model Mechanisms

In this section we analyse the importance for this midline model of all the mechanisms. The adaptation mechanism allows the GC to cross the midline after being desensitized to the Netrin ligand and to regain sensitivity after having crossed it. The intersection mechanisms prevent the GC from leaving the three-dimensional model turning this 3D model into a set of adhesive cells. The receptor limits mechanism could be useful in decision points such as the center of the midline, where a definition of a certain saturation level for receptor activation could help the axons to continue growing towards the end of the midline. Nonetheless in the experimentation here presented the best model found did not consider this mechanism.

Table 6.12: Structural results of midline model without the Robo receptor. The directions expressed in the decisions points mean: Continues - Continues to grow in the same direction, Up - Grows up, Down - Grows down, Right - Grows by turning right

| | **Commissural Neuron** | **Ipsilateral Neuron** |
|---|---|---|
| Midline1 (begin) | Up | Up |
| Midline1 (center) | Up | Up |
| Midline1 (end) | Right | Right |
| Midline2 (begin) | Down | Down |
| Midline2 (center) | Down | Down |
| Midline2 (end) | Right | Right |
| Target | No | Yes |
| Axon Length ($\mu$m) | | 12.2628 |
| Total simulation steps | | 2816 |

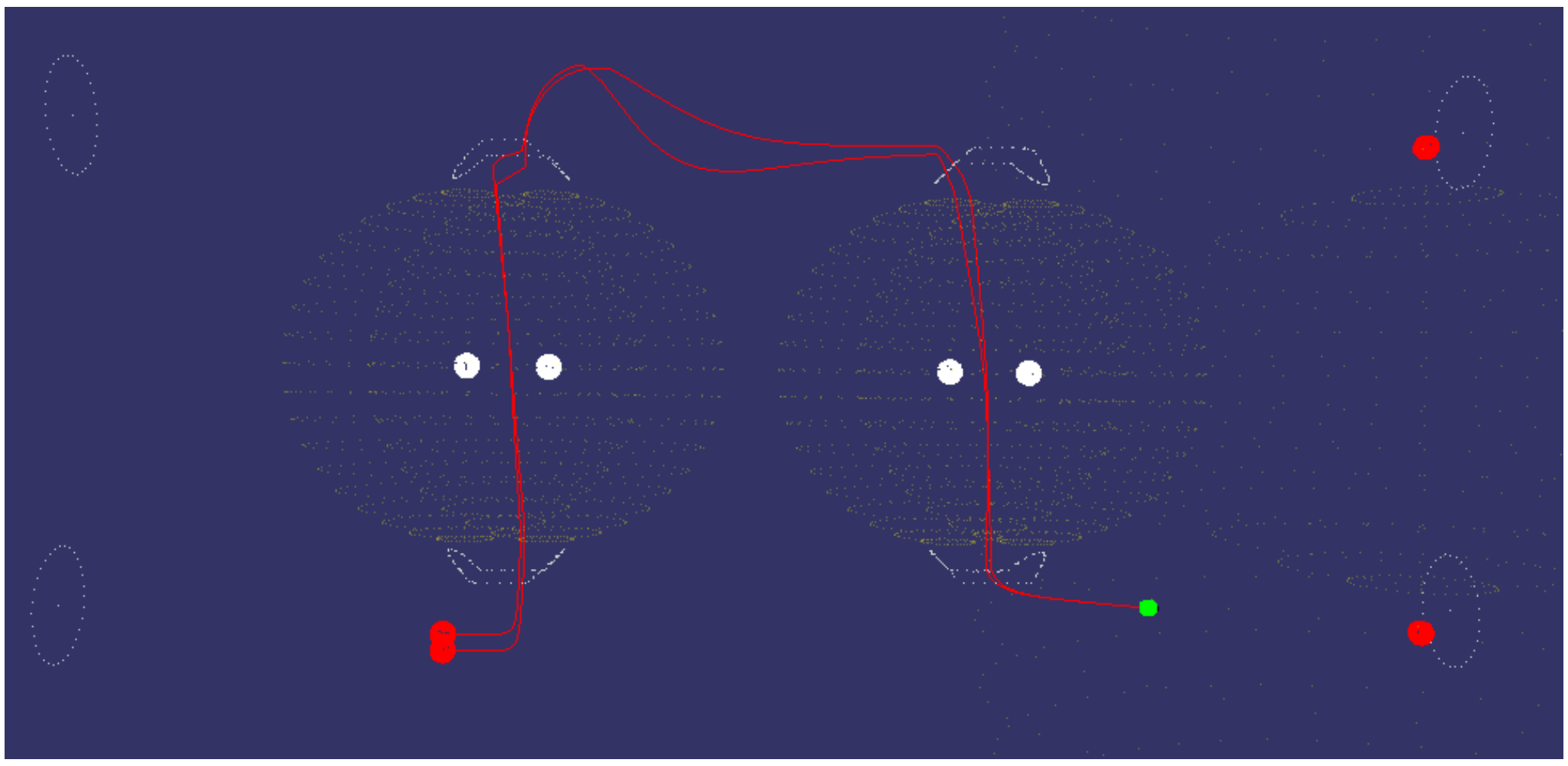

Figure 6.19: Simulation step from the Robo mutant simulation.

Interestingly both the axonal transport and the GC complexification mechanisms led to shorter axons. (see the next sections for further details).

### Axonal transport

Table 6.13 shows that commissural axons are shorter when axonal transport is enabled. This suggests that this mechanism allows the nervous system to transmit information more efficiently. The longer the axon is the slower it is in reacting to new extracellular cues. In this specific system this may help the growth cone to keep focused in its target.

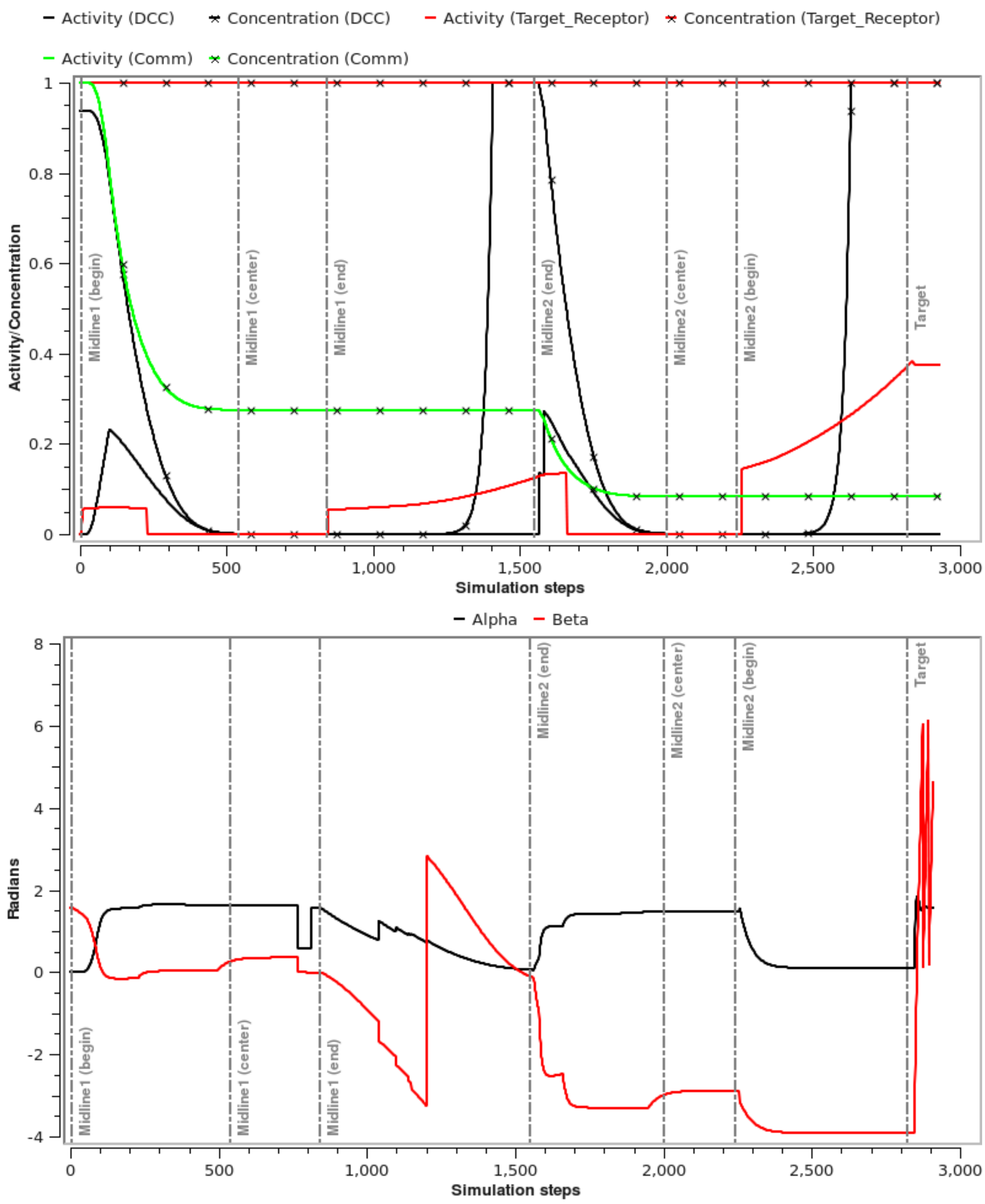

Figure 6.20: Robo mutant simulation: Forces (Comm, DCC and Target Receptor), concentrations (Comm, DCC and Target Receptor) and angles (Alpha and Beta) of Commissural neuron.

### Growth cone complexification

Surprisingly, the deactivation of the growth cone complexification mechanism leads the GC to turn down in the Midline2 (end) region (Figure 6.22). This happens because it grows too fast and in that region it collides with the midline tissue growing along with it (due to the second intersection mechanism). On the other hand when it is activated the axonal growth is slower, leading to more accurate decisions and to avoid entering in the Midline2 as it was demonstrated by the normal pathfinding. These results reinforce the importance of this mechanism either in simulation or in *in vivo*.

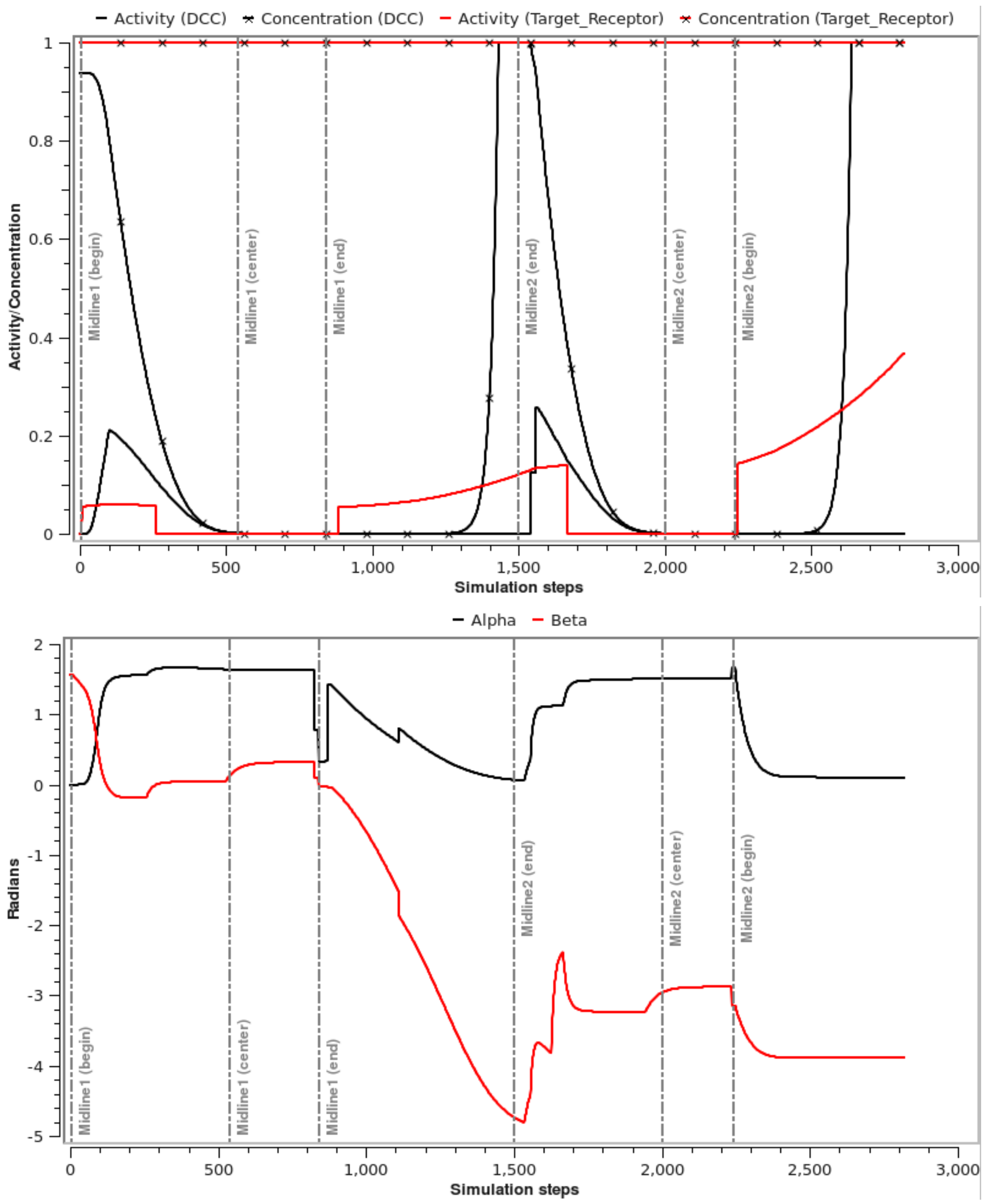

Figure 6.21: Robo mutant simulation: Activities (DCC and Target Receptor), concentrations (DCC and Target Receptor) and angles (Alpha and Beta) of ipsilateral neuron.

Table 6.13: Effect of axonal transport in the midline crossing.

| Neuron | Enabled? | Axonal length ($\mu$m) | Simulation steps |
|---|---|---|---|
| Ipsi | Yes | 8.325 | 1832 |
| Ipsi | No | 8.34628 | 1841 |
| Comm | Yes | 11.104 | 2753 |
| Comm | No | 10.3894 | 2626 |

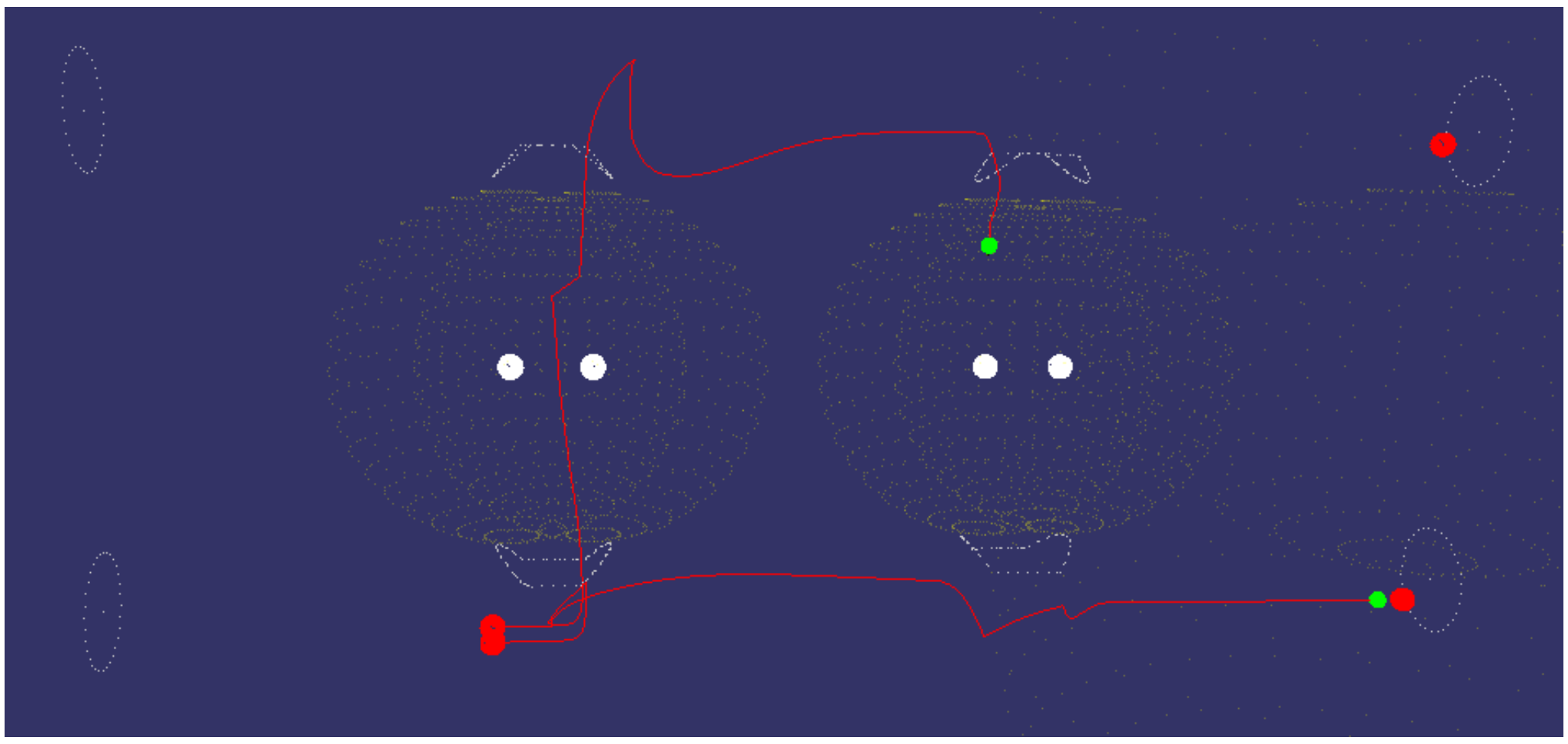

Figure 6.22: Simulation step with the growth cone complexification disabled. Unexpectedly the commissural axon turns down in the Midline2 (end).

### Fasciculation

Two new neurons were added to the simulation to evaluate the fasciculation mechanism in the midline model, one that should grow with the commissural neuron and other that should grow with the ipsilateral neuron. In a first test the same ligand was added to the pioneer commissural and pioneer ipsilateral axons, along with the respective receptors in the two new neurons. However, the new commissural axon was not able to follow its pioneer, due to the proximity of the attractive ligand of the ipsilateral neuron (Figure 6.23(a)).

Afterwards different ligands were added to the commissural and ipsilateral neurons together with their receptors. This approach leads to the expected fasciculation behavior. The path developed with fasciculation by the new neurons was smoother (Figure 6.23(b)) and shorter, but requires more simulation steps (see Table 6.14). Surprisingly, with the fasciculation disabled (Figure 6.23(c)) the new commissural neuron finds its target first than the old one that starts growing closer to the midline (i.e. the initial pioneer); nonetheless its length is higher than when fasciculation is enable. This suggests that the initial position of the new neurons is more suitable for midline crossing.

Once again it is interesting to observe the flexibility of our model, in which the new neurons with new starting positions are still able to find their targets.

This study suggests that for fasciculation to occur at least two different ligands are needed, one for each type of neuron. Nevertheless this evidence requires further stud-ies.

Table 6.14: Effect of fasciculation in midline crossing.

| Neuron | Enabled? | Axonal length ($\mu$m) | Simulation steps |
|---|---|---|---|
| Ipsi Follower | Yes | 7.88833 | 2007 |
| Ipsi Follower | No | 8.75731 | 1956 |
| Comm Follower | Yes | 10.2457 | 2793 |
| Comm Follower | No | 10.4459 | 2614 |

### 6.2.6 Discussion

Although most of the parameters of the midline model used throughout this chapter are based on experimental evidences, almost all the specific values such as the ranges or the forces of the guidance complexes are not known. Therefore, it was necessary to search empirically for these values confronting the simulation results with the behavior of biological systems. As a result, several values were used that are not directly based on experimental evidences. In this section we discuss this issue, providing direct or at least indirect relations between the results obtained by our model and the state of the art knowledge in developmental neuroscience.

All the simulations made with this model allowed concluding that the physical barrier imposed by the midline tissue plays a very important role, allowing axons with slightly different parameters to still find their targets. These effects at the midline are the same performed by the activity of cell adhesion molecules (CAMs) that mediate contact dependent axonal growth and guidance, which is consistent with biological studies based on the analysis of the *Drosophila* midline [46, 4].

With this *Drosophila* midline model we give for the first time a detailed computational model that works in three different scenarios and raised hypotheses that could help to explain how the system works. These findings should drive further *in vivo* or *in vitro* experiments. Moreover, with these results we prove the usefulness and the realism of our model.

In this section we address the same questions presented in [22].

#### Why commisural axons cross the midline and ipsilateral do not cross?

The available experimental evidences show that commissural axons cross the midline due to the expression of the Comm protein that targets Robo receptors to endosomes, thereby down-regulating the repulsive gradients released by the midline glial cells [48, 22]. This is incorporated in our model by adding the Comm protein into the Commissural neurons. The presence of this protein inhibits the action of the Robo receptors, which is defined by a regulatory link of -0.05 between them. This element allows the Commissural neuron to enter the midline while the Ipsilateral neuron which lacks the Comm protein is repulsed by it.

In our model as observed *in vivo* the concentration levels of Robo are initially small (at 0.1). In the Ipsilateral axons this concentration sharply increases due to the lack of the Comm protein while the concentration of DCC receptors decreases (due to the removal rate). These two behaviors lead this axon to be repulsed by the midline.

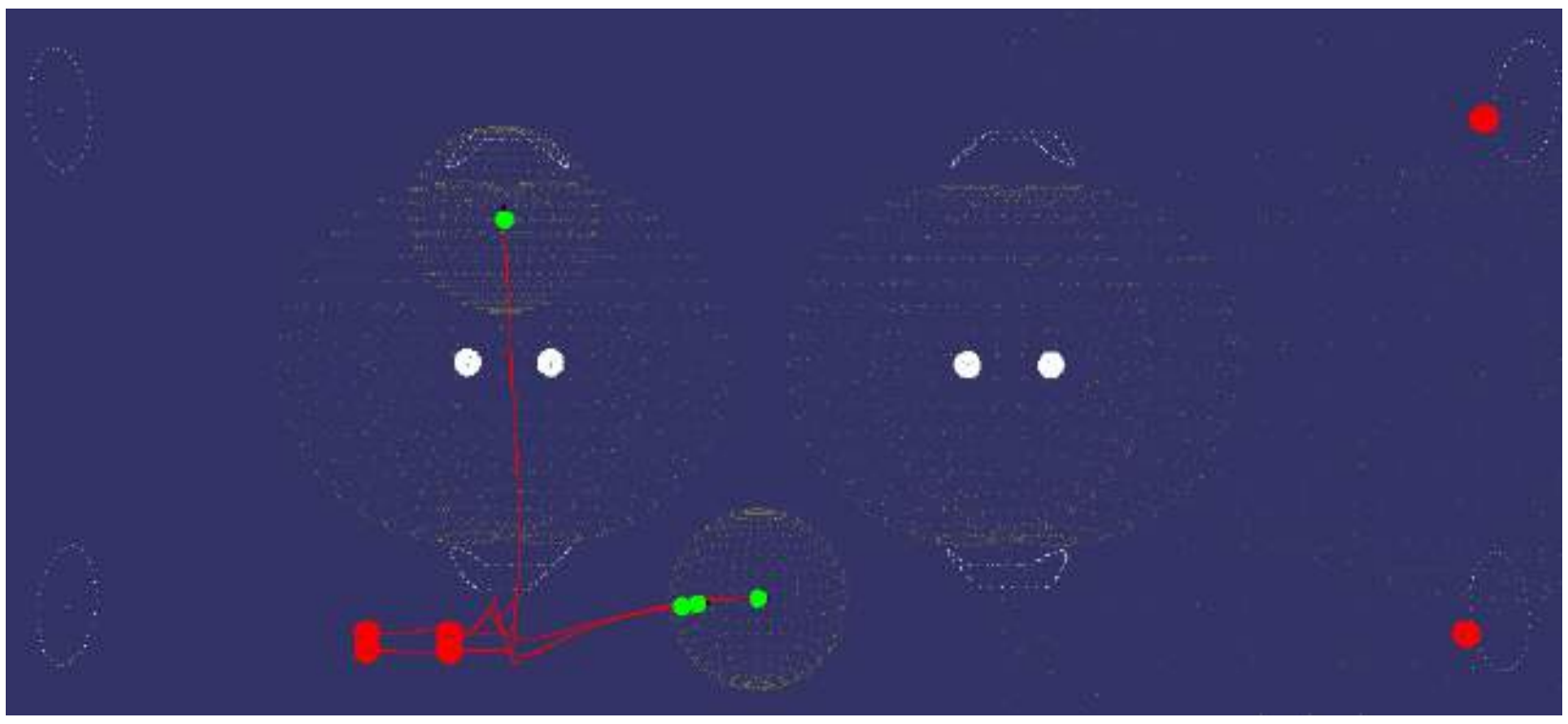

(a) Fasciculation with the same ligand in both pioneer neurons

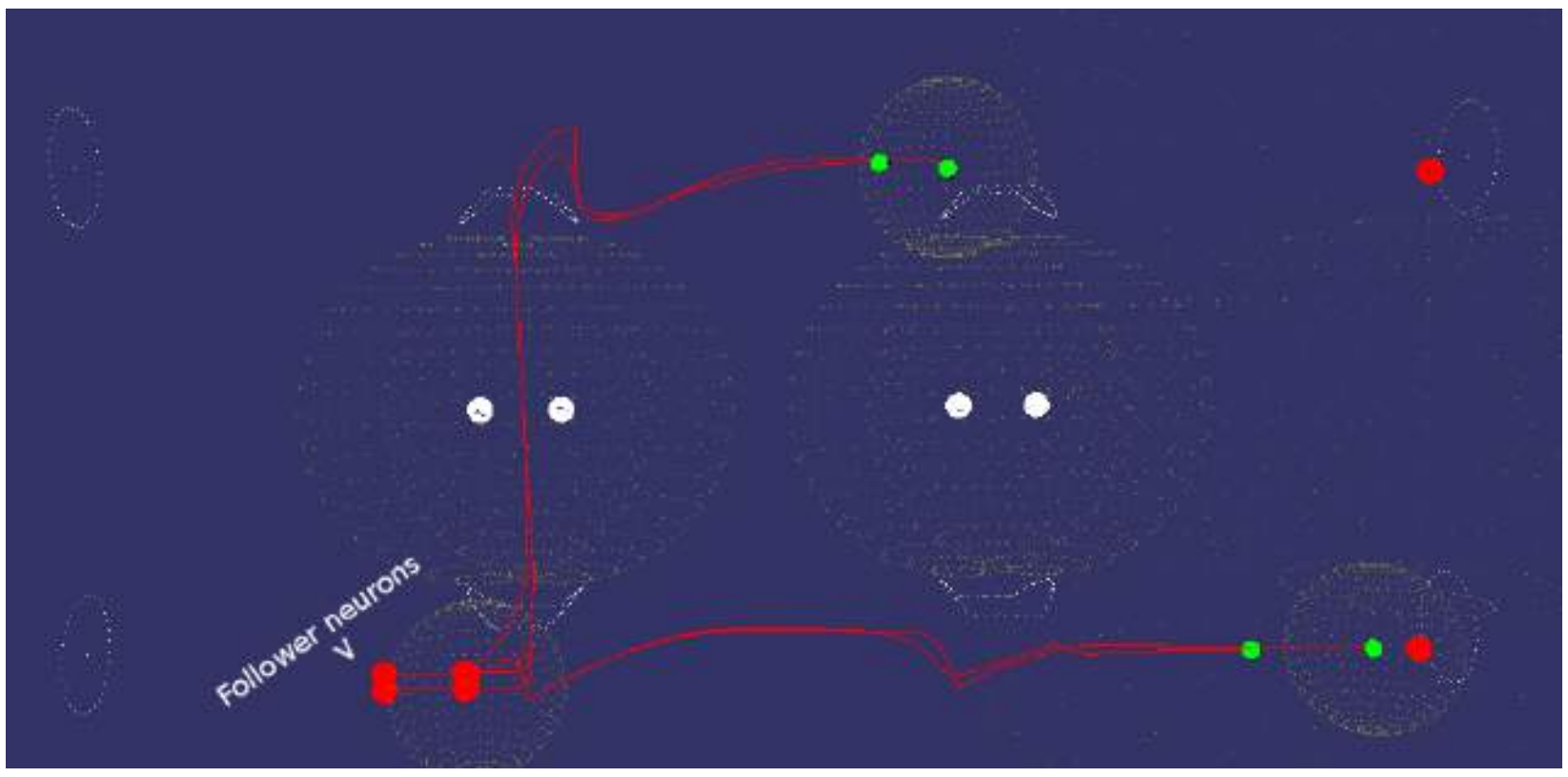

(b) Fasciculation with different ligands for the pioneer neurons

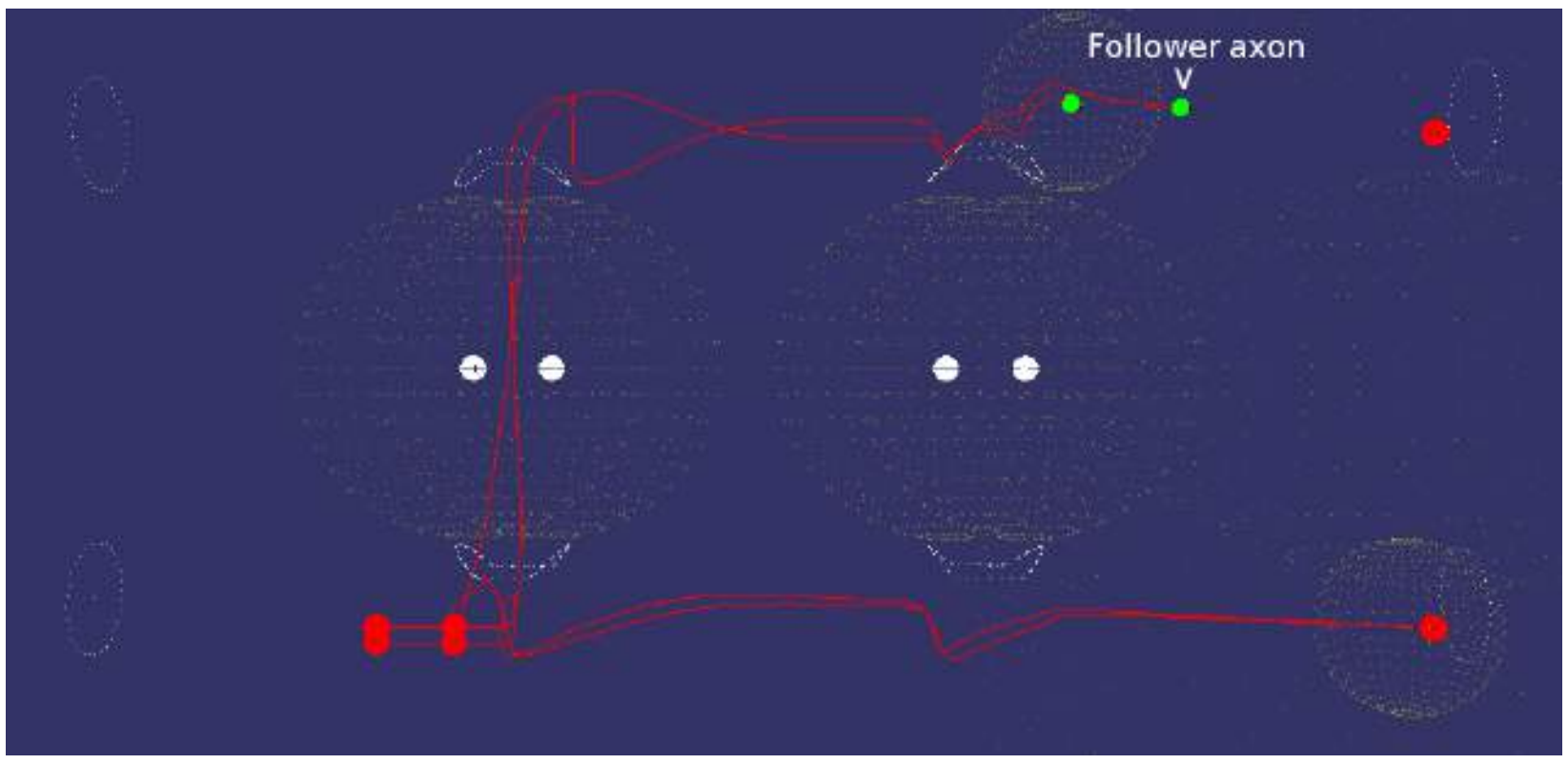

(c) Fasciculation disabled

Figure 6.23: Visual results of the fasciculation mechanism in the midline model.

On the other hand the concentration of the Robo receptors in the Commissural axons decreases until zero due to the existence of the Comm protein. Therefore the Robo receptor does not contribute to the first stage of the midline crossing of comissural axons (Midline1 (begin)).

**What prevents commissural axons from lingering at the Midline?**

An interesting question that neuroscientists have been trying to address is, how can the commissural axons pass through the attractive midline? If it is attractive what allows them to ignore this attraction and keep growing towards the end of the midline?

It is believed that this happens for two main reasons: first, the desensitization of the GC to the attractive Netrin ligands and second, the desensitization to the repulsive Slit ligands [22]. This behavior was observed in our model, as soon as the axon enters the midline the Netrin activates the DCC receptors which are removed from the GC at a rate of 6.16% (e.g. through endocytosis) while being produced at a slower rate (4.25%). The higher removal rate leads to a drop in the concentration of the DCC receptors which in turn decreases the activation of these receptors. As a result, when the GC reaches the center of the midline it does not react to the Netrin gradient and keeps growing towards the end of the midline. The existence of two equal attractive glial cells, one in each side of the midline tube, allows the axon to grow between them (i.e. axons are attracted exactly to the middle of the midline).

Additionaly, after crossing the midline center the axons start reacting to the repulsive ligand (Slit) that is diffused by the two gradients, which helps the axon to grow towards the contralateral side. The concentration level of the Robo receptors slightly increases between the simulation step $600^{th}$ and $1100^{th}$ after crossing the midline center (Figure 6.15), because the concentration of the Comm protein drops due to its down-regulation performed by the activation of the DCC receptor (at a rate of 0.03 per simulation step). This new regulatory link between the DCC receptor and the Comm protein was added in order to obtain the results observed *in vivo* (see section 6.2.6 for a detailed discussion about this topic).

Our model is consistent with what is observed *in vivo* at the *Drosophila* midline, where after the midline the expression of the Comm protein drops and the concentration levels of the Robo receptor increases.

Interestingly, a down-regulation of the attractive effect of the DCC by the Robo receptor was observed in the growth cones of the embryonic *Xenopus* spinal axons [82]. If this principle is added in our model it could increase the efficiency of the midline crossing and at the same time its realism. The desensitization would occur not only due to the removal of receptors but also due to DCC inhibition through the activity of Robo receptors.

**Lateral crossing after the midline**

By the end of the midline the commissural axon should starts growing towards the *Drosophila* brain. In our model we define a target neuron that diffuses an attractive gradient that guides the commissural neuron towards it. This gradient and the repulsive effect of the midline allows the axon to cross laterally by turning right. *In vivo* this target gradient can be a neurotrophic factor that stimulates the axon outgrowth or/and guidance (see [4], pages 81-90). This behavior is supported both by other mathematical models [31] and experimental work [88, 97].

Apart from the Robo receptor two other Robo receptors (Robo2 and Robo3) have been identified, which may be involved in the formation of specific bundles after midline crossing [32]; together they seem to form a codification that specify in which bundle the axon should grow. The influence of this selective mechanism in our model should be studied in the future.

#### Preventing commissural axons to not recross

After crossing the midline the DCC levels return to their normal levels (due to the production rate) and the Comm protein is at a medium concentration level. Near the second crossing point (Midline2) the DCC is again stimulated, which decreases its concentration and the concentration of the Comm protein, which in turn leads to a rise in the Robo concentration levels. Consequently the GC is repulsed by the second midline due to: (i) the high levels of Robo and the (ii) low levels of DCC. These two factors maintain the Commissural axon in the contralateral side. This behavior should happen independently of the number of midlines, because the level of the Comm protein will not increase, keeping the Robo levels at its maximum concentration level.

The decrease of Comm protein and increase of Robo receptors are consistent with the experimental observations [48, 22], but the exact relation between them is unknown as well as this decrease/increase behavior in the DCC concentration.

#### Comm Mutant

In Comm mutants most axons are not able to cross the midline [48, 22]. This is due to the lack of Comm in Commissural neurons because this protein blocks the activity of the Robo receptors. Without Comm the concentration of Robo receptors in the GC increases and it is repulsed by the midline.

We replicated a Comm mutant by removing the Comm protein from our model, and with this change the simulations showed a response similar to that observed *in vivo* with the mutant Drosophila. However some sudden perturbations arise in the decision points (Midline1 (begin) and Midline2 (begin)), being these effects due to the diffusion of both attractive and repulsive cues by the midline.

#### Robo Mutant

When Robo is removed from the midline neurons, both axons (ipsilateral and commissural) cross and then recross the midline, one or more times. In our model when Robo receptors are removed, both neurons cross and recross the midline once, reflecting the behavior of some of axons *in vivo* [76, 22].

Although in our simulations the axons just crossed and recross once, if the midline had more crossing points the axons would have crossed back and forth.

In order to keep the axons crossing in circles we believe that the midline tissue should be modelled as a dynamic tissue with a variable size. This could lead the attractive force of the targets not to be strong enough to prevent axons from returning back and starting to cross in circles.

Furthermore, a different pattern of glial cells may also contribute to this behavior. However, this putative new pattern must work in at least two other scenarios (normal and comm pathfindings).

Finally is interesting to note that unlike the Comm mutant and normal midline the simulation results Robo mutant did not reveal sudden perturbations on the decision points. This is caused by the lack of repulsive forces.

#### Down-regulation of Comm protein by DCC

All the fine-tuning processes done during this experimentation revealed new values, but the most surprinsing and important is probably the down-regulation link between the DCC receptor and the Comm protein.

It is believed that the Comm protein is down-regulated once the axon has crossed the midline. With this new regulatory link the Comm protein drops gradually during the midline crossing. In our model this decline is accentuated when the DCC receptors start being activated by the Netrin ligand on the second midline. Meanwhile the Robo concentration increases allowing the axon to be repulsed by the midline once it has crossed it for the first time.

Interestingly, a recent study [99] reported that the DCC receptor activates the transcription of the Comm protein in a process that is independent of the Netrin stimulation, playing a very important role in the regulation of the first phase of the midline crossing. Therefore this study helps to support the midline simulations presented in this chapter that suggest a regulation between the DCC receptor and the Comm protein. We further suggest based on the simulation results that in addition to activating Comm transcription the DCC receptor could inhibit this transcription when stimulated by the Netrin ligand.

Yang et al. [99] also suggest that a third unknown signal that activates the Comm transcription should exist. This signal can also be involved in this down-regulation of the Comm transcription in the *Drosophila* midline, and this may occur under different physiological contexts or receptor stimulation levels.

This putative third sinal must be attractive because Netrin mutants show many axons that are still attracted by the midline [11]. The activation of the Comm transcription found by Yang et al. is considered in our model as the initial concentrarion value of the Comm protein (predefined at 1); thus our model does not consider this first step of dynamic transcriptional regulation (i.e. it assumes that it already happened).

Furthermore, this new relation is also supported by another recent study [57] that states that the midline is responsable for regulating part of the genetic transcription, being decisive in the midline crossing.

## Chapter 7

# Conclusions and future directions

The main aim of this thesis was to study the neuroscience and state of the art knowledge of AG, and then develop a computational model based on this knowledge. As a result, we proposed a computational model of AG based on simple elements and mechanisms that are known to be relevant for the AG dynamics.

This model was the basis for yArbor, the first AG simulator, and one of the results of this thesis. yArbor is a three-dimensional, user-friendly and flexible simulator, so that can be used by scientists interested in this topic.

The following parameters are considered as evaluation measures: (i) decisions on critical points, (ii) the concentrations and activities of receptors and proteins, (iii) the total axon length and number of simulations, (iv) the final topographic map and finally (v) the visual results in three dimensions. These metrics are then compared with experimental evidences.

Two groups of experiments were performed in order to evaluate our model The first was an independent evaluation of each element and mechanism, with which was possible to demonstrate the relevance of all components. The second was based on a biological system, the Drosophila midline. Using our model as a basis, it was possible to develop the first computational model of the midline crossing in Drosophila that focus all the decision points. This model was tested in three different pathfinding scenarios: normal, comm mutant and robo mutant, returning for all results that are coincident with experimental studies. In addition, new hypotheses arose with this model: first, all the parameters are hypotheses because none of them was quantified so far (to the best of our knowledge), second, a glial cell pattern with a stronger concentration at the center of the midline and third, the new evidence towards a inhibitory link between the DCC receptor and the Comm protein that can be Netrin-mediated or mediated by a third midline unknown signal. This last hypothesis is obtaining support from the neuroscience with two works published recently [99, 57]. The glial cell pattern reinforces the stereotropism hypothesis (see section 2.5.1) that states that the formation of tunnels of glial cells help AG.

All these hypotheses prove the usefulness of our model by providing clues that can lead to new studies in experimental and theoretical neuroscience.

To the best of our knowledge this is also a new theoretical model of midline crossing in Drosophila that explains what the system in three different scenarios.

The simulator allowed us to tune the several parameters and come up with a realistic model consistent with the literature for three different scenarios.

During the parameter tuning process several variations were done in which the axon was still able to navigate successfully, at least, part of the system. This level of flexibily reinforces our model, because experimental works usually state that despite the mutations a reasonable number of axons are still able to navigate with success, what is also truth in natural systems in general.

It is important to note that during the experiments the only changes made in the models are made before the simulation starts, i.e. these changes only affect the initial state. What differentiates the various types of axons are the receptors and proteins that they express and their initial positions. Considering this and the results obtained so far we suggest this phenomenon seems to have general principles that work for all type of axons.

The fact that a simple guidance model that maps directly the receptor activation into a turning angle can achieve realistic results in a simulation of a real system, may suggest that the extracellular environment has a much more important role in AG than the intracellular.

The high number of papers published last year about AG (more than 2000), reveals that this topic is far from been completely understood. More specifically, the number of papers related with AG and the drosophila midline is about 400 published in journals, what also reveals that even this apparently simple system has a lot of unknowns.

Inspired by this thesis Costa et al. [15] proposed a new multi-objective algorithm. This algorithm adapts Dijkstra algorithm to deal with environmental information such as natural spaces, pollution, traffic and weather that are modelled as exponential gradients. As in AG the gradients can be either attractive or repulsive and each gradient type has a certain force that is defined by the user preference for that specific type. Preliminary results demonstrate that this simple mechanism of attraction/repulsion lead to routes that avoid repulsive gradients and are attracted by the attractive ones, taking into account the surrounding gradients and the user profile.

During this thesis emerged many possible further improvements, here we present some of these improvements by dividing them into four categories: neuro-inspired, other models, artificial intelligence/complex system ideas and simulator features.

**Neuro-inspired** Although some of the most relevant elements to AG were included in our model, several new bio-inspired components can be added. Some of these elements are introduced in the following list (sorted by importance):

1. Branching in *decision points*, including back-branching and pruning [62];
2. Model lipid rafts and explicitly co-receptors, both should increase the GC sensibility.
3. Add channels (e.g. Calcium, Sodium and Fosfatum) and rules that affect intracellular concentrations;
4. Include extracellular molecular concentrations (e.g. sodium and calcium);

5. Model the internal state of glial cells;
6. Consider a diffusion function that has different characteristics depending on dimension (x,y or z);
7. Extend the adaptation mechanism to include axonal transport effects (production rate $\times$ axonal tranport or removal rate $\times$ axonal transport);
8. Include inhibitory interactions between ligands and receptors;
9. Differenciate the axonal transport into anterograde and retrogade transport. They have difference transport speeds, which could affect the the regulation between receptors and proteins;
10. Detect intersections between the different elements (e.g. between the GC and glial cells).

A great step to our model would be its application to study AG in the midline and the optic pathway of vertebrates (see section 2.6.2 for a brief description of these systems).

The experimental studies keep showing new evidences about the intracellular pathways that influences the GC decisions. Using this knowledge a new guidance model that considers the different molecular components and how these are changed by the extracellular gradients could be proposed, with or without explicit pathways. This set of rules that maps the receptor activation into changes in the intracellular environment is beginning to emerge.

**Other models** An extension of the receptor limits mechanism could be to include the minimum change detectable by the GC [31, 5] and a response to the saturarion threshold that decreases the GC reaction to a specific gradient, rather than just keeping it. Another one is to identify what could be the relevance of two new Robo receptors, Robo2 and Robo3 with repulsive effects in our midline model [32].

The bayesian model proposed by Mortimer et al. [66] for gradient detection was proved to be optimal. The effect of this gradient detection model could be added in our model as an alternative receptor distribution function. Furthermore, the effect on the turning angles of this and our distribuction function must be addressed. However should be considered that this Bayesian model is analytical and based on one dimension, which could lead to some adaptations. Other possible distribution functions that could be tested are: Winner-take-all, spatial average and symmetric [66].

The probabilistic computational model [33] proposed by Goodhill et al. to deal with the direction of filopodia could be also used as an alternative distribution function in order to obtain a smoother axon turning.

**Artificial Intelligence/Complex Systems** In mutants some axons are still able to find the correct path, how can such a level of flexibility be obtained? A study about the system dynamics in light of complex systems would be important to understand how sensible the system is to variations.

The manual optimization process done in order to find a model that works for the crossing of drosophila midline was difficult. Using an optimization technique inspired in the principles of natural evolution known as genetic algorithms is possible to automatically optimize complex systems towards a single or multiple objectives. In genetic algorithms these objectives are called fitness functions, that in this problem could be the euclidian difference to the real axon path.

A new guidance model could be develop based on computational intelligence approaches in order to learn the decisions made by the GC. The idea is to let the technique learn the GC decisions based on the receptors activation. Thus, the input is the activation force and the output the new direction (angle) that the axon should follows. A possible approach are evolutionary neural networks (e.g. NEAT [81]). The evolved network could then be compared with known intracellular pathways.

**Simulator features** Concerning yArbor further work should be done in order to make possible interfaces with other simulators such as Topographica or Neuron. A possible path is to export the topographic network developed in the yArbor and then study its organization and activity in Topographica. For interaction with other tools the language NeuroML (XML specification that support several computational neuroscience tools) could be used.

An easy interface with Python (one of the most used programming languages in Neuroscience), mediated by the Qt framework should also be implemented. In order to improve the performance, namely in the optimization process could be necessary to use high performance technologies such as OpenMP[1], Open MPI[2] or CUDA[3].

[1]http://openmp.org/wp
[2]http://www.open-mpi.org
[3]http://www.nvidia.com/cuda

# List of Figures

# List of Tables